\documentclass[preprint,12pt,3p]{elsarticle}

\usepackage{lineno,hyperref}
\usepackage{amsmath}
\usepackage{subfigure}
\usepackage{caption}
\usepackage{xcolor}
\usepackage{amssymb}
\usepackage{lipsum}
\usepackage{amsfonts}
\usepackage{graphicx}
\usepackage{epstopdf}
\usepackage{algorithmic}
\usepackage{bm}
\usepackage{mathrsfs}
\usepackage{amsthm}
\usepackage{algorithm}
\usepackage{algorithmic}
\usepackage{amsfonts}
\usepackage{soul}
\usepackage{mathtools}

\usepackage{color}
\newcommand{\HighlightA}{\color{black}}
\definecolor{DarkMagenta}{RGB}{139,0,139}
\newcommand{\HighlightB}{\color{DarkMagenta}\it}
\definecolor{DarkGreen}{RGB}{0,100,0}
\newcommand{\HighlightC}{\color{DarkGreen}\bf}
\definecolor{DarkBlue}{RGB}{0,0,139}
\newcommand{\HighlightD}{\color{DarkBlue}\it}
\newcommand{\HighlightE}{\color{blue}\bf}
\newcommand{\HighlightF}{\color{blue}}
\newcommand{\HighlightG}{\color{black}\bf}

\newtheorem{theorem}{Theorem}

\newtheorem{problem}[theorem]{Problem} 
  
\modulolinenumbers[5]

\journal{CMAME}

\biboptions{numbers,sort&compress}

\begin{document}

\begin{frontmatter}

\title{A monolithic one-velocity-field optimal control formulation for fluid-structure interaction problems with large solid deformation} 


\author[add0]{Yongxing Wang\corref{mycorrespondingauthor}}
\cortext[mycorrespondingauthor]{Corresponding author}
\ead{scsywan@leeds.ac.uk/yongxingwang6@gmail.com}


\address[add0]{School of Computing, University of Leeds, Leeds, UK.}

\begin{abstract}
In this article, we formulate a monolithic optimal control method for general time-dependent Fluid-Structure Interaction (FSI) systems with large solid deformation. We consider a displacement-tracking type of objective with a constraint of the solid velocity, tackle the time-dependent control problems by a piecewise-in-time control, cope with large solid displacement using a one-velocity fictitious domain method, and solve the fully-coupled FSI and the corresponding adjoint equations in a monolithic manner. We implement the proposed method in the open-source software package FreeFEM++ and assess it by three numerical experiments, in the aspects of stability of the numerical scheme for different regularisation parameters, and efficiency of reducing the objective function with control of the solid velocity.
\end{abstract}


\end{frontmatter}


\section{Introduction}
\label{sec_introduction}
Fluid-Structure Interaction (FSI) problems arise from aerodynamics \cite{morgenthal2000fluid,bazilevs2013computational,mohammadi2010applied}, ocean mechanics \cite{mccormick2009ocean,bai2009fully,finnegan2012numerical}, hemodynamics \cite{vcanic2014fluid,deparis2016facsi,piatti2015hemodynamic}, and so on. For most FSI problems, analytical solutions of the controlling equations are impossible to obtain, whereas laboratory experiments are complex, expensive and limited in scope. Therefore, numerical simulations play an important role in order to understand the fundamental physics involved in the complex interaction between fluids and structures. Computational methods for FSI problems have developed rapidly in past decade and reached a significant level of maturity. A brief review and broad categorisation of the exiting FSI methods can be based upon three questions: first, what kind of mesh do we use (one interface-fitted mesh, one interface-unfitted mesh or two meshes)? second, which variables do we solve (fluid velocity, pressure and solid displacement; or one velocity for both fluid and solid)? third, what type of coupling strategies do we use (monolithic/fully-coupled, or partitioned/segregated)? Therefore, a combination of the answers to these three questions would produce different types of numerical methods. In particular, we have classical partitioned/segregated methods \cite{kuttler2008fixed,Degroote_2009,bazilevs2013computational,degroote2013partitioned} using one interface-fitted mesh and solving for both the fluid velocity and solid displacement; monolithic methods \cite{Heil_2004,Heil_2008,Muddle_2012,Wang_2017,wang2020energy} using one interface-fitted mesh and solving for fluid velocity, pressure, solid displacement and a Lagrange multiplier to enforce the continuity at the fluid-solid interface; immerse methods \cite{peskin2002immersed,zhang2004immersed,baaijens2001fictitious} or fictitious domain methods \cite{Muddle_2012,Boffi_2016,Boffi_2015} use two meshes, the former solve for one velocity field, and the latter solve for the fluid velocity, pressure, solid displacement and a Lagrange multiplier; recent developed one-velocity methods \cite{Wang_2017,wang2020energy,Hecht_2017} solve for one-velocity field in a monolithic manner using either one interfaced-fitted mesh or two meshes; there are also fully Eulerian methods \cite{wick2013fully,Richter_2010,Rannacher_2010,schott2019monolithic} using one interface-unfitted mesh and solve for both velocity and displacement in a monolithic manner.

Optimal control is a branch of mathematical optimization which seeks to optimise an objective of a stationary or dynamical system by a control variable of the system \cite{troltzsch2010optimal}. We focus on reviewing a fluid dynamical system, in which case the objective could be reduction of the drag force by shape optimisation \cite{pironneau1974optimum,glowinski1975numerical,mohammadi2010applied,montenegro2015other,henrot2010optimal,dapogny2018geometrical,jenkins2016immersed,gunzburger2000shape} or by active turbulence control at the boundary layer \cite{mohammadi2010applied,choi1994active,kim2011physics,jeon2004active,dong2020influences,mcnally2015drag}; it could also be velocity tracking (or steering velocity) by controlling a distributed body force \cite{lions1988exact,hou1997dynamics,hou1997numerical,gunzburger1998computations,gunzburger2002perspectives,povsta2007optimal,attavino2017adjoint,manservisi2016numerical,gunzburger2000analysis,manservisi2016optimal} or boundary force \cite{gunzburger1991analysis,fattorini1992existence,gunzburger2000velocity,gunzburger2002perspectives,fursikov2005optimal,aulisa2006multigrid,attavino2017adjoint}; there are also objective of reducing vorticity \cite{abergel1990some,povsta2007optimal,attavino2017adjoint} or matching a turbulence kinetic energy \cite{manservisi2016numerical,attavino2017adjoint,manservisi2016optimal} by controlling a distributed body or boundary force. Velocity-tracking type of optimal distributed control has a rigorous mathematical theory for its solution existence \cite{abergel1990some,fattorini1992existence,gunzburger1991analysis,gunzburger2012flow}, and convergence and stability of the its numerical algorithm \cite{gunzburger2012flow,hou1997dynamics,hou1997numerical,gunzburger2000velocity}.

In the context of optimal control for fluid-structure systems, the research remains limited and publications can be found in the past two decades. The earlies study of FSI control could be found in \cite{moubachir2002optimal} where the sensitivity of a rigid body's movement inside a fluid with respect to a boundary velocity is analysed. This method was extended to another fluid-structure interaction case (fluid inside an elastic solid) in \cite{moubachir2006moving,bociu2013sensitivity}, which is a pioneering work for FSI control through shape analysis. A general quadratic objective functional is minimised by a boundary control for a fixed solid inside a fluid, and the well-posedness of this FSI control problem is established in \cite{lasiecka2008boundary,bucci2010optimal,lasiecka2009riccati}. A velocity-tracking objective is optimised by controlling a boundary pressure, and the optimal control problem is first formulated and solved using Newton method in \cite{richter2013optimal} for a static FSI problem, and latter extended to time-dependent case in \cite{wick2020optimization}. Recently, a linearised distributed FSI control problem is analysed, and the solution existence is proved in \cite{peralta2020analysis}. The existence of an optimal FSI control for the problem of minimizing flow turbulence, by controlling a distributed force, is established in \cite{bociu2015optimal}. In recent years, many studies of the optimal FSI control focus on numerical methods and implementations: monolithic formulation and Newton multigrid method are presented in \cite{failer2016optimal,failer2020newton}; a velocity-tracking objective is considered by controlling either a distributed body force \cite{chierici2019distributed} or a boundary pressure \cite{chirco2017optimal,chirco2019adjoint,chirco2020optimal}.

In this paper we shall apply the distributed control method to fluid-structure interaction problems, and study a displacement-tracking type of objective which has not been fully studied in literature, especially in the case of large solid deformation. To the best of the author's knowledge, up to now there is no publication concerning the control of large-deformed solid interacting with fluids. However, this displacement-tracking FSI control has potentials to be applied to accurate designing and controlling a range of biologically inspired robots, such as swimming robots (observing secretive sea life or carrying out a search-and-rescue mission \cite{crespi2008controlling}) or micro medical robots (crawling through the human body to perform an operation or deliver a medicine \cite{xiao2019classifications}).

The paper is organized as follows. The control Partial Differential Equations (PDE) for the FSI problem are introduced in Section \ref{sec_pde}, followed by time discretisation of these PDEs and the optimisation problem in Section \ref{sec_optimisation_problem}. The main deduction of the optimality system using the Lagrange multiplier method is presented in Section \ref{sec_lm}, and a monolithic scheme of the primal and adjoint equations is formulated in Section \ref{bounary_neumann}. Numerical experiments are carried out in Section \ref{sec_numerical_exs}, with conclusions drawn in Section \ref{sec_conclusion}.

\section{Control partial differential equations for the fluid-structure interaction problems}
\label{sec_pde}
\begin{figure}[bt]
	\centering
	\includegraphics[width=3.5in,angle=0]{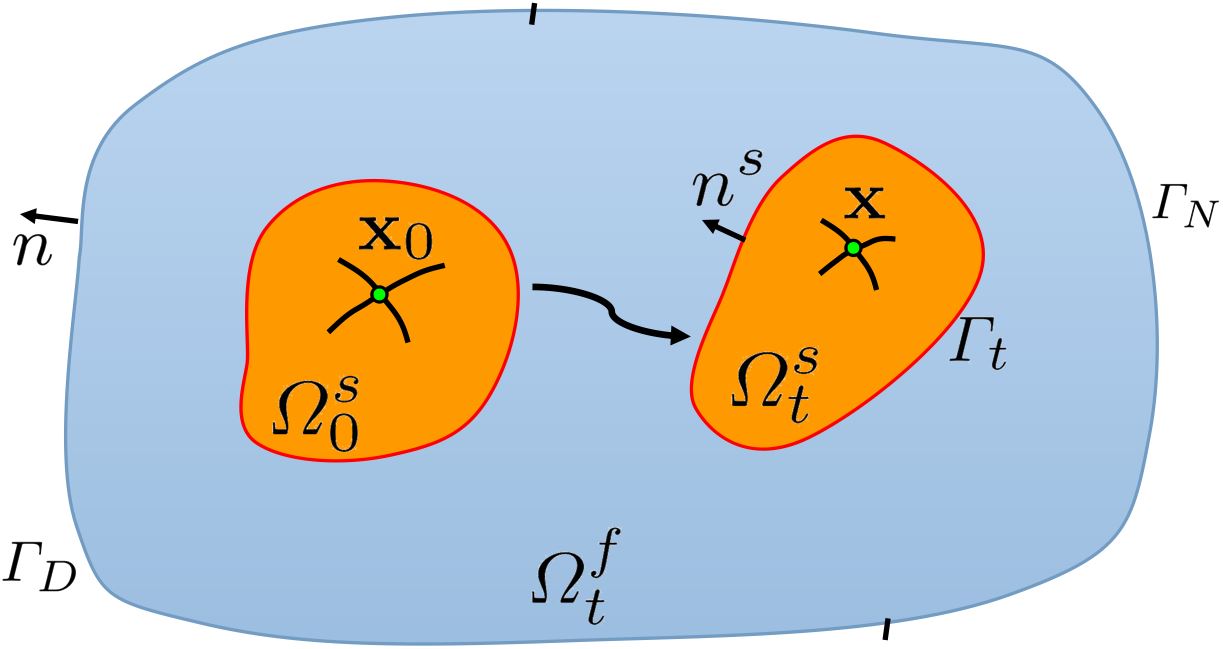}
	\captionsetup{justification=centering}
	\caption {\scriptsize A sketch of FSI problems in which $\Gamma_t=\overline{\Omega}_t^f \cap \overline{\Omega}_t^s$ and $\Gamma_D\cap\Gamma_N=\Gamma$.} 
	\label{fsi_diagram}
\end{figure}

We consider general fluid-structure interaction problems sketched in Figure \ref{fsi_diagram}, in which $\Omega_t^f\subset\mathbb{R}^d$ ($d=2,3$) and $\Omega_t^s\subset\mathbb{R}^d$ are the fluid and solid domain respectively (which are time dependent regions), and $\Gamma_t=\overline{\Omega}_t^f \cap \overline{\Omega}_t^s$ is the moving interface between the fluid and solid. The superscripts $f$ and $s$ denote fluid and solid respectively, and the subscript $t$ explicitly highlights when regions are time dependent. $\Omega=\overline{\Omega}_t^f \cup \overline{\Omega}_t^s$ is a fixed domain with an outer boundary $\Gamma=\Gamma_{D}+\Gamma_N$, where $\Gamma_D$ is the Dirichlet boundary and $\Gamma_N$ is the Neumann boundary on which the zero-normal stress is enforced in this article.

We consider both an incompressible fluid and an incompressible hyperelastic solid in this paper, and we shall only solve for one velocity field in the whole domain. The conservation of momentum and conservation of mass take the same form in the fluid and solid, which just differs in the specific expressions of the stress tensor. Therefore, it is convenient to introduce an indicator function ${\it 1}_{\omega}({\bf x})=1$ if ${\bf x}\in\omega$ and ${\it 1}_{\omega}({\bf x})=0$ otherwise, and let $\rho=\rho^f{\it 1}_{\Omega_t^f}+\rho^s{\it 1}_{\Omega_t^s}$, ${\bf u}={\bf u}^f{\it 1}_{\Omega_t^f}+{\bf u}^s{\it 1}_{\Omega_t^s}$, ${\bm{\sigma}}={\bm\sigma}^f{\it 1}_{\Omega_t^f}+{\bm\sigma}^s{\it 1}_{\Omega_t^s}$ denote the density, velocity vector and stress tensor respectively. The control partial differential equations, with initial and boundary conditions, for the FSI problem can then be expressed as follows.
\begin{eqnarray}
	\label{momentum_equation}
	\text{Momentum equation:} && \rho\frac{\partial{\bf u}}{\partial t}
	+\rho\left(\left({\bf u}-{\bf w}\right)\cdot\nabla\right){\bf u}
	-\nabla \cdot {\bm\sigma}
	={\bf f}{\it 1}_{\Omega_t^s},\\
	\label{continuity_equation}
	\text{Continuity equation:}  &&  \nabla \cdot {\bf u}=0,\\
	\label{initial_condition}
	\text{Initial condition:} && \left. {\bf u}\right|_{t=0}={\bf u}_0, \\
	\label{boundary_dirichlet}
	\text{Dirichlet BC:} && \left.{\bf u}\right|_{\Gamma_D}=\bar{\bf u}{\it 1}_{\Gamma_{D}}, \\
	\label{bounary_neumann}
	\text{Neumann BC:} && \left.{\bm \sigma}{\bf n}\right|_{\Gamma_N}={\bf 0}, \\
	\label{interface_velocity}
	\text{Continuity of velocity:} && \left.\left({\bf u}^s-{\bf u}^f\right)\right|_{\Gamma_t}={\bf 0}, \\
	\label{interface_stress}
	\text{Continuity of normal stress:} && \left.\left({\bm\sigma}^s-{\bm\sigma}^f\right){\bf n}^s\right|_{\Gamma_t}={\bf 0}.
\end{eqnarray}
We shall use the body force ${\bf f}$ in (\ref{momentum_equation}) as a control variable in the following. The stress tensor of an incompressible Newtonian flow is expressed as:
\begin{equation} \label{constitutive_fluid}
	{\bm\sigma}^f=\mu^f{\rm D}{\bf u}^f-p^f{\bf I},
\end{equation}
with ${\rm D}(\cdot)=\nabla(\cdot)+\nabla^{\scriptsize T}(\cdot)$, and  $\mu^f$ being the viscosity parameter. The stress tensor of an incompressible neo-Hookean solid is expressed as \cite{Hecht_2017,wang2020energy}:
\begin{equation} \label{constitutive_solid}
	{\bm\sigma}^s=c_1\left({\rm D}{\bf d}-\nabla^T{\bf d}\nabla{\bf d}\right)-p^s{\bf I},
\end{equation}
where $c_1$ is the elasticity parameter and ${\bf d}$ is the solid displacement. Notice that although the solid stress tensor is expressed as a function of displacement ${\bf d}$, we shall not solve for ${\bf d}$ as an independent variable. Instead we view it as a function of velocity, and solve the whole FSI problem based upon a one-field-velocity method \cite{Wang_2017}. In the above equation (\ref{momentum_equation}), ${\bf w}$ is an arbitrary velocity field of the moving frame in order to describe the FSI system; ${\bf w}={\bf 0}$ in the case of Eulerian description and ${\bf w}={\bf u}$ in the case of Lagrangian description. In the following sections, we shall use the Eulerian description for the background fluid (including the fictitious fluid covered by the solid domain) and Lagrangian description for the moving solid.

\section{Time discretisation and the piecewise control problem}
\label{sec_optimisation_problem}
In order to introduce the piecewise-in-time control problem, we first disretise the control PDEs at a sequence of time points: $t_0=0, t_1, t_2 \ldots$, with $t_{n+1}-t_n=\Delta t$ ($n\in\mathbb{N}_0$ is a non-negative integer). We then solve for ${\bf u}_{n+1}$, ${\bf d}_{n+1}$ and ${\bm\sigma}_{n+1}$ given ${\bf u}_n$, ${\bf d}_n$ and ${\bm\sigma}_n$ using the backward-Euler scheme. The disretised momentum equation (\ref{momentum_equation}) is
\begin{equation}\label{momentum_discretisation}
	\rho\frac{{\bf u}_{n+1}-{\bf u}_n}{\Delta t}
	+\rho\left(\left({\bf u}_{n+1}-{\bf w}_{n+1}\right)\cdot\nabla\right){\bf u}_{n+1}
	-\nabla \cdot {\bm\sigma}_{n+1},
	={\bf f}_{n+1},
\end{equation}
and the disretised version of all the equations from (\ref{continuity_equation}) to (\ref{interface_stress}) would take the same form except introducing a subscript $n+1$ to corresponding variables. Therefore, it is convenient to omit the subscript $n+1$ in the rest of this paper. We shall focus on computing ${\bf u}={\bf u}_{n+1}$ given ${\bf u}_n$ in the time interval $\left[t_n, t_{n+1}\right]$, on which we shall also formulate an optimal control problem. Notice that ${\bf d}_{n+1}$ and ${\bm\sigma}_{n+1}$ are not explicit unknowns based upon the one-velocity-field formulation we shall introduce.

Let $L^2(\omega)$ be the square integrable functions in domain $\omega$ with inner product $\left(u,v\right)_\omega=\left(\int_\omega uvdx\right)$,  $\forall u, v\in L^2(\omega)$, and the induced norm $\|v\|_{L^2\left(\omega\right)}=\left(v,v\right)_\omega^{1/2}$, $\forall v\in L^2(\omega)$. For a vector function ${\bf v}\in L^2(\omega)^d$, the norm is defined component-wise as $\|{\bf v}\|_{L^2\left(\omega\right)^d}^2=\sum_{i=1}^{d}\|v_i\|_{L^2\left(\omega\right)}^2$. Then let $H^1(\omega)=\left\{v: v, \nabla v\in L^2(\omega)^d\right\}$, and denote by $H_{u\left(\gamma\right)}^1(\omega)$ the subspace of $H^1(\omega)$, which has the boundary data ${u}$ on $\gamma$. We also denote by $L_0^2(\Omega)$ the subspace of $L^2(\Omega)$ whose functions have zero mean values.

We consider the following optimisation problem: reducing the discrepancy between the solid displacement ${\bf d}$ and an objective displacement ${\bf d}_g$ profile, with constraint of the solid velocity, by controlling a distributed force ${\bf f}$ on the solid body. 

\begin{problem}[piecewise-in-time control]\label{problem_optimisation}
	Given an objective displacement profile ${\bf d}_g(t)$ and objective velocity norm $u_c(t)$ of the solid,
	\begin{equation}\label{original_optimisation_problem}
		\underset{{\bf f}\in L^2\left(\Omega_n^s\right)^d}{\text{minimise}} \quad 
		J({\bf u},{\bf f})
		=\frac{1}{2}\|{\bf d}-{\bf d}_g(t_{n+1})\|_{\Omega_n^s}^2
		+\frac{\alpha}{2}\|{\bf f}\|_{\Omega_n^s}^2,
	\end{equation}
	subject to 
	\begin{equation}\label{inequality_constraint}
		\|{\bf u}\|_{\Omega_n^s} \le u_c(t_{n+1}),
	\end{equation}
	and equations (\ref{momentum_discretisation}); (\ref{continuity_equation}) to (\ref{interface_stress}) after time discretisation (omitting the subscript $n+1$ of ${\bf u}_{n+1}$ and ${\bf d}_{n+1}$ for notation simplicity).
\end{problem}
In the above Problem \ref{problem_optimisation}, we consider an optimisation problem integrated in the old time domain $\Omega_n^s$, and we shall also solve our FSI problem using this explicit formulation. It is not significant to iteratively construct $\Omega_{n+1}^s$ and perform integration on it using a small time step as pointed out in \cite{Hecht_2017,wang2020energy}. The first term in (\ref{original_optimisation_problem}) is the real objective to be minimised, and the second term is a regularisation term with a regularisation parameter $\alpha$. A too large $\alpha$ would make it difficult to achieve the real objective, while a too small $\alpha$ may cause convergence issues for the numerical scheme. The inequality constraint (\ref{inequality_constraint}) provides an upper bound for the magnitude of the velocity.

\section{The Lagrange multiplier method}
\label{sec_lm}
In this section, we introduce the Lagrange multipliers (or adjoint variables) $\hat{\bf u}\in H_{0\left(\Gamma_D\right)}^1(\Omega)$ and $\hat{p}\in L^2(\Omega)$  to eliminate the equality constraints of Problem \ref{problem_optimisation}. For the inequality constraint (\ref{inequality_constraint}) we simply introduce a penalty (or barrier) parameter $\lambda$ to be included in the Lagrangian functional. Other methods, such as active-set or trust-region algorithm \cite{bertsekas2014constrained,el2018active}, may be used to deal with inequality constraints as well, which however would not be the main focus of this paper.
\begin{equation}\label{lagrangian_functional_0}
	\begin{split}
		& L\left({\bf u}, p, \hat{\bf u}, \hat{p}, {\bf f}\right) 
		= J({\bf u},{\bf f}) +\frac{\lambda}{u_c^2-\|{\bf u}\|_{\Omega_{n}^s}^2}\\
		+&\rho^f\int_{\Omega_{n}^f}\frac{{\bf u}-{\bf u}_n}{\Delta t}\cdot\hat{\bf u}
		+\rho^s\int_{\Omega_{n}^s}\frac{{\bf u}-{\bf u}_n}{\Delta t}\cdot\hat{\bf u} \\
		+&\rho^f\int_{\Omega_{n}^f}\left(\left({\bf u}-{\bf w}\right)\cdot\nabla\right){\bf u}\cdot\hat{\bf u}
		+\rho^s\int_{\Omega_{n}^s}\left(\left({\bf u}-{\bf w}\right)\cdot\nabla\right){\bf u}\cdot\hat{\bf u}\\
		-&\int_{\Omega_{n}^f}\left(\nabla\cdot{\bm{\sigma}^f}\right)\cdot\hat{\bf u} 
		-\int_{\Omega_{n}^s}\left(\nabla\cdot{\bm{\sigma}^s}\right)\cdot\hat{\bf u} \\
		-&\int_{\Omega} \hat{p}\nabla \cdot {\bf u}
		- \int_{\Omega_n^s}{\bf f}\cdot\hat{\bf u}
		+\int_{\Gamma_{n}}\left({\bm\sigma}^s-{\bm\sigma}^f\right){\bf n}^s\cdot\hat{\bf u}.
	\end{split}
\end{equation}
We integrate stress term by part and the last term in (\ref{lagrangian_functional_0}) would be cancelled out thanks to the interface condition (\ref{interface_stress}). We also rearrange all the integrals such that the integrations only exist in the whole domain $\Omega$ and the solid domain  $\Omega_n^s$. In this case, we shall use an Eulerian framework to describe the background fluid (including the fictitious fluid covered by the solid domain) on $\Omega$ and an updated Lagrangian framework to describe the solid on $\Omega_n^s$, i.e.: ${\bf w}={\bf 0}$ on $\Omega$ and ${\bf w}={\bf u}$ on $\Omega_n^s$. Substituting the constitutive equations (\ref{constitutive_fluid}) and (\ref{constitutive_solid}) into (\ref{lagrangian_functional_0}) and discretising the solid displacement ${\bf d}={\bf d}_n+\Delta t{\bf u}$, equation (\ref{lagrangian_functional_0}) can then be expressed as: 
\begin{equation}\label{lagrangian_functional}
	\begin{split}
		& L\left({\bf u}, p, \hat{\bf u}, \hat{p}, {\bf f}\right) 
		= J({\bf u},{\bf f})  +\frac{\lambda}{u_c^2-\|{\bf u}\|_{\Omega_{n}^s}^2}\\
		+&\rho^f\int_{\Omega}\frac{{\bf u}-{\bf u}_n}{\Delta t}\cdot\hat{\bf u}
		+(\rho^s-\rho^f)\int_{\Omega_{n}^s}\frac{{\bf u}-{\bf u}_n}{\Delta t}\cdot\hat{\bf u} \\
		+&\frac{\mu^f}{2}\int_{\Omega}{\rm D}{\bf u}:{\rm D}\hat{\bf u} 
		+\frac{\Delta tc_1-\mu^f}{2}\int_{\Omega_{n}^s}{\rm D}{\bf u}:{\rm D}\hat{\bf u} 
		+\frac{c_1}{2}\int_{\Omega_{n}^s}{\rm D}{\bf d}_n:{\rm D}\hat{\bf u}\\
		+&\rho^f\int_{\Omega}\left({\bf u}\cdot\nabla\right){\bf u}\cdot\hat{\bf u}
		-\int_{\Omega}p\nabla\cdot \hat{\bf u}
		-\int_{\Omega} \hat{p}\nabla \cdot {\bf u}
		- \int_{\Omega_n^s}{\bf f}\cdot\hat{\bf u}\\
		-& c_1\Delta t\int_{\Omega_{n}^s}\left(\nabla^T{\bf u}\nabla{\bf d}_n+\nabla^T{\bf d}_n\nabla{\bf u}\right):\nabla\hat{\bf u}
		-c_1\int_{\Omega_{n}^s}\nabla^T{\bf d}_n\nabla{\bf d}_n:\nabla\hat{\bf u}.
	\end{split}
\end{equation}
The following Karush-Kuhn-Tucker (KKT) conditions are the first-order necessary conditions in order to minimise (\ref{lagrangian_functional}):
\begin{eqnarray}
	\label{optimality_1}
	\frac{\partial{L}\left({\bf u}, {p}, {\bf f}, \hat{\bf u}, \hat{p} \right)}{\partial \left(\hat{\bf u}, \hat{p}\right)}\left[\delta\hat{\bf u}, \delta\hat{p}\right]=0, \\
	\label{optimality_2}
	\frac{\partial{L}\left({\bf u}, {p}, {\bf f}, \hat{\bf u}, \hat{p} \right)}{\partial \left({\bf u}, {p}\right)}\left[\delta{\bf u}, \delta{p}\right]=0, \\
	\label{optimality_3}
	\frac{\partial{L}\left({\bf u}, {p}, {\bf f}, \hat{\bf u}, \hat{p} \right)}{\partial {\bf f}}\left[\delta{\bf f}\right]=0,
\end{eqnarray}
where
\begin{equation}
	\frac{\partial L(\cdot)}{\partial{\bf q}}[\delta{\bf q}]=\left.\frac{d}{d\epsilon}L\left({\bf q}+\epsilon\delta{\bf q}\right)\right|_{\epsilon=0}
\end{equation} 
is the G${\rm\hat{a}}$teaux derivative with respective to variable ${\bf q}$ along the direction $\delta{\bf q}$ \cite{bazilevs2013computational,rall2014nonlinear}.

\subsection{Primal equation}
\label{section_discretization}
The optimality condition (\ref{optimality_1}) gives us the primal equation in a weak form as follows. Given ${\bf u}_n$ and ${\bf d}_n$, find ${\bf u}\in H_{{\bar{\bf u}}\left(\Gamma_D\right)}^1(\Omega)^d$ and $p\in L_0^2(\Omega)$, such that $\forall \delta\hat{\bf u}\in {H}_{0\left(\Gamma_D\right)}^1(\Omega)^d$ and $\forall \delta\hat{p}\in L^2(\Omega)$:
\begin{equation}\label{weak_form_state_time}
	\begin{split}
		&\rho^f\int_{\Omega}\frac{{\bf u}-{\bf u}_n}{\Delta t}\cdot\delta\hat{\bf u}
		+(\rho^s-\rho^f)\int_{\Omega_n^s}\frac{{\bf u}-{\bf u}_n}{\Delta t}\cdot\delta\hat{\bf u} \\
		+&\frac{\mu^f}{2}\int_{\Omega}{\rm D}{\bf u}:{\rm D}\delta\hat{\bf u} 
		+\frac{\Delta tc_1-\mu^f}{2}\int_{\Omega_n}{\rm D}{\bf u}:{\rm D}\delta\hat{\bf u} \\
		+&\rho^f\int_{\Omega}\left({\bf u}\cdot\nabla\right){\bf u}\cdot\delta\hat{\bf u}
		-\int_{\Omega}p\nabla\cdot \delta\hat{\bf u}
		-\int_{\Omega} \delta\hat{p}\nabla \cdot {\bf u}\\
		-& c_1\Delta t\int_{\Omega_{n}^s}\left(\nabla^T{\bf u}\nabla{\bf d}_n+\nabla^T{\bf d}_n\nabla{\bf u}\right):\nabla \delta\hat{\bf u}\\
		=& 
		\int_{\Omega_n^s}{\bf f}\cdot\delta\hat{\bf u}
		+c_1\int_{\Omega_{n}^s}\nabla^T{\bf d}_n\nabla{\bf d}_n:\nabla\delta\hat{\bf u}
		-\frac{c_1}{2}\int_{\Omega_{n}^s}{\rm D}{\bf d}_n:{\rm D}\delta\hat{\bf u}.
	\end{split}
\end{equation}
The solid domain is updated by $\Omega_{n+1}^s=\left\{{\bf x}: {\bf x}={\bf x}_n+\Delta t{\bf u}, \forall {\bf x}_n\in\Omega_{n}^s\right\}$ after solving the above primal equation.

\subsection{Adjoint equation}\label{sec_adjoint}
The optimality condition (\ref{optimality_2}) gives us the adjoint equation in a weak form as follows. Given ${\bf u}$ and ${\bf d}_n$, find $\hat{\bf u}\in H_{0\left(\Gamma_D\right)}^1(\Omega)^d$ and $\hat{p}\in L_0^2(\Omega)$, such that $\forall \delta{\bf u}\in {H}_{0\left(\Gamma_D\right)}^1(\Omega)^d$ and $\forall \delta{p}\in L^2(\Omega)$:
\begin{equation}\label{weak_form_dual_time_up}
	\begin{split}
		& \Delta t\int_{\Omega_n}\left({\bf d}-{\bf d}_g\right)\cdot\delta{\bf u}
		+2\lambda\int_{\Omega_n^s}{\bf u}\cdot\delta{\bf u}/\left(\|{\bf u}\|_{\Omega_n^s}^2-u_g^2(t_{n+1})\right)^2\\
		+&\frac{\rho^f}{\Delta t}\int_{\Omega}\delta{\bf u}\cdot\hat{\bf u}
		+\frac{\rho^s-\rho^f}{\Delta t}\int_{\Omega_n^s}\delta{\bf u}\cdot\hat{\bf u} \\
		+&\frac{\mu^f}{2}\int_{\Omega}{\rm D}\delta{\bf u}:{\rm D}\hat{\bf u} 
		+\frac{\Delta tc_1-\mu^f}{2}\int_{\Omega_n}{\rm D}\delta{\bf u}:{\rm D}\hat{\bf u} \\
		+&\rho^f\int_{\Omega}\left(\delta{\bf u}\cdot\nabla\right){\bf u}\cdot\hat{\bf u}
		+\rho^f\int_{\Omega}\left({\bf u}\cdot\nabla\right)\delta{\bf u}\cdot\hat{\bf u}\\
		-&\int_{\Omega}\delta p\nabla\cdot \hat{\bf u}
		-\int_{\Omega} \hat{p}\nabla \cdot \delta{\bf u} \\
		-& c_1\Delta t\int_{\Omega_{n}^s}\left(\nabla^T \delta{\bf u}\nabla{\bf d}_n+\nabla^T{\bf d}_n \nabla\delta{\bf u}\right):\nabla \hat{\bf u}
		=0.
	\end{split}
\end{equation}
In the above, the first order variation of the displacement ${\bf d}$ is approximated as $\delta{\bf d}=\Delta t\delta{\bf u}$.
\subsection{Optimality equation}
The optimality condition (\ref{optimality_3}) gives the relation between the control force and adjoint variable:
\begin{equation}\label{optimality_condition_time}
	\alpha\int_{\Omega_n^s}\delta{\bf f}\cdot{\bf f}
	=\int_{\Omega_n^s}\delta{\bf f}\cdot\hat{\bf u}.
\end{equation}
\section{A monolithic scheme}
\label{sec_monolithic}
Substituting the optimality condition (\ref{optimality_condition_time}) into equation (\ref{weak_form_state_time}), we have a monolithic scheme:
\begin{problem}[monolithic formulation for FSI control]\label{problem_monlithic}
Given ${\bf u}_n$ and ${\bf d}_n$, find ${\bf u}\in H_{{\bar{\bf u}}\left(\Gamma_D\right)}^1(\Omega)^d$, $\hat{\bf u}\in H_{0\left(\Gamma_D\right)}^1(\Omega)^d$ and $p$, $\hat{p}\in L_0^2(\Omega)$, such that $\forall \delta{\bf u}$, $\delta\hat{\bf u} \in {H}_{0}^1(\Omega)^d$ and $\forall \delta{p}$, $\delta\hat{p}\in L^2(\Omega)$:
\begin{equation}\label{monolithic_scheme}
\begin{split}
&\frac{\rho^f}{\Delta t}\int_{\Omega}\left({\bf u}\cdot\delta\hat{\bf u}+\delta{\bf u}\cdot\hat{\bf u}\right)
+\frac{\rho^s-\rho^f}{\Delta t}\int_{\Omega_n^s}\left({\bf u}\cdot\delta\hat{\bf u}+\delta{\bf u}\cdot\hat{\bf u}\right) \\
+&\frac{\mu^f}{2}\int_{\Omega}\left({\rm D}{\bf u}:{\rm D}\delta\hat{\bf u}+{\rm D}\delta{\bf u}:{\rm D}\hat{\bf u}\right) 
+\frac{\Delta tc_1-\mu^f}{2}\int_{\Omega_n^s}\left({\rm D}{\bf u}:{\rm D}\delta\hat{\bf u}+{\rm D}\delta{\bf u}:{\rm D}\hat{\bf u}\right)  \\
+&\rho^f\int_{\Omega}\left[\left({\bf u}\cdot\nabla\right){\bf u}\cdot\delta\hat{\bf u}
+\left(\delta{\bf u}\cdot\nabla\right){\bf u}\cdot\hat{\bf u}
+\left({\bf u}\cdot\nabla\right)\delta{\bf u}\cdot\hat{\bf u}\right] \\
-&\int_{\Omega}p\nabla\cdot \delta\hat{\bf u}
-\int_{\Omega} \delta\hat{p}\nabla \cdot {\bf u}
-\int_{\Omega}\delta p\nabla\cdot \hat{\bf u}
-\int_{\Omega} \hat{p}\nabla \cdot \delta{\bf u}\\
-& c_1\Delta t\int_{\Omega_{n}^s}\left[\left(\nabla^T{\bf u}\nabla{\bf d}_n+\nabla^T{\bf d}_n\nabla{\bf u}\right):\nabla \delta\hat{\bf u}+\left(\nabla^T \delta{\bf u}\nabla{\bf d}_n+\nabla^T{\bf d}_n \nabla\delta{\bf u}\right):\nabla \hat{\bf u}\right]\\
-& \frac{1}{\alpha}\int_{\Omega_n^s}\hat{\bf u}\cdot\delta\hat{\bf u}
+ 2\lambda\int_{\Omega_n^s}{\bf u}\cdot\delta{\bf u}/\left(\|{\bf u}\|_{\Omega_n^s}^2-u_g^2(t_{n+1})\right)^2 \\
= & \frac{\rho^f}{\Delta t}\int_{\Omega}{\bf u}_n\cdot\delta\hat{\bf u}
+\frac{\rho^s-\rho^f}{\Delta t}\int_{\Omega_n^s}{\bf u}_n\cdot\delta\hat{\bf u} \\
+& c_1\int_{\Omega_{n}^s}\nabla^T{\bf d}_n\nabla{\bf d}_n:\nabla\delta\hat{\bf u}
-\frac{c_1}{2}\int_{\Omega_{n}^s}{\rm D}{\bf d}_n:{\rm D}\delta\hat{\bf u}
-\Delta t\int_{\Omega_n^s}\left({\bf d}_n-{\bf d}_g\right)\cdot\delta{\bf u}.
\end{split}
\end{equation}
\end{problem}
We use the mixed finite elements $\left(P_2, P_2, P_1, P_1\right)$ to disretise space $\left(H^1, H^1, L^2, L^2 \right)$ of the solution pair ${\bf z}=\left({\bf u}, \hat{\bf u}, p, \hat{p}\right)$. Based upon the fictitious domain method, an Eulerian mesh is used to discretise the integrations in the augmented fluid domain $\Omega$, and an updated Lagrangian mesh to discretise the integrations in the moving solid domain $\Omega_n^s$. We then have the following linear equation system after space discretisation:
\begin{equation}\label{space_discretisation}
\left({\bf K}+{\bf P}^T{\bf K}^s{\bf P}\right){\bf z}={\bf g}+{\bf P}^T{\bf g}^s,
\end{equation}
where ${\bf K}$ and ${\bf K}^s$ are the system matrices from discretisation of the integrations, on the left-hand side of equation (\ref{monolithic_scheme}), in domain $\Omega$ and $\Omega_n^s$ respectively, and ${\bf g}$ and ${\bf g}^s$ are vectors from discretisation of the integrations, on the right-hand side of equation (\ref{monolithic_scheme}), in domain $\Omega$ and $\Omega_n^s$ respectively. ${\bf P}$ is the finite element interpolation matrix from the background mesh and to solid mesh. Notice that the proposed monolithic scheme has similar features with the our previous one-field monolithic fictitious domain method for FSI problems \cite{Wang_2017}. In this paper, we develop the previous monolithic scheme to include both the state variables $({\bf u}, p)$ and the adjoint variables $(\hat{\bf u}, \hat{p})$ in order to solve FSI control problems with large solid deformation.

In the rest of this section, we present a reduced version the above monolithic formulation in order to solve a pure flow control problem: a monolithic method for velocity-tracking type of flow control by a body force in $\Omega$. This can be achieved by first, replacing the last term in (\ref{monolithic_scheme}) by a velocity objective: $\int_{\Omega}\left({\bf u}-{\bf u}_g\right)\cdot\delta{\bf u}$; second, changing the domain of integration of term $\frac{1}{\alpha}\int_{\Omega_n^s}\hat{\bf u}\cdot\delta\hat{\bf u}$ in (\ref{monolithic_scheme}) to $\Omega$, which is related to the control force; third, removing all the other integrations in the solid domain in equation (\ref{monolithic_scheme}) (correspondingly solid matrix ${\bf K}^s$ and vector ${\bf g}^s$ in (\ref{space_discretisation})). Finally, we have the following monolithic formulation for a flow control problem.

\begin{problem}[monolithic formulation for flow control]\label{problem_monlithic_fluid}
Given ${\bf u}_n$, find ${\bf u}\in H_{{\bar{\bf u}}\left(\Gamma_D\right)}^1(\Omega)^d$, $\hat{\bf u}\in H_{0\left(\Gamma_D\right)}^1(\Omega)^d$ and $p$, $\hat{p}\in L_0^2(\Omega)$, such that $\forall \delta{\bf u}$, $\delta\hat{\bf u} \in {H}_{0}^1(\Omega)^d$ and $\forall \delta{p}$, $\delta\hat{p}\in L^2(\Omega)$:
\begin{equation}\label{monolithic_scheme_fluid}
\begin{split}
&\frac{\rho^f}{\Delta t}\int_{\Omega}\left({\bf u}\cdot\delta\hat{\bf u}+\delta{\bf u}\cdot\hat{\bf u}\right) 
+\frac{\mu^f}{2}\int_{\Omega}\left({\rm D}{\bf u}:{\rm D}\delta\hat{\bf u}+{\rm D}\delta{\bf u}:{\rm D}\hat{\bf u}\right) \\
+&\rho^f\int_{\Omega}\left[\left({\bf u}\cdot\nabla\right){\bf u}\cdot\delta\hat{\bf u}
+\left(\delta{\bf u}\cdot\nabla\right){\bf u}\cdot\hat{\bf u}
+\left({\bf u}\cdot\nabla\right)\delta{\bf u}\cdot\hat{\bf u}\right] \\
-&\int_{\Omega}p\nabla\cdot \delta\hat{\bf u}
-\int_{\Omega} \delta\hat{p}\nabla \cdot {\bf u}
-\int_{\Omega}\delta p\nabla\cdot \hat{\bf u}
-\int_{\Omega} \hat{p}\nabla \cdot \delta{\bf u}\\
-& \frac{1}{\alpha}\int_{\Omega}\hat{\bf u}\cdot\delta\hat{\bf u}
+\int_{\Omega}{\bf u}\cdot\delta{\bf u}
= \frac{\rho^f}{\Delta t}\int_{\Omega}{\bf u}_n\cdot\delta\hat{\bf u} 
+\int_{\Omega}{\bf u}_g\cdot\delta{\bf u}.
\end{split}
\end{equation}
\end{problem}

\section{Numerical experiments}
\label{sec_numerical_exs}
In this section, we assess and validate the proposed method using three numerical tests implemented using FreeFEM++ \cite{MR3043640}. We first validate the scheme using a flow control problem which is widely studied in literature. The second numerical test is a benchmark FSI problem whose controllability is studied by an ALE formulation in \cite{wangoptimal2021}, and we will show that the proposed two-mesh method will achieve the same goal of reduction of the objective. Our third numerical experiment involves controlling a large-deformed solid; this problem is widely studied as a forward FSI problem in literature, which however has not been considered as an inverse control problem up to now. We hope our result will provide a potential benchmark for other researches in the area of optimal FSI control in the future.
\subsection{Cavity flow}
\label{sec_cavity_intial}
In this example, we solve the reduced version of the proposed monolithic scheme formulated in Problem \ref{problem_monlithic_fluid}, and we consider control of a dynamic cavity pure fluid flow: steering the velocity to be a complicated predefined velocity profile with some vortices, which was studied in \cite{hou1997dynamics,hou1997numerical,gunzburger1998computations,gunzburger2000velocity}. We shall demonstrate that the proposed monolithic scheme can efficiently and accurately tracking the fluid field for a long time, with many vortices being developed (previous publications studied the case of less vortices). The computational domain is a $[0,1]\times[0,1]$ unit square. A wall boundary condition is prescribed for all the four sides of the cavity, and the fluid with $\rho^f=1$ and $\mu^f=0.1$ is initially stationary.
The goal velocity
\begin{equation}
	{\bf u}_g(x, y, t)=\left(\frac{\partial}{\partial y}\Psi(x,y,t), -\frac{\partial}{\partial x}\Psi(x,y,t)\right),
\end{equation}
is derived from the following stream function:
\begin{equation}
	\Psi(x, y, t)=\psi(x,t)\psi(y,t)
\end{equation}
with
\begin{equation}
	\psi(s,t)=\left(1-s\right)^2\left(1-cos(4\pi st)\right), \quad s\in [0, 1].
\end{equation}

\begin{figure}[h!]
	\centering  
	\includegraphics[width=2.5in,angle=0]{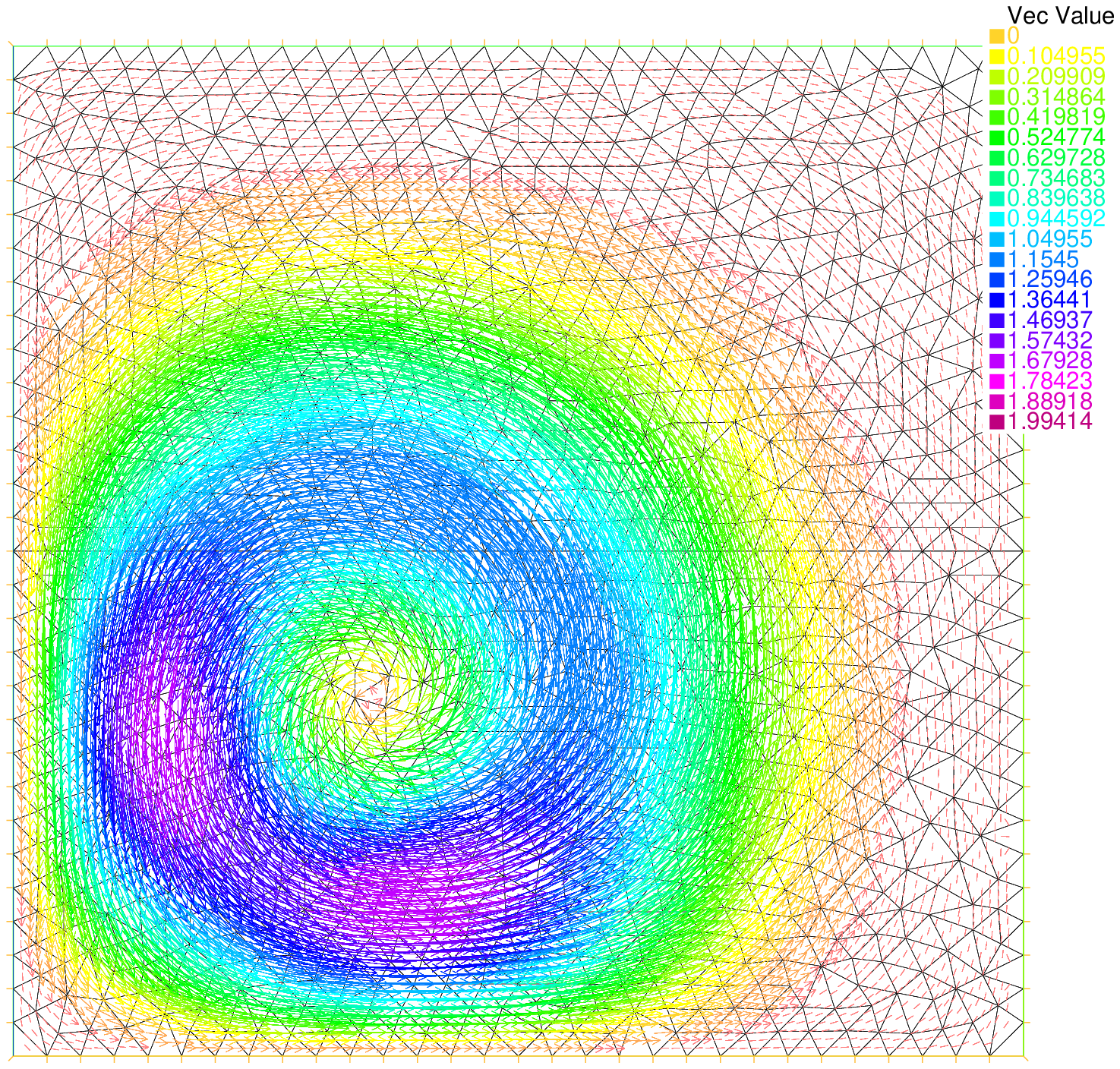}	
	\includegraphics[width=2.5in,angle=0]{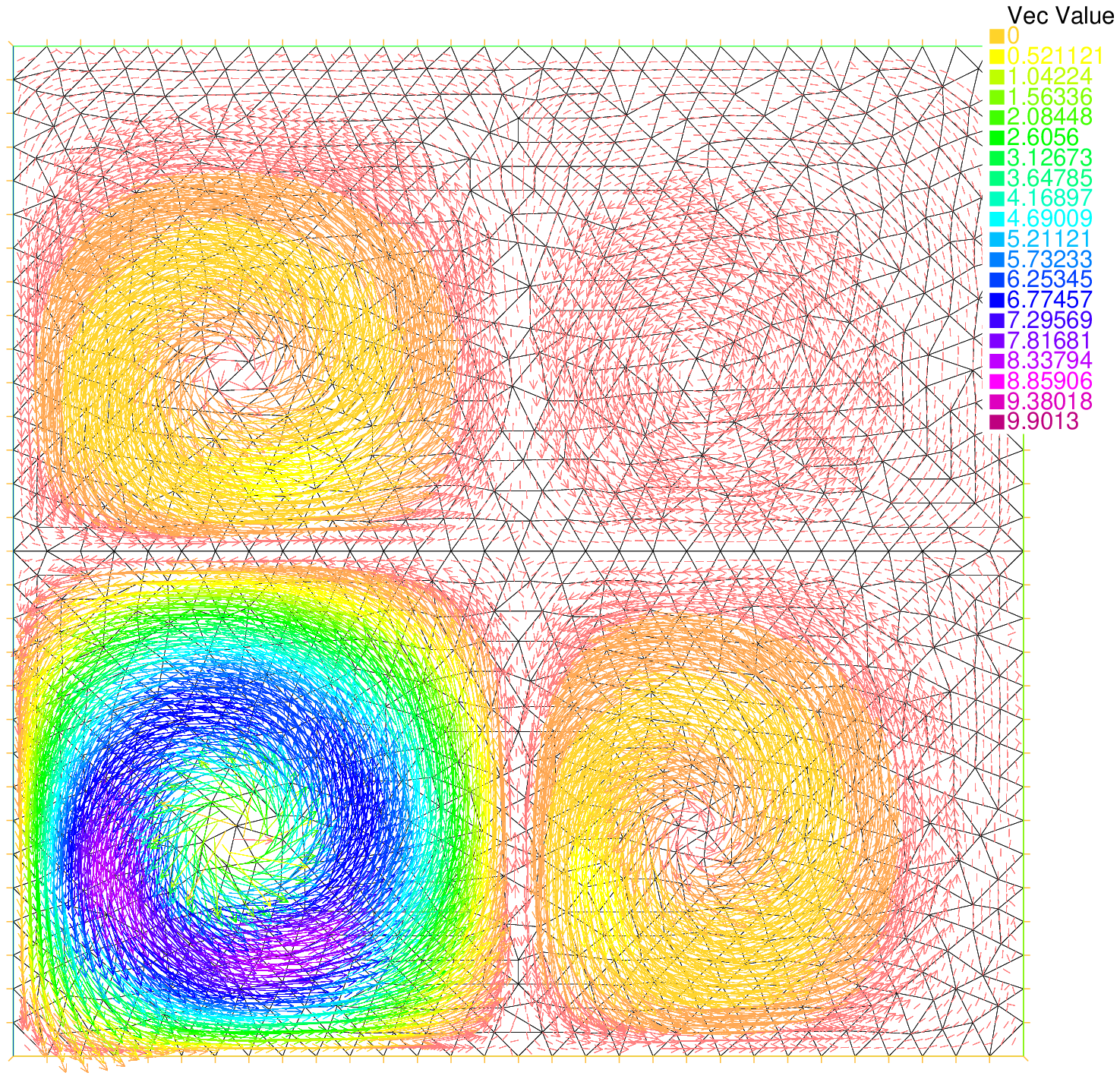}	
	\includegraphics[width=2.5in,angle=0]{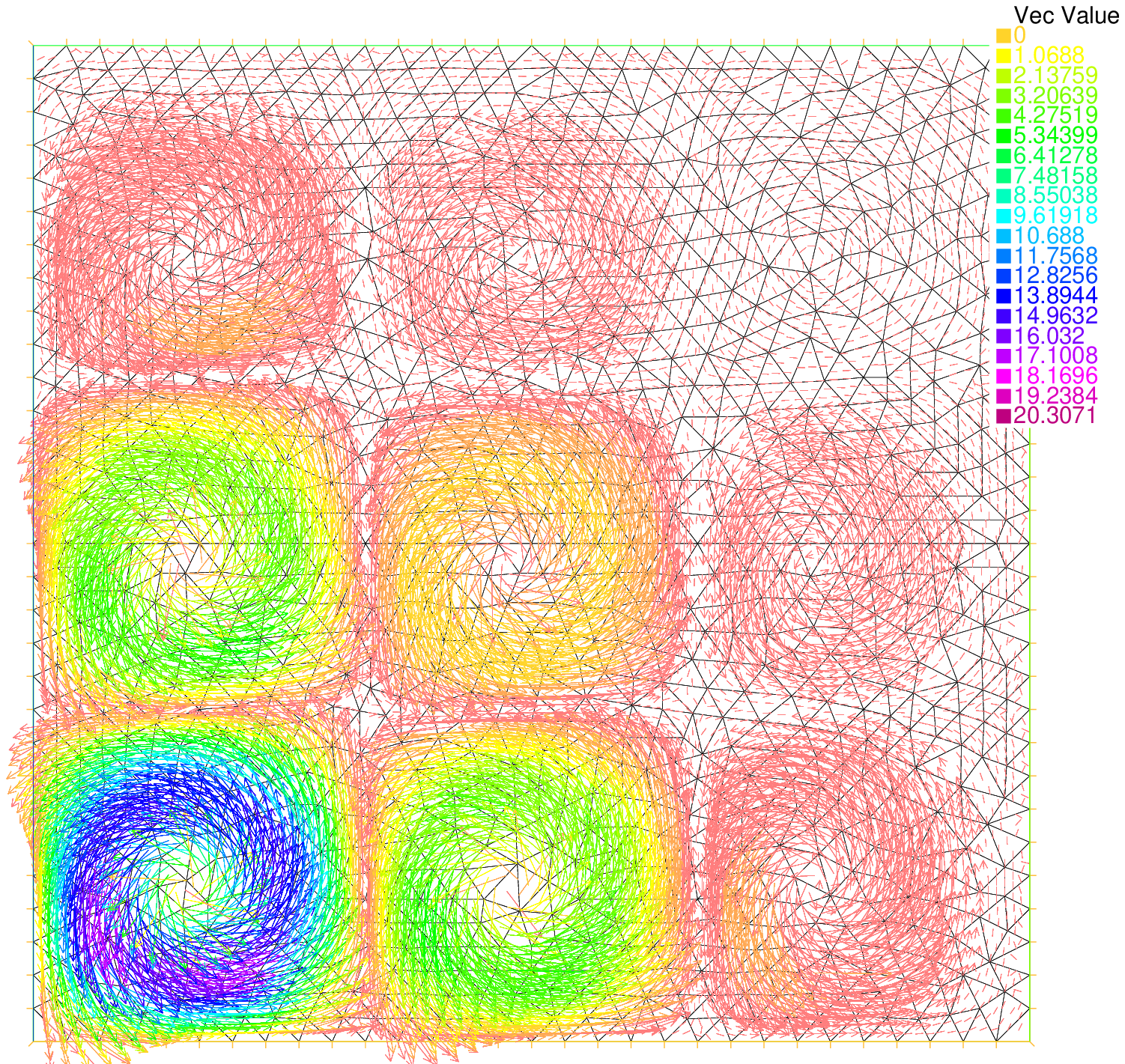}	
	\includegraphics[width=2.5in,angle=0]{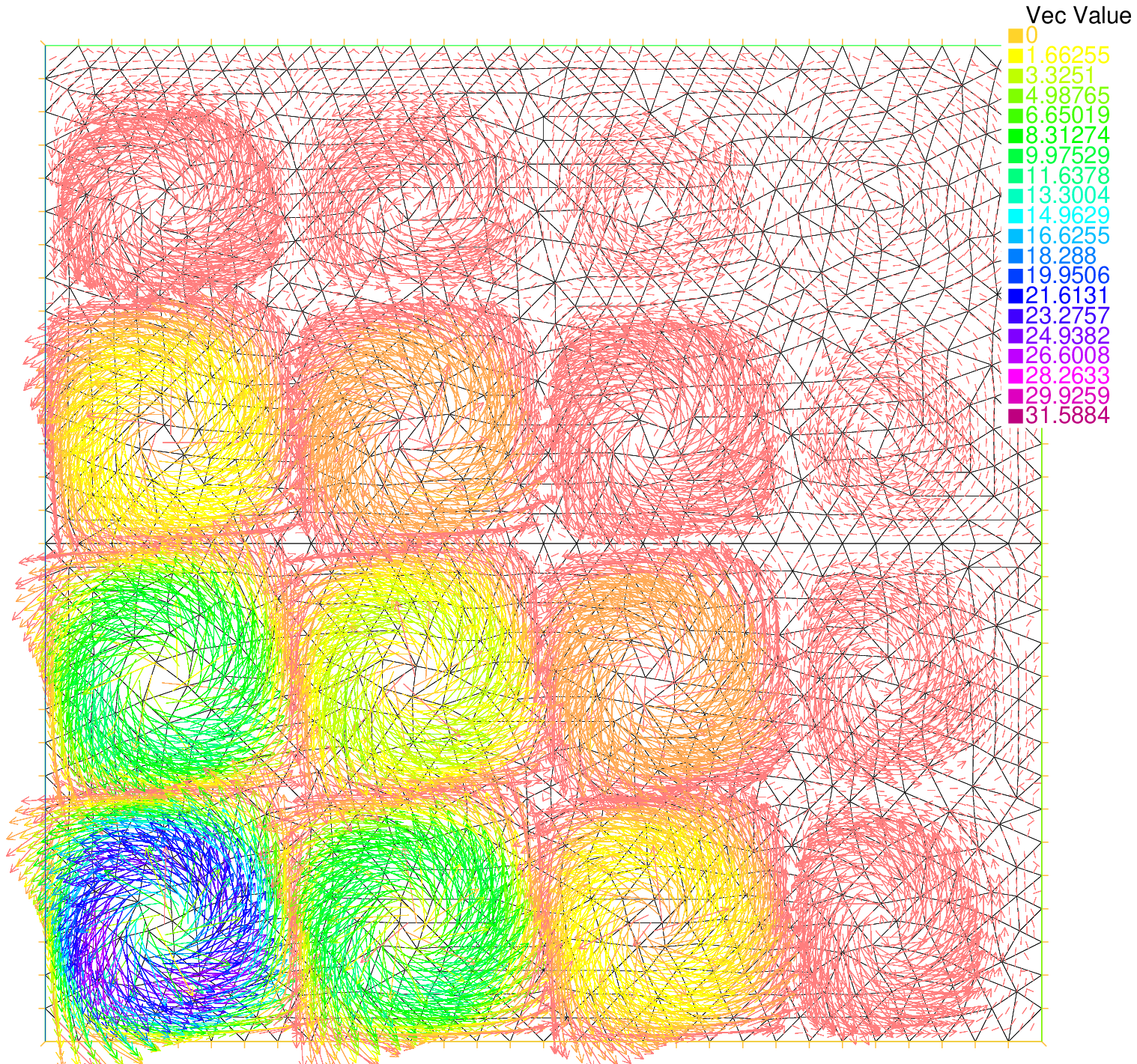}	
	\captionsetup{justification=centering}
	\caption {\scriptsize Velocity field at different times: $t=0.5$, $t=1$, $t=1.5$ and $t=2$ (from top to bottom and left to right).} 
	\label{cavity_goal}
\end{figure}

\begin{figure}[h!]
	\centering  
	\includegraphics[width=2.5in,angle=0]{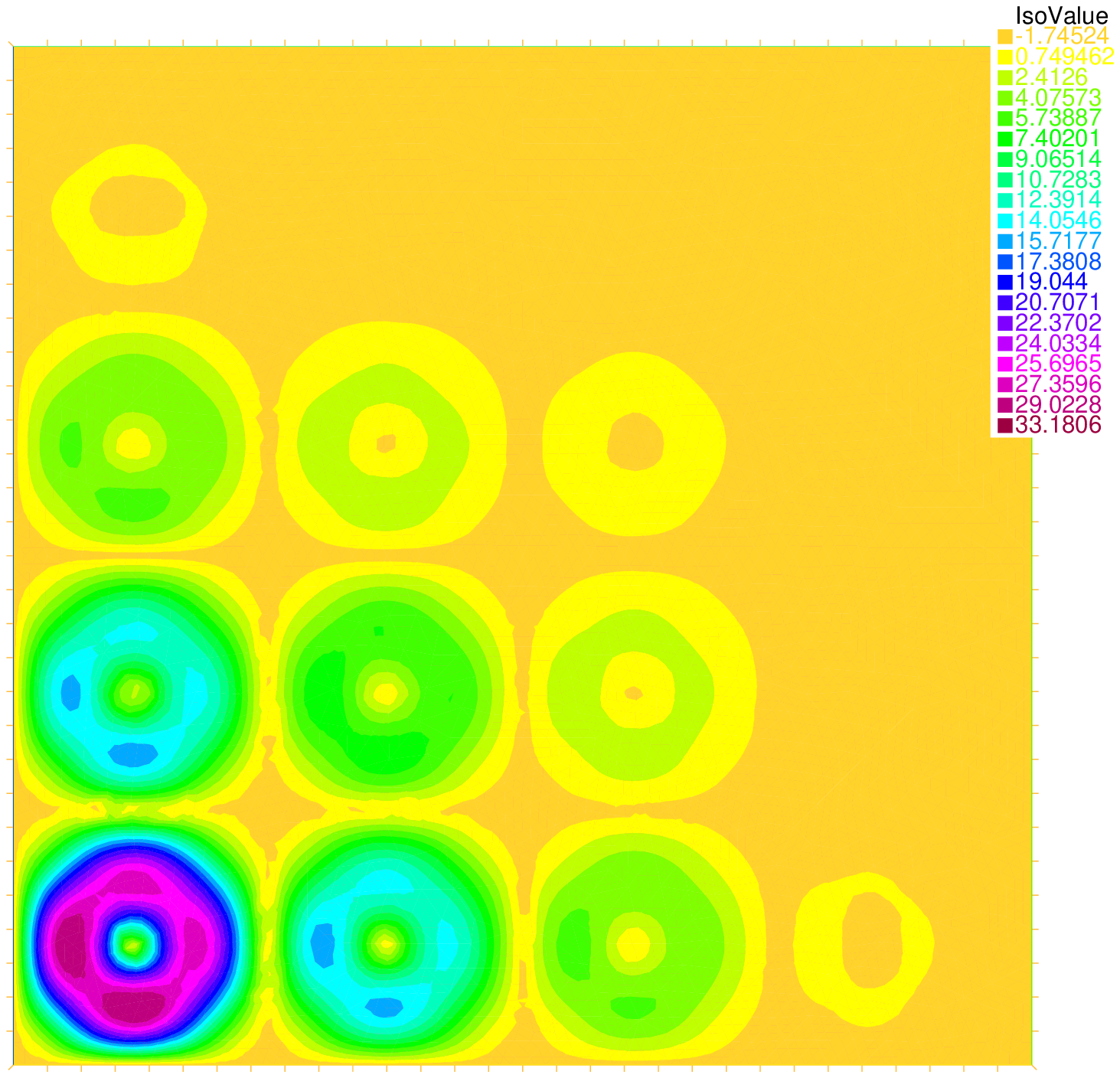}	
	\includegraphics[width=2.5in,angle=0]{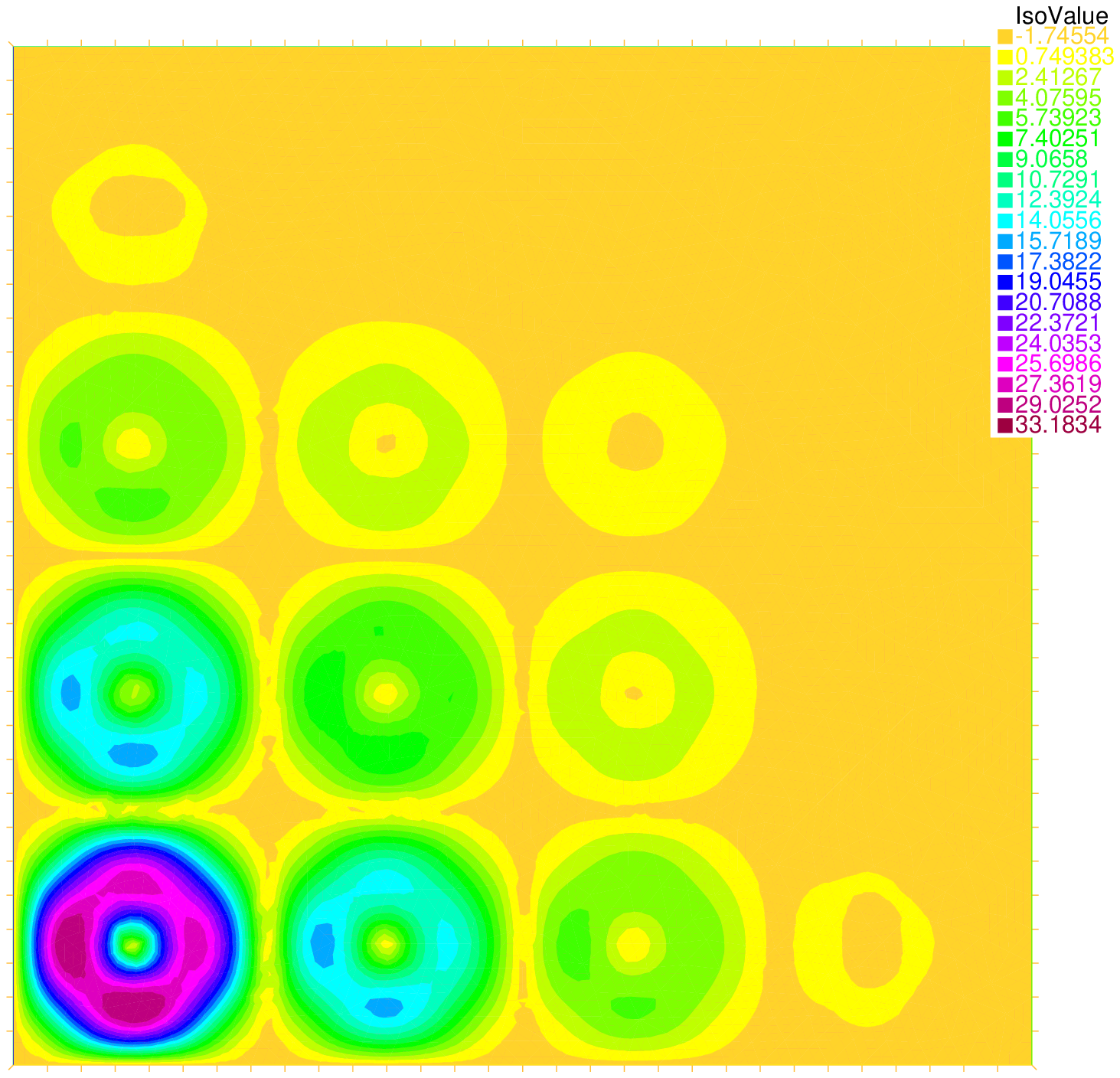}	
	\captionsetup{justification=centering}
	\caption {\scriptsize Velocity norm of the objective (left) and controlled (right) flow at $t=2$ with $\alpha=10^{-10}$; $\|{\bf u}-{\bf u}_g\|/\|{\bf u}_g\|<10^{-3}$.} 
	\label{cavity_conpare}
\end{figure}

To get an intuition of the flow field, we visualise the objective flow at different times in Figure \ref{cavity_goal}, from which it can be seen that more and more vortices are developed as times evolves, and the magnitude of the velocity increase at the same time. These figures are plotted on a mesh of 2138 triangles with 1130 vertices and 4394 degrees of freedom. Using the same mesh to carry out the simulation, we find that the controlled flow field can almost duplicate the objective flow to a very high accuracy. A typical comparison is shown in Figure \ref{cavity_conpare}, from which we cannot distinguish the objective and controlled flow field by a naked eye -- the $L^2$ error is less than $0.001$. 

\begin{figure}[h!]
	\begin{minipage}[t]{0.5\linewidth}
		\centering  
		\includegraphics[width=2.8in,angle=0]{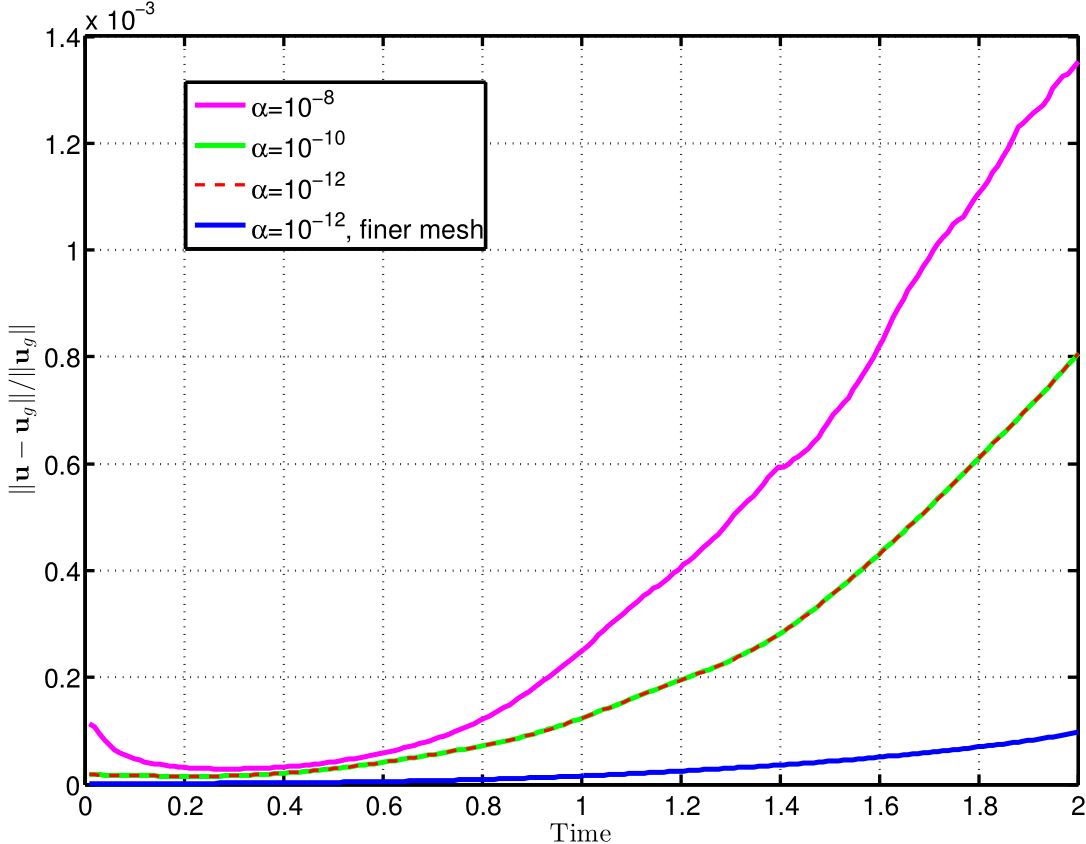}			
		\captionsetup{justification=centering}
		\caption {\scriptsize Convergence of the objective.} 
		\label{cavity_j}
	\end{minipage}
	\begin{minipage}[t]{0.5\linewidth}
		\centering  
		\includegraphics[width=2.8in,angle=0]{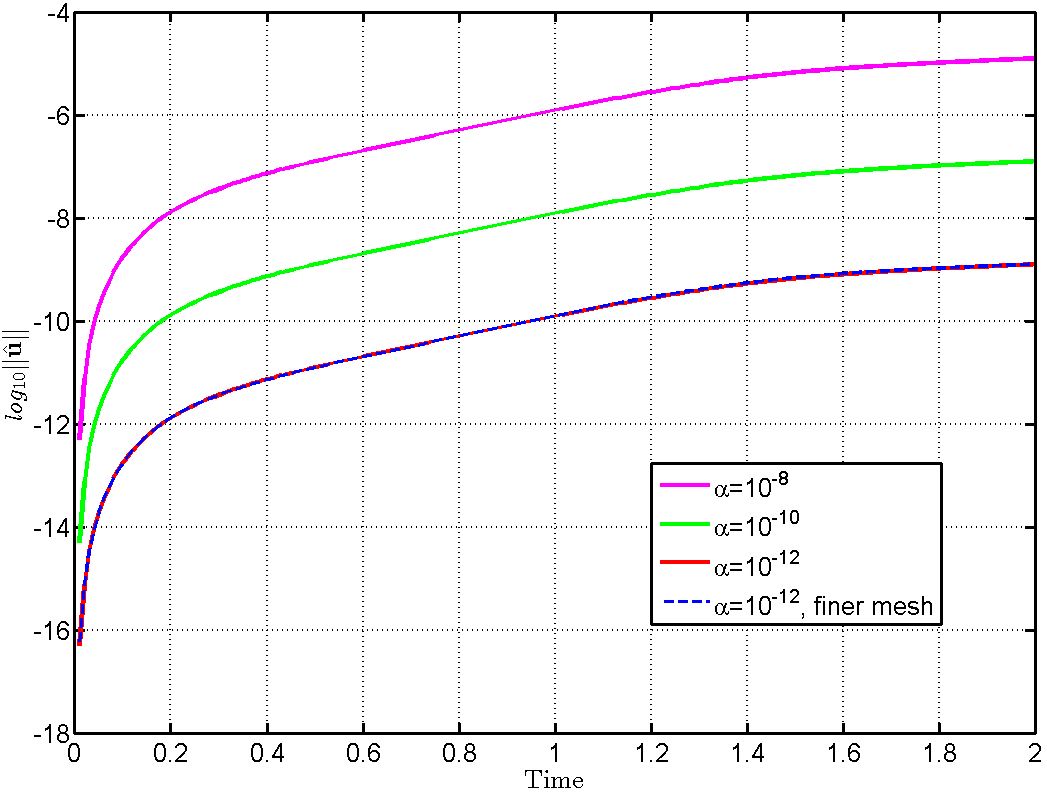}			
		\captionsetup{justification=centering}
		\caption {\scriptsize $L^2$ norm of the adjoint velocity $\hat{\bf u}=\alpha{\bf f}(t)$.} 
		\label{cavity_f}
	\end{minipage}     		
\end{figure}

Convergence of the objective function with respective to the regularisation parameter $\alpha$, using a converged time step $\Delta t=0.01$, is presented in Figure \ref{cavity_j}, from which it can be seen that the error between the state velocity and the objective velocity gradually increase as time involves. This is not surprising because the flow filed becomes more complicated and the control is more difficult as time increases. However, the accuracy can be further improved by refining the mesh in order to capture more details of the vortices. The convergence of the control force is presented in Figure \ref{cavity_f}, from which it can be seen that the same force, which cannot improve the accuracy on a coarse mesh, does improve the accuracy on a finer mesh.

\subsection{Oscillating leaflet in a fluid channel}
\label{subsec_flag}
\begin{figure}[h!]
	\centering  
	\includegraphics[width=5in,angle=0]{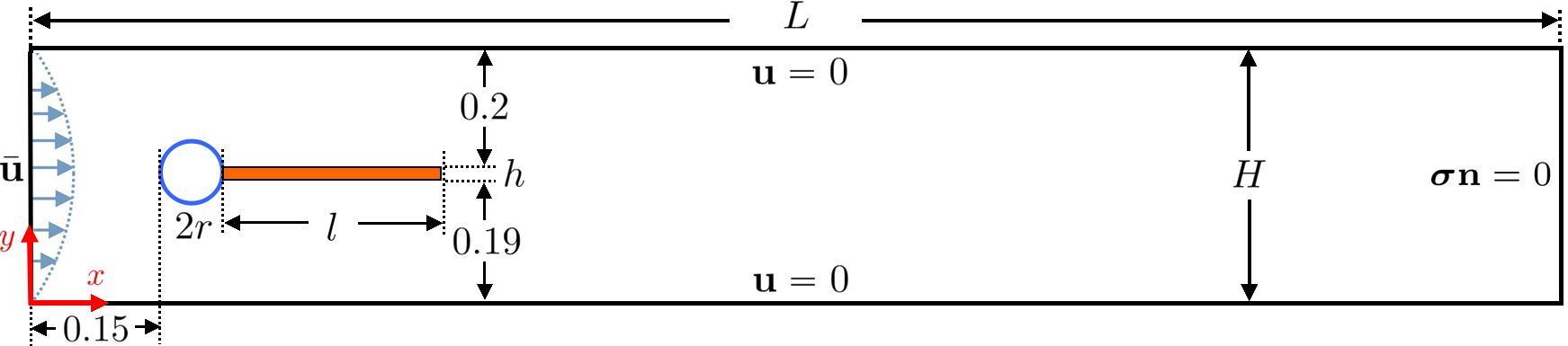}			
	\captionsetup{justification=centering}
	\caption {\scriptsize Computational domain and boundary conditions for the oscillating leaflet.} 
	\label{thick_flag_sketch}
\end{figure}

In this test, we consider a benchmark FSI problem of an oscillating leaflet attached to a cylinder \cite{turek2006proposal,Hecht_2017,wangoptimal2021}, and our objective is to minimise the solid deflection through an activation force on the solid leaflet. The computational domain is a rectangle ($L\times H$) with a cut hole of radius $r$ and center $(c,c)$ as shown in Figure \ref{thick_flag_sketch}. The geometry parameters are: $L=2.5$, $H=0.41$, $l=0.35$, $h=0.02$, $c=0.2$ and $r=0.05$. The fluid and solid parameters are: $\rho^f=\rho^s=10^3$, $\mu^f=1$ and $c_1=2.0\times 10^6$. The inlet flow is prescribed as:
\begin{equation}
	\bar{u}_x=\frac{12y}{H^2}\left(H-y\right),\quad \bar{u}_y=0.
\end{equation}

\begin{figure}[h!]
	\centering  
	\includegraphics[width=5 in,angle=0]{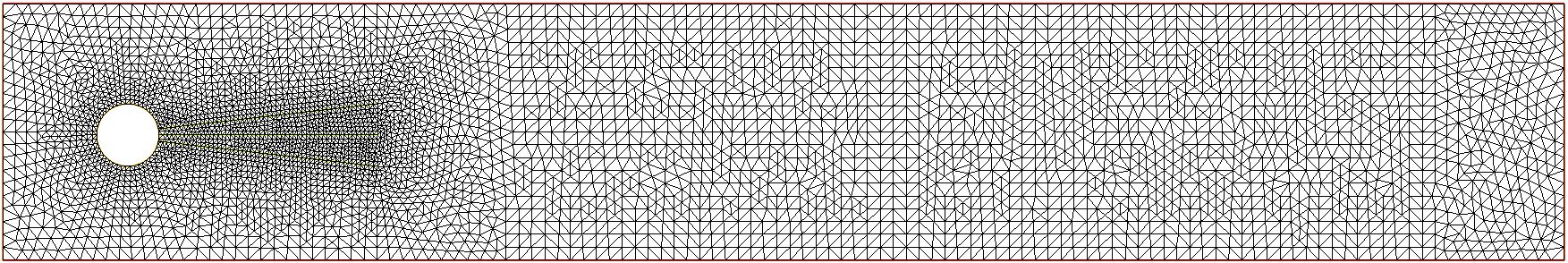}			
	\captionsetup{justification=centering}
	\caption {\scriptsize Background fluid mesh.} 
	\label{flag_fluid_mesh}
\end{figure}

\begin{figure}[h!]
	\centering  
	\includegraphics[width=4.8in,angle=0]{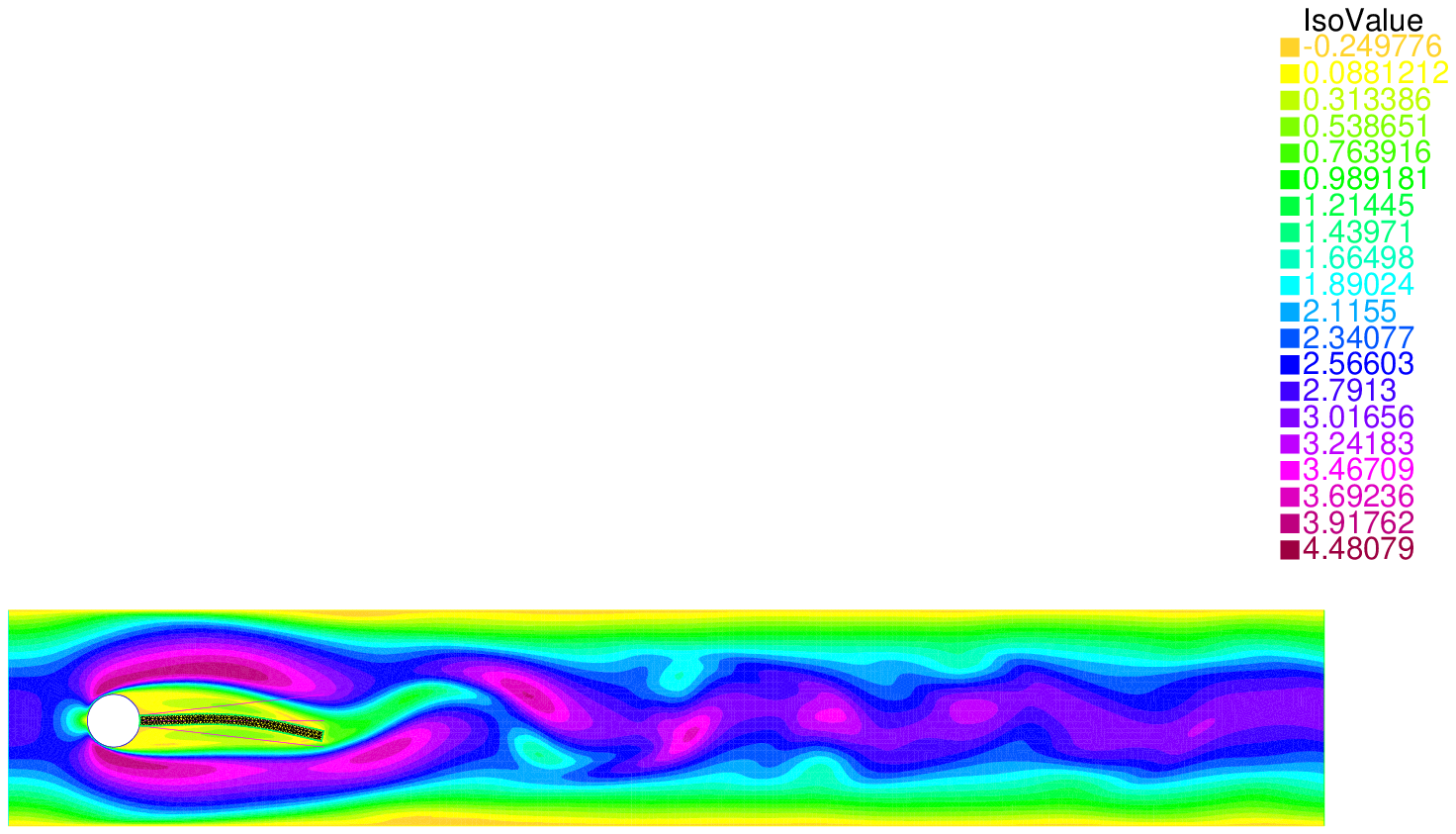}	
	\includegraphics[width=0.32in,angle=0]{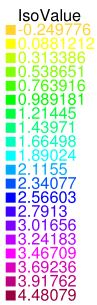}			
	\captionsetup{justification=centering}
	\caption {\scriptsize A snap shot of the velocity norms at $t=4$ when the leaflet is maximally deformed.} 
	\label{flag}
\end{figure}

\begin{figure}[h!]
	\centering  
	\includegraphics[width=4.0in,angle=0]{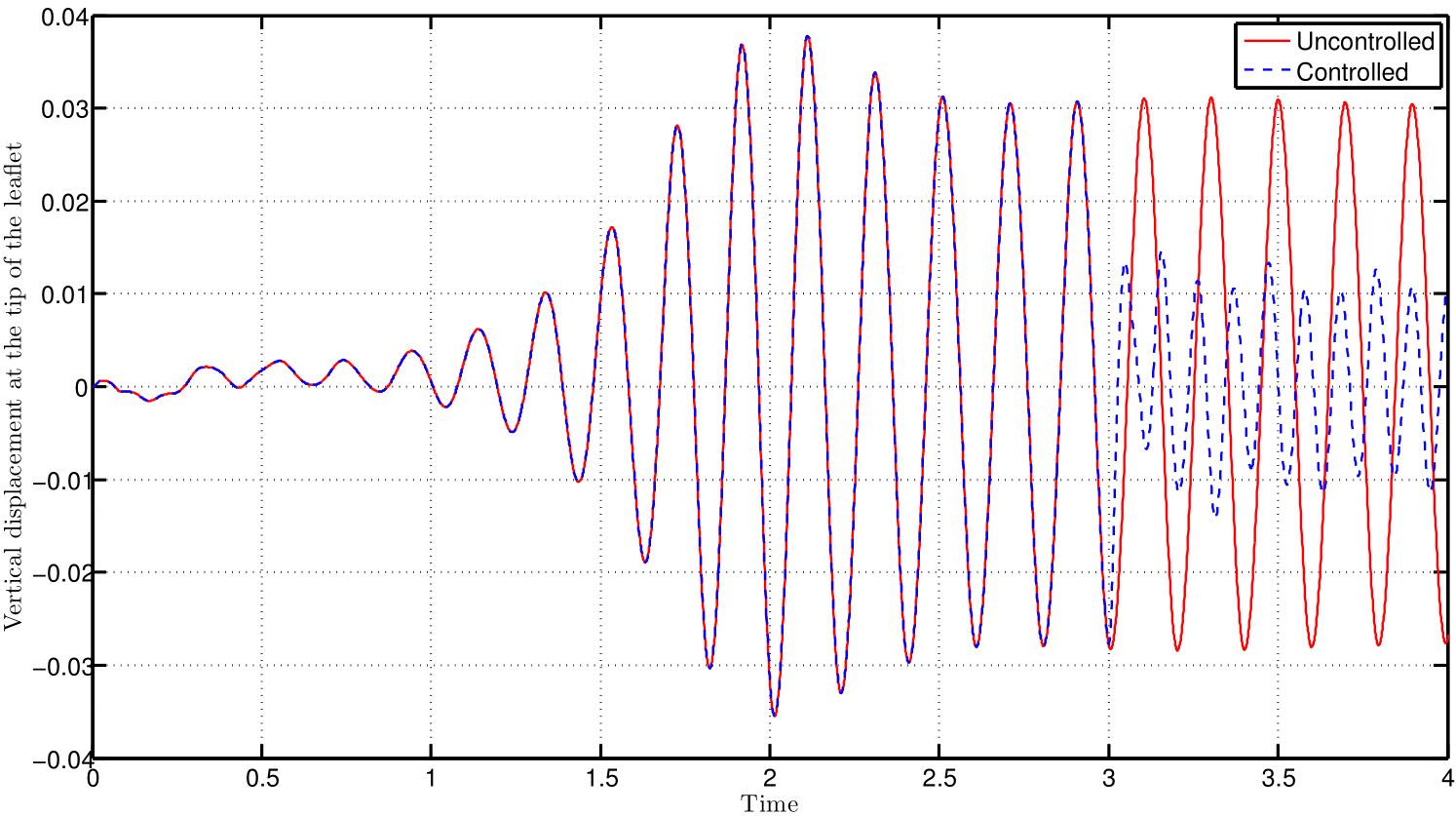}			
	\captionsetup{justification=centering}
	\caption {\scriptsize Vertical displacement at the tip of the leaflet. $\alpha=10^{-17}$ for the controlled case.} 
	\label{disp_time}
\end{figure}

\begin{figure}[h!]
	\centering  
	\includegraphics[width=4in,angle=0]{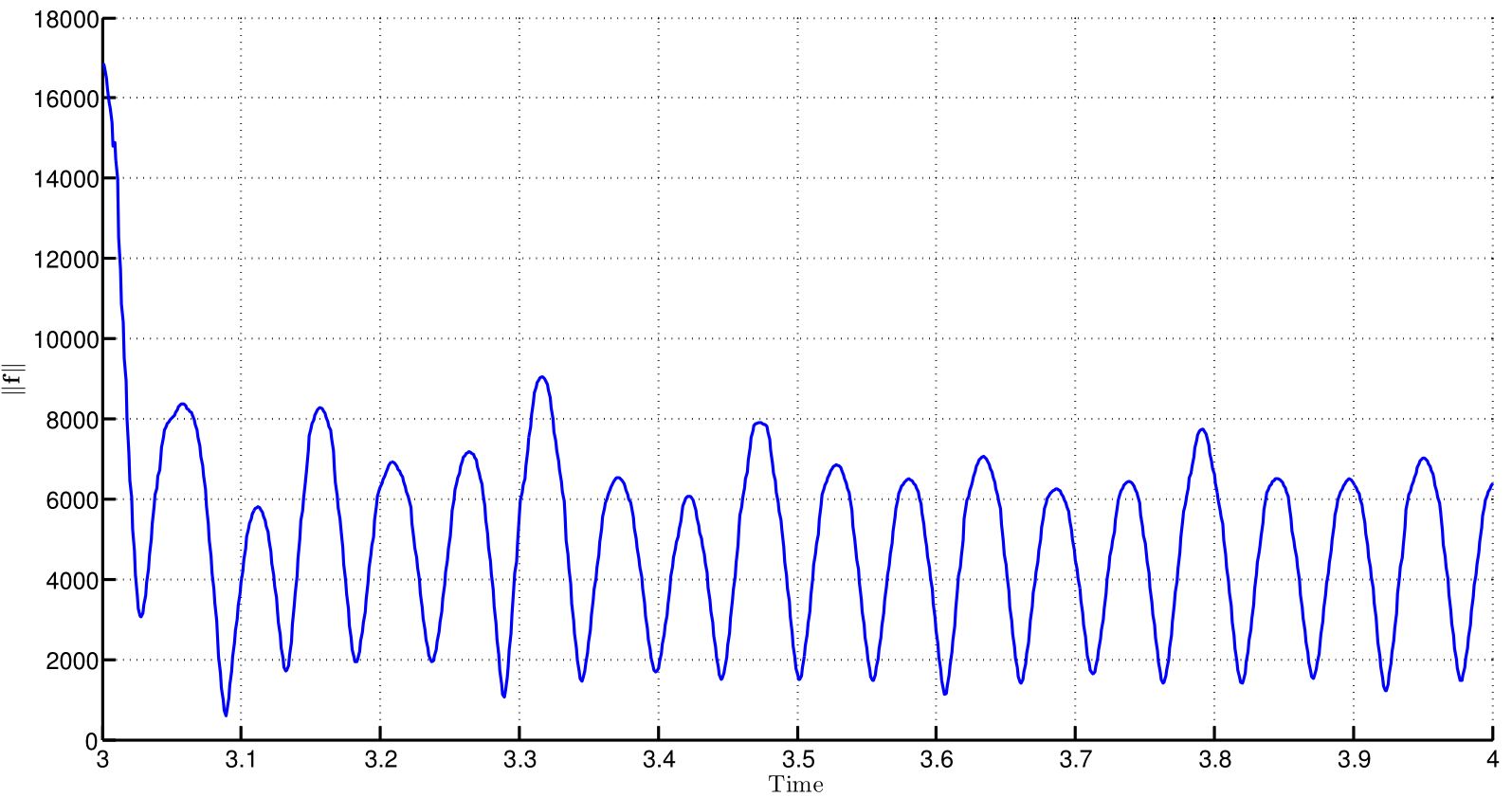}			
	\captionsetup{justification=centering}
	\caption {\scriptsize $L^2$ norm of the control force ${\bf f}(t)$,  $\alpha=10^{-17}$.} 
	\label{flag_f}		
\end{figure}

\begin{figure}[h!]
	\centering  
	\includegraphics[width=4.0in,angle=0]{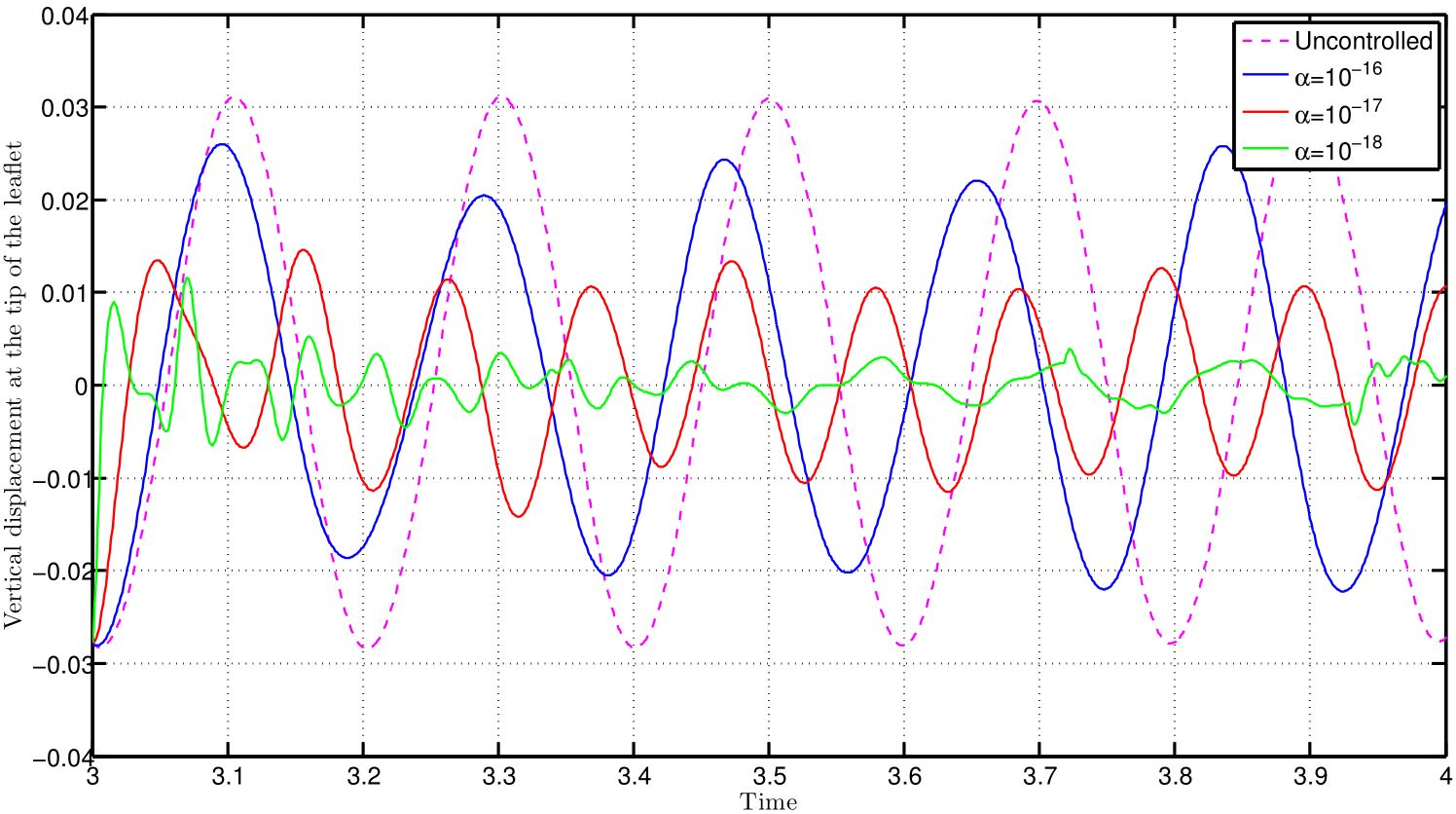}			
	\captionsetup{justification=centering}
	\caption {\scriptsize Vertical displacement at the tip of the leaflet for different regularisation parameters.} 
	\label{disp_time_all}
\end{figure}

\begin{figure}[h!]
	\centering  
	\includegraphics[width=4.0in,angle=0]{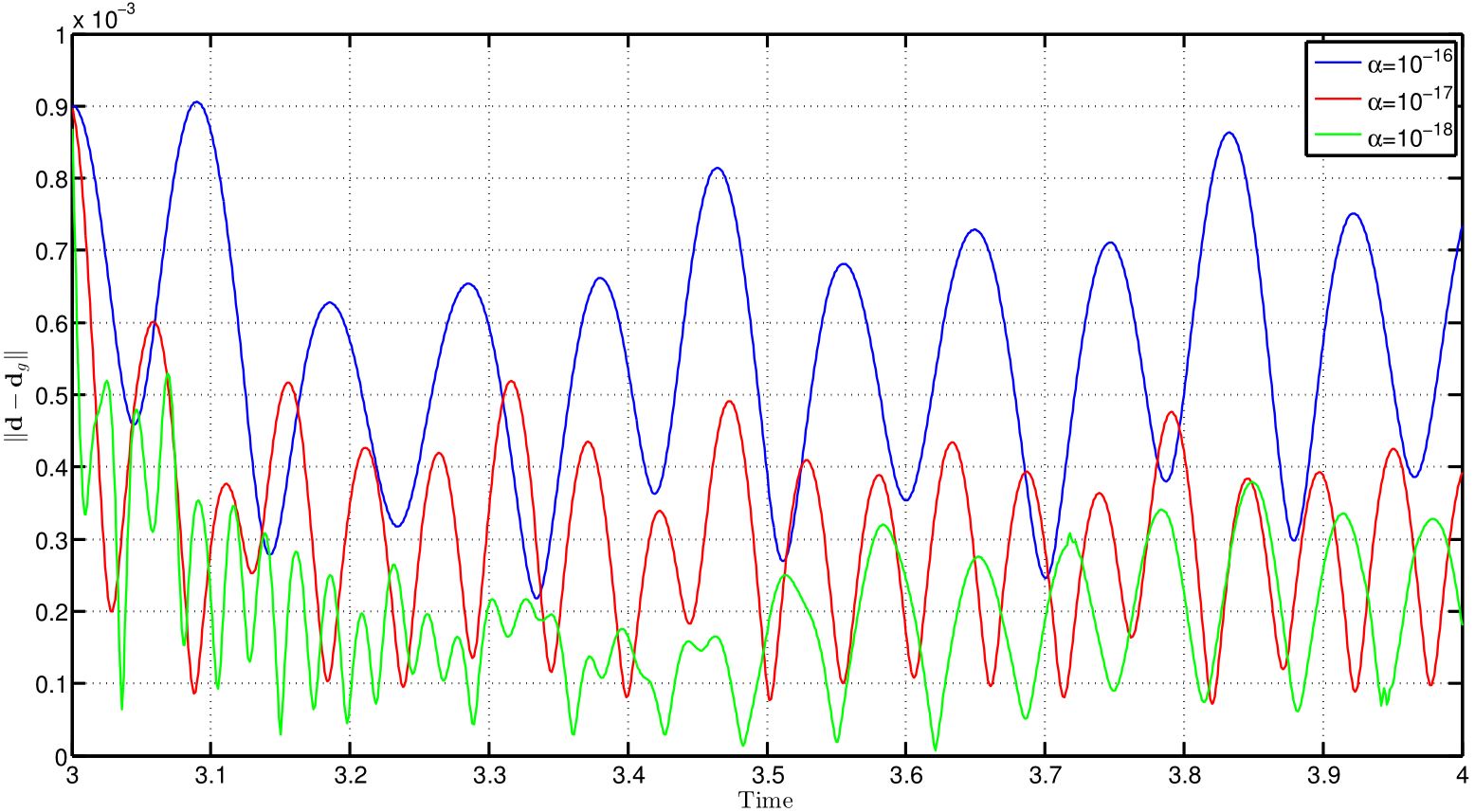}			
	\captionsetup{justification=centering}
	\caption {\scriptsize Objective redcution for different regularisation parameters.} 
	\label{err_time_all}
\end{figure}

We use a mesh of $9008$ elements with $4668$ vertices for the background fluid (see Figure \ref{flag_fluid_mesh}), and a mesh of $314$ elements with $213$ vertices for the solid leaflet (see Figure \ref{flag}). A converged time step of $\Delta t=10^{-3}$ is used for this test, and our two-mesh method presents the same accuracy as the fitted-mesh method \cite{Hecht_2017} with an oscillation period and amplitude being $0.530$ and $0.03$ respectively as shown in Figure \ref{disp_time} (red curve). We then focus on the control of this FSI system, and start to add an activation force ${\bf f}$ on the solid leaflet from $t=3$ by solving the Problem \ref{problem_monlithic} ($\lambda=0$, constraint (\ref{inequality_constraint}) is turned off). The overall control is tractable, and the vertical displacement at the tip of the leaflet is plotted in Figure \ref{disp_time} (dashed blue curve) from which it can be seen that the deflection of the leaflet is reduced around $50\%$. It is interesting to notice that the frequency of the leaflet's oscillation increases as its amplitude decreases after the control. The magnitude of the corresponding activation force is plotted in Figure \ref{flag_f}, from which it can be seen that a large control force is computed at the very beginning of the control at $t=3$, and then it decreases rapidly and responses periodically to the oscillation of the leaflet and keeps its deflection down. We also study the effect the regularisation parameter $\alpha$ on the reduction of the objective as shown in Figure \ref{disp_time_all} and \ref{err_time_all}. It is clear that the smaller $\alpha$ is the more the objective can be reduced, but the more instability it could introduce to the algorithm as shown for the case of $\alpha=10^{-18}$ in Figure \ref{disp_time_all} and \ref{err_time_all} (green curve).

\subsection{Solid disc within a lid-driven cavity flow}
\label{cavity_disc}
\begin{figure}[h!]
	\centering  
	\includegraphics[width=2.3in,angle=0]{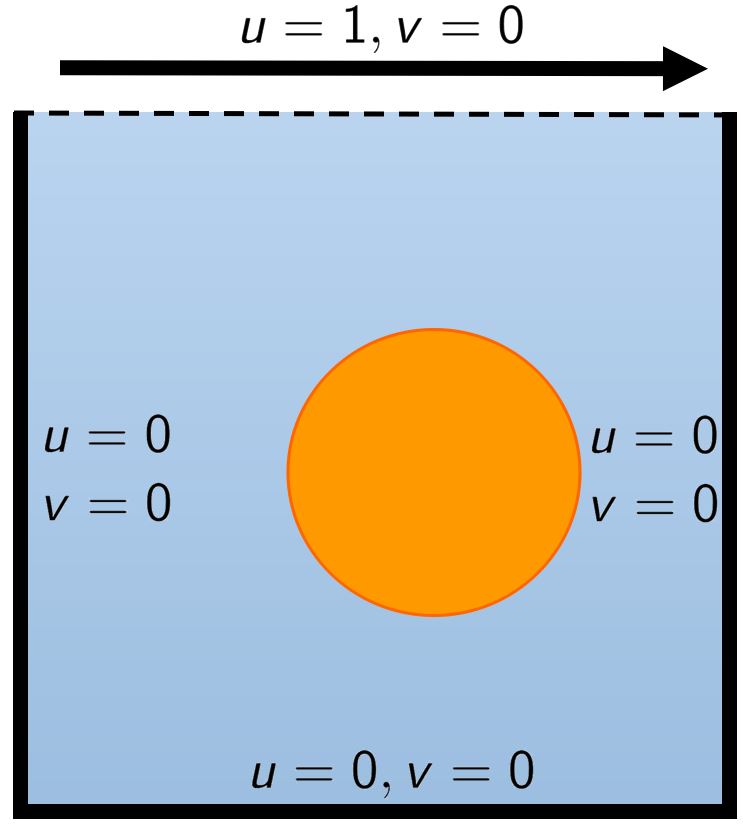}	
	\captionsetup{justification=centering}
	\caption {\scriptsize Sketch of a solid disc within a lid-driven cavity flow.}
	\label{lid_driven_cavity_disc}
\end{figure}

This FSI problem is considered in many publications \cite{Zhao_2008,Wang_2017,wang2019theoretical,roy2015benchmarking} as a forward FSI benchmark problem, whose controllability however has not be studied due to the complex movement and large deformation of the solid disc. The computational domain is a unit square $[0,1]\times[0,1]$ and a solid disc of radius $r=0.2$ is initially located at $(x_0,y_0)=(0.6, 0.5)$ as shown shown in Figure \ref{lid_driven_cavity_disc}. The fluid and solid material parameters are: $\rho^f=\rho^s=1$, $\mu^f=0.01$ and $c_1=1$. Due to drag prescribed at the top of the cavity, the solid gradually moves and rotates inside the cavity. We use a stable time step of $\Delta t=0.005$, background fluid mesh of 2404 triangles with 1273 vertices, and solid mesh of 642 triangles with 352 vertices.

\begin{figure}[h!]
	\centering  
	\includegraphics[width=2.5in,angle=0]{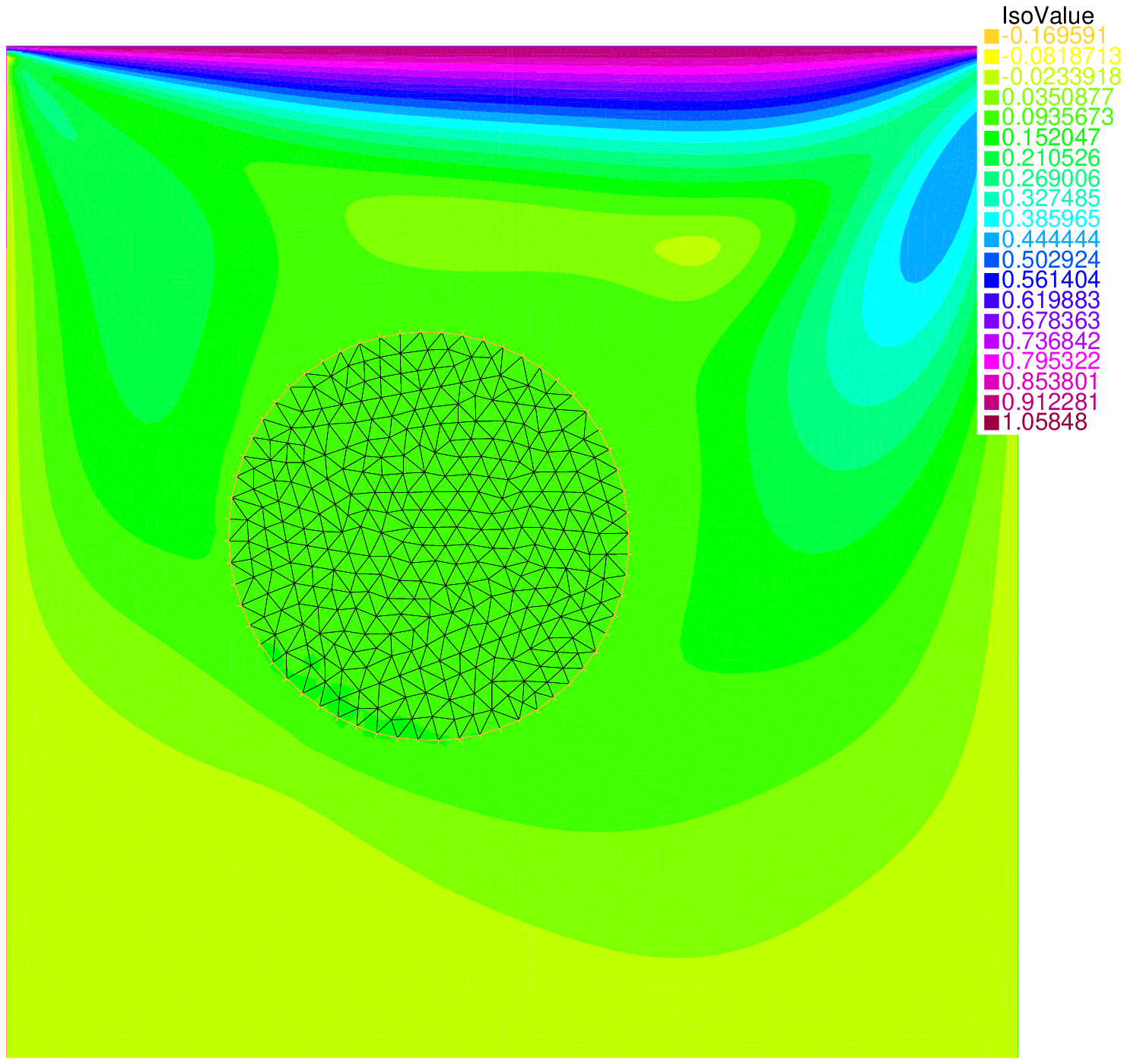}	
	\includegraphics[width=2.5in,angle=0]{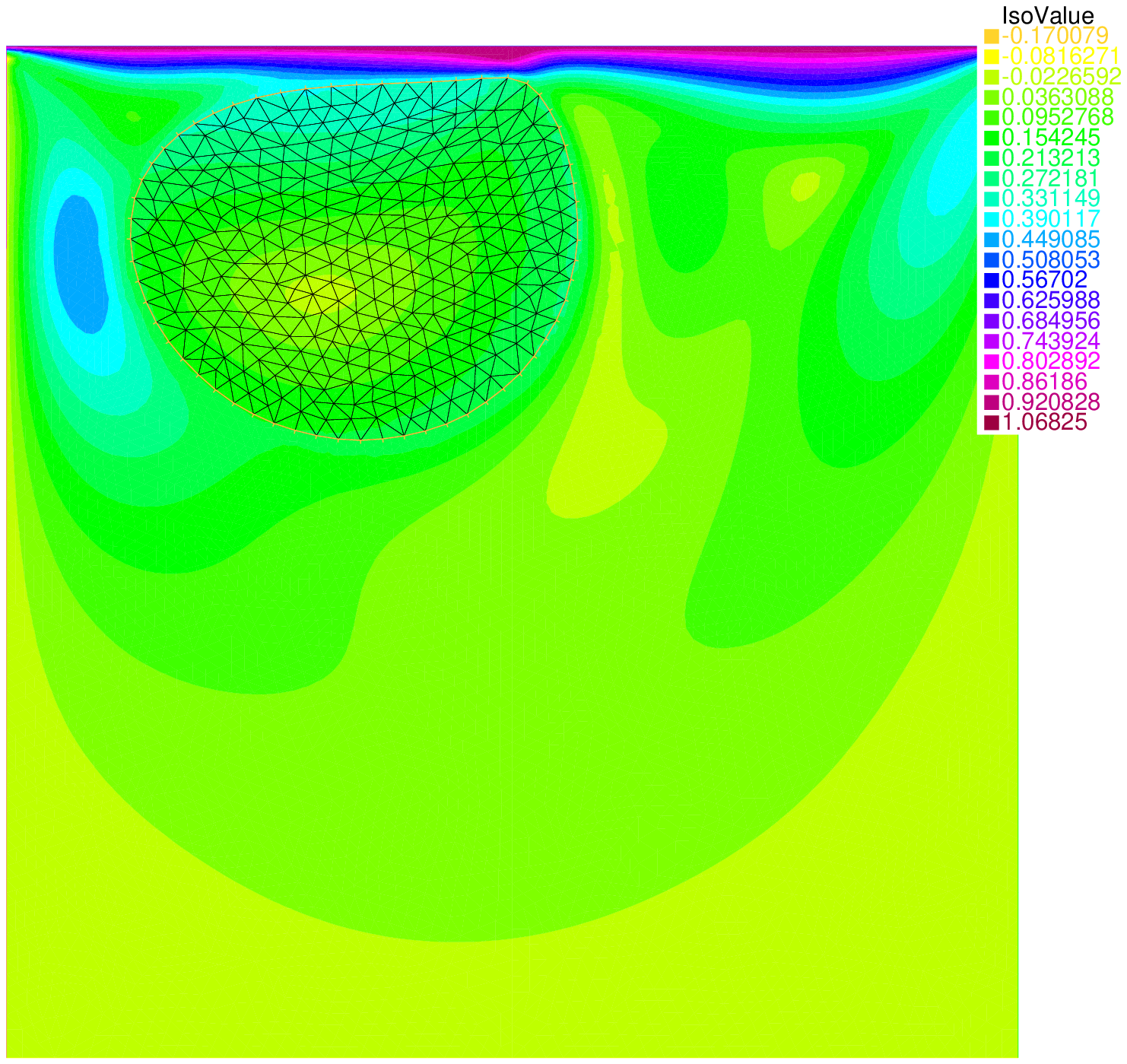}	
	\includegraphics[width=2.5in,angle=0]{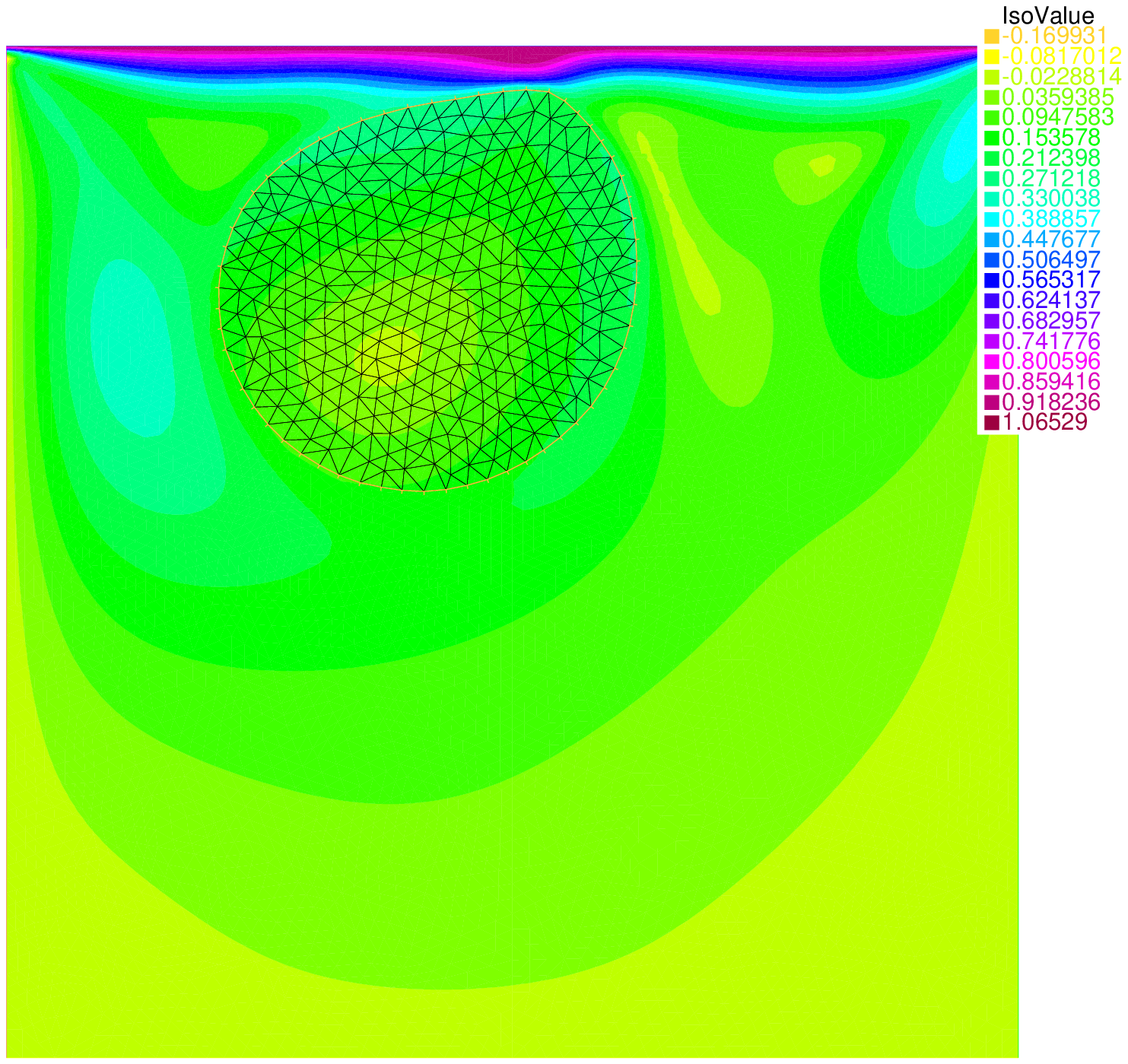}	
	\includegraphics[width=2.5in,angle=0]{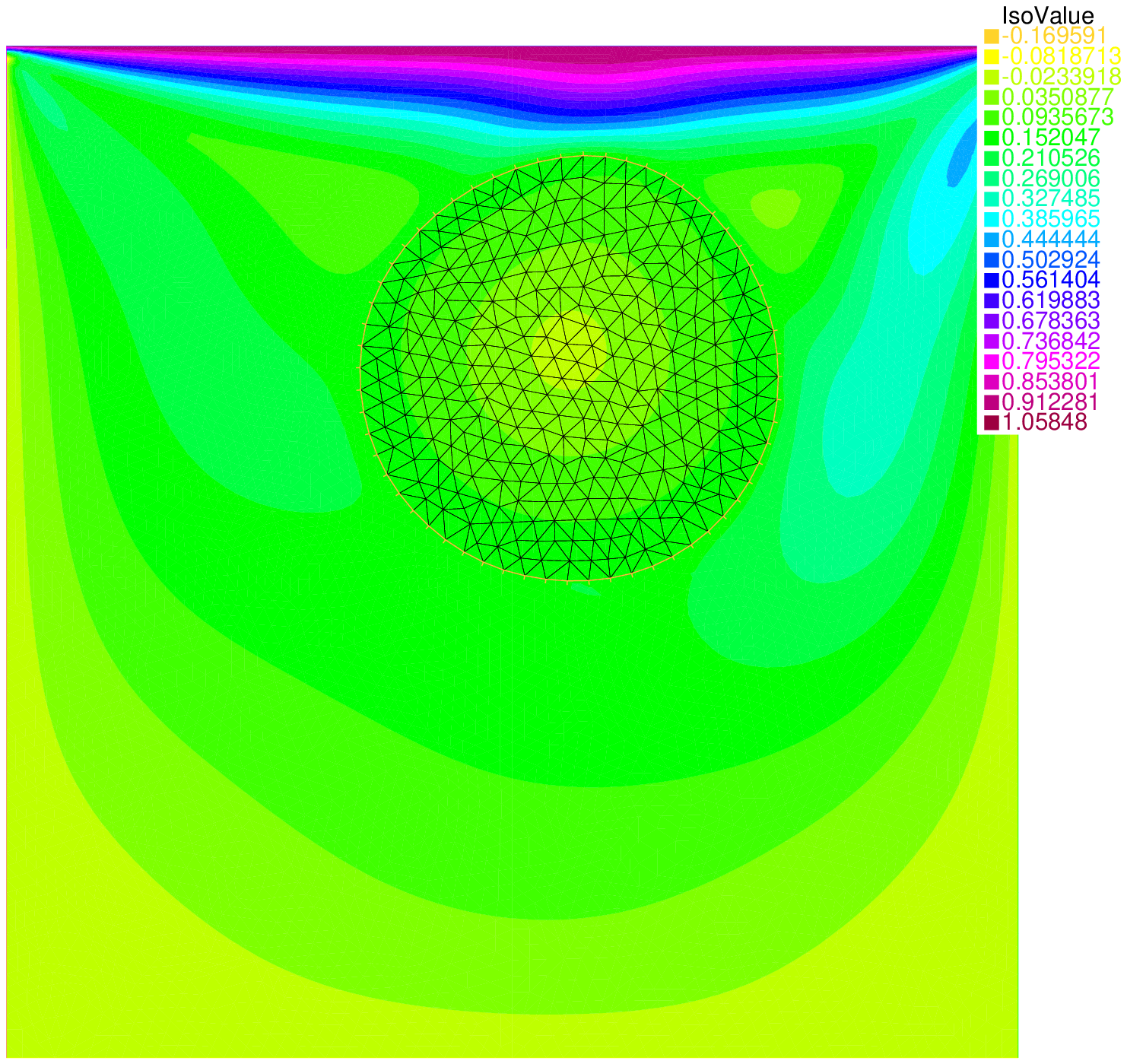}	
	\captionsetup{justification=centering}
	\caption {\scriptsize Velocity norm at different times: $t=2$, $t=5$, $t=6$ and $t=20$ (from top to bottom and left to right).} 
	\label{cavity_disc_uncontrolled}
\end{figure}

\begin{figure}[h!]
	\centering  
	\includegraphics[width=4in,angle=0]{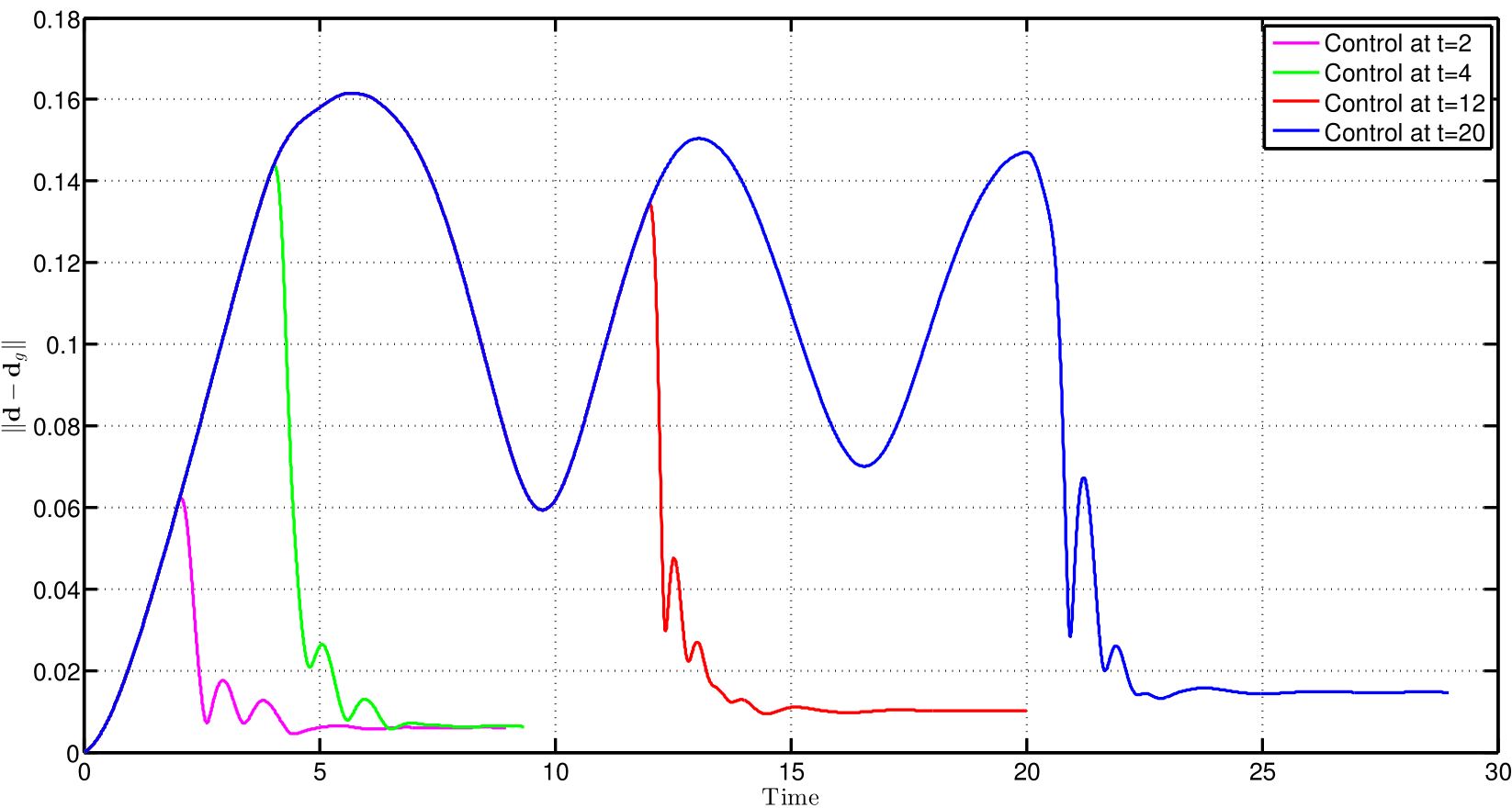}	
	\captionsetup{justification=centering}
	\caption {\scriptsize Reduction of the objective at different times using $\alpha=2.5\times10^{-7}$.} 
	\label{cavity_disc_err}
\end{figure}

\begin{figure}[h!]
	\begin{minipage}[t]{0.5\linewidth}
		\centering  
		\includegraphics[width=2.8in,angle=0]{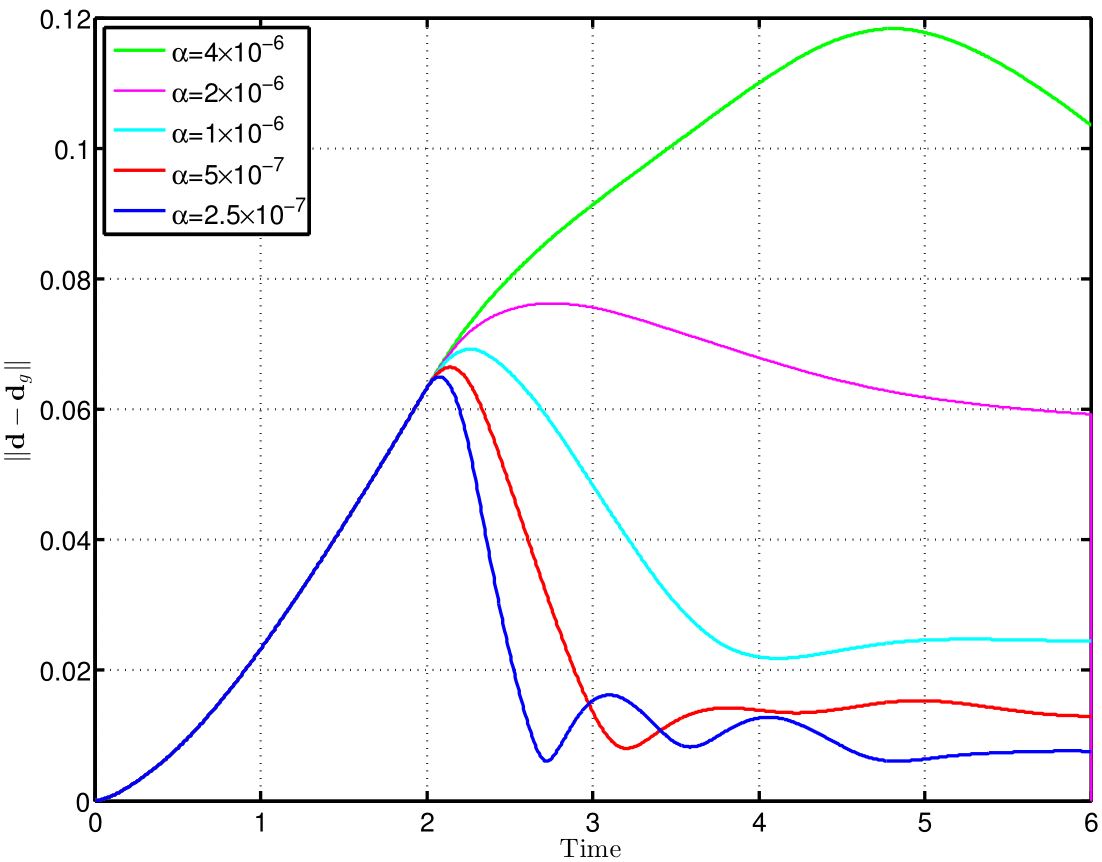}	
		\captionsetup{justification=centering}
		\caption* {\scriptsize (a) Reduction of the objective using different regularisation parameter $\alpha$.}
	\end{minipage}
	\begin{minipage}[t]{0.5\linewidth}
		\centering  
		\includegraphics[width=2.8in,angle=0]{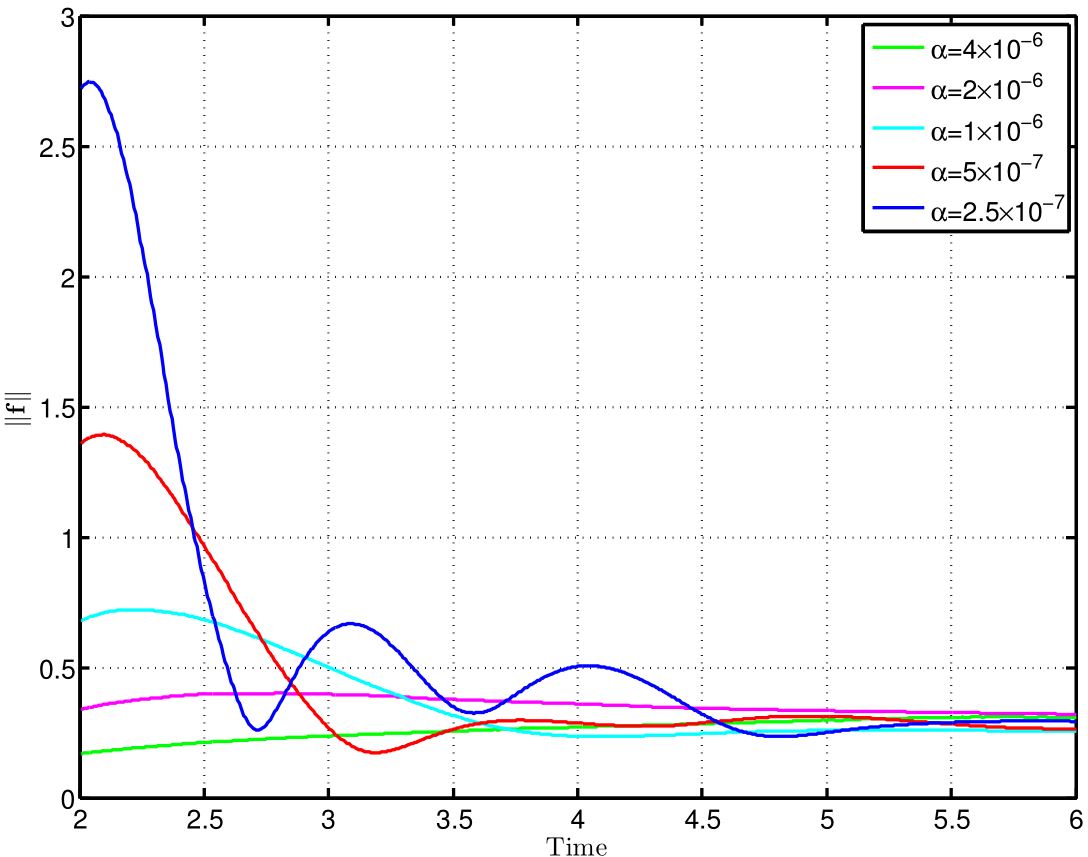}
		\captionsetup{justification=centering}
		\caption* {\scriptsize (b) Control force for different cases of regularisation parameter $\alpha$.}
	\end{minipage}     		
	\captionsetup{justification=centering}
	\caption {\scriptsize Apply the control from $t=2$ using different regularisation parameter $\alpha$.} 
	\label{disc_err_f}
\end{figure}

We first consider a case of pulling/pushing the solid back to the original position at different times, i.e., we solve Problem \ref{problem_monlithic} with ${\bf u}_g={\bf 0}$ and penalty parameter $\lambda=0$ (constraint (\ref{inequality_constraint}) is turned off). Figure \ref{cavity_disc_uncontrolled} shows the solid disc at different stages without control and Figures \ref{cavity_disc_err} shows that the proposed control method can successfully pull the solid back to the original position at different control times. We test effect of the regularisation parameter $\alpha$ on the control results as shown in Figure \ref{disc_err_f}, from which it can be seen that larger $\alpha$ would not reduce the objective sufficiently, and smaller $\alpha$ can reduce the objective more while it also introduces slight oscillations for both the objective function and the control force.

\begin{figure}[h!]
	\begin{minipage}[t]{0.5\linewidth}
		\centering  
		\includegraphics[width=2.8in,angle=0]{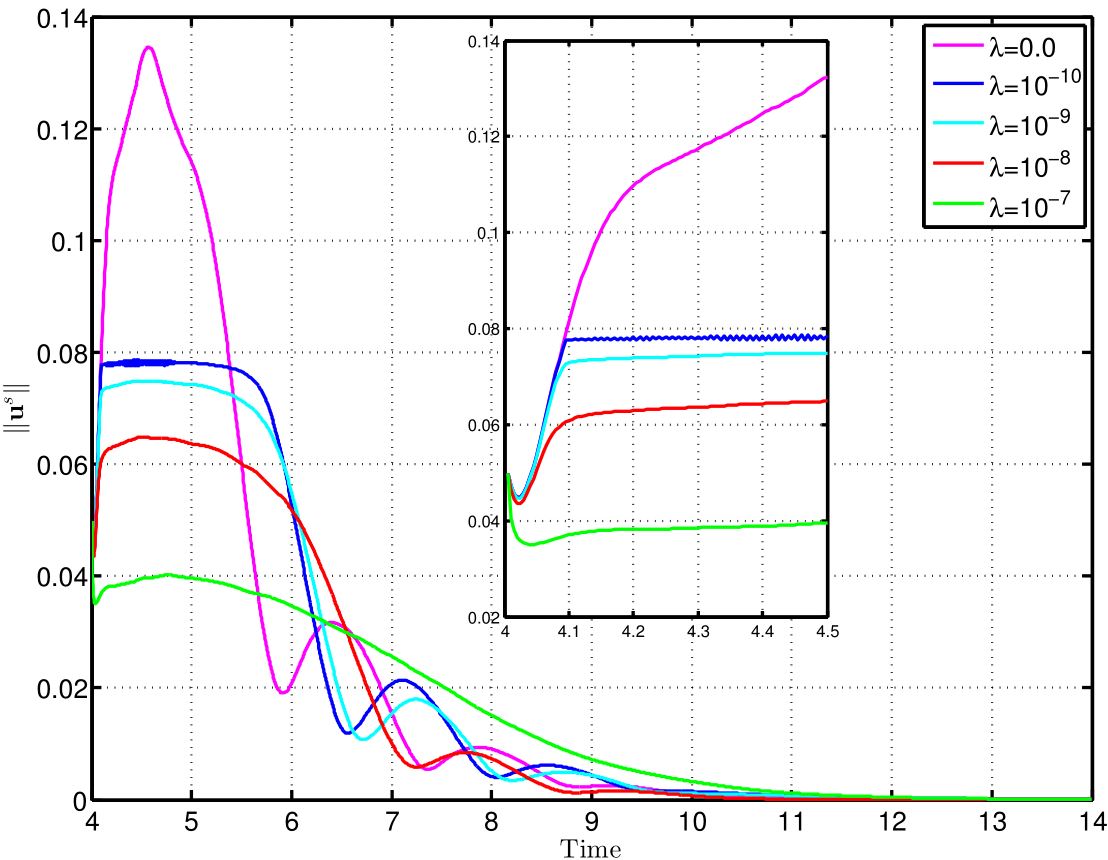}
		\captionsetup{justification=centering}
		\caption* {\scriptsize (a) Velocity norm of the solid disc for different penalty $\lambda$ and objectives.}
	\end{minipage}     	
	\begin{minipage}[t]{0.5\linewidth}
		\centering  
		\includegraphics[width=2.8in,angle=0]{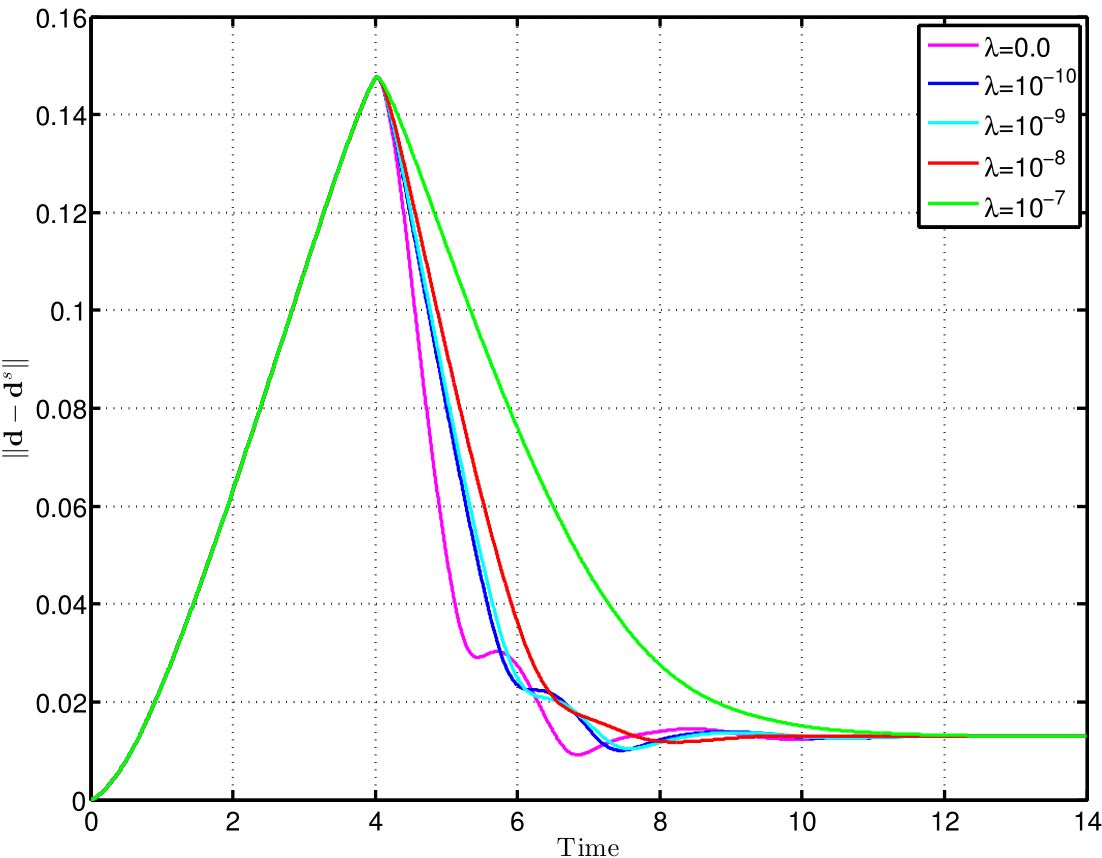}	
		\captionsetup{justification=centering}
		\caption* {\scriptsize (b) Reduction of the objective for different penalty $\lambda$ and objectives.}
	\end{minipage}	
	\captionsetup{justification=centering}
	\caption {\scriptsize Reducing the solid diplacement with a control of the solid's speed, starting from $t=4$.} 
	\label{disc_err_speed}
\end{figure}

In the above control, we have no control of the velocity of the solid disc by setting $\lambda=0$. We now start to control the movement of the solid at $t=4$ using $\alpha=5\times10^{-7}$ and $u_c=0.08$ in problem \ref{problem_optimisation}, and investigate the speed of the solid body. It can be seen from Figure \ref{disc_err_speed} (a) that the solid speed can be reduced bellow the predefined upper bound by choosing a reasonable penalty parameter $\lambda$. Notice that the solid speed would not approach to the case of $\lambda=0$ (i.e.: constraint (\ref{inequality_constraint}) is inactive) as $\lambda\rightarrow 0$, instead, it approaches to the equality scenario of constraint of (\ref{inequality_constraint}). This is a feature of discontinuity of the penalty (barrier) method \cite{bertsekas2014constrained}. We also notice that too small $\lambda$ would cause instability issue as can be observed from the blue curve in Figure \ref{disc_err_speed} (a). Therefore, a reasonable penalty parameter $\lambda$ should be used in order to control the speed of the solid disc. All these control of the solid speed does not have a significant influence of the reduction of the real objective as shown in Figure \ref{disc_err_speed} (b).

\begin{figure}[h!]
	\begin{minipage}[t]{0.5\linewidth}
		\centering  
		\includegraphics[width=2.8in,angle=0]{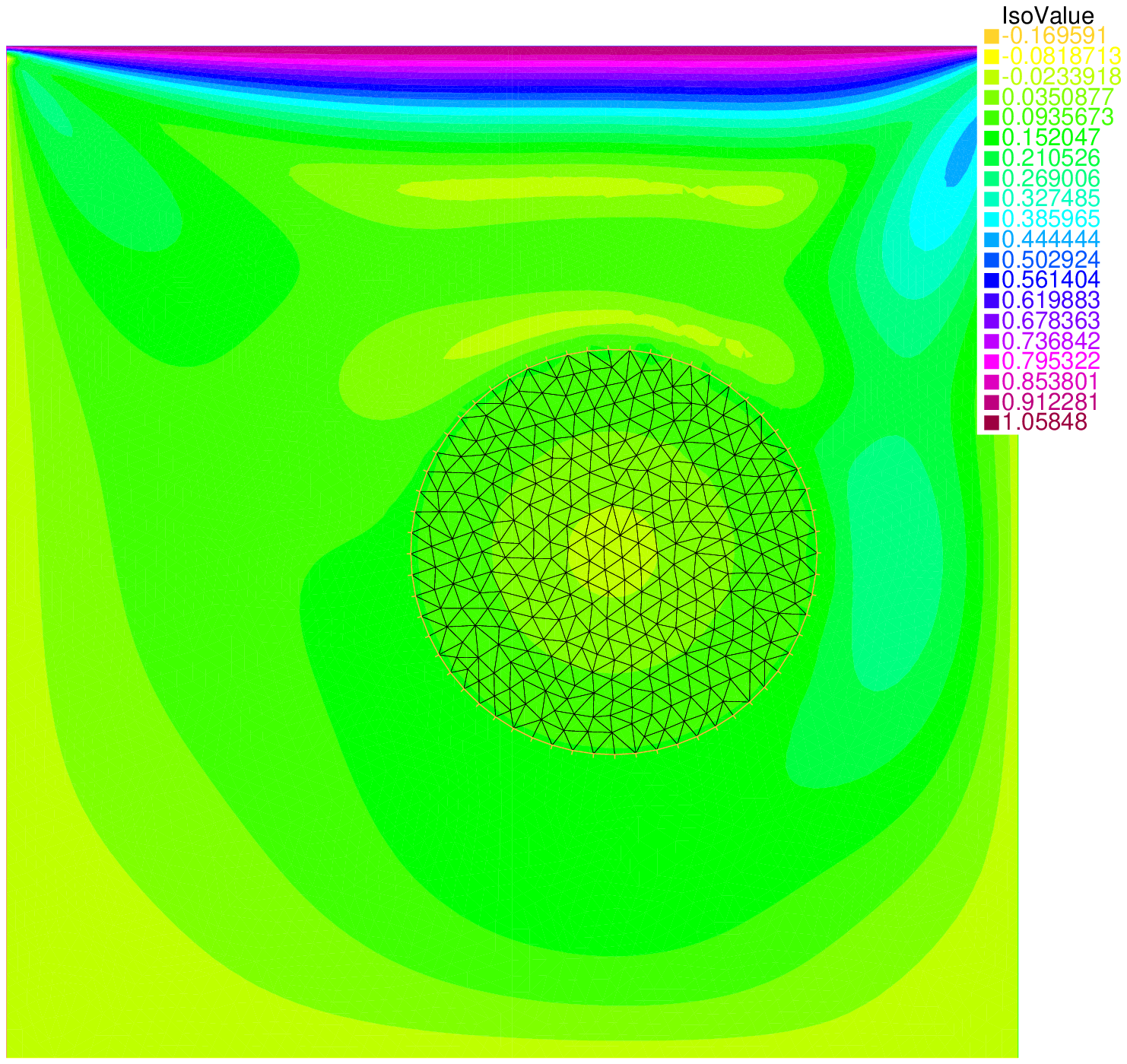}
		\captionsetup{justification=centering}
		\caption* {\scriptsize (a) Velocity norm and the solid mesh.}
	\end{minipage}     	
	\begin{minipage}[t]{0.5\linewidth}
		\centering  
		\includegraphics[width=2.8in,angle=0]{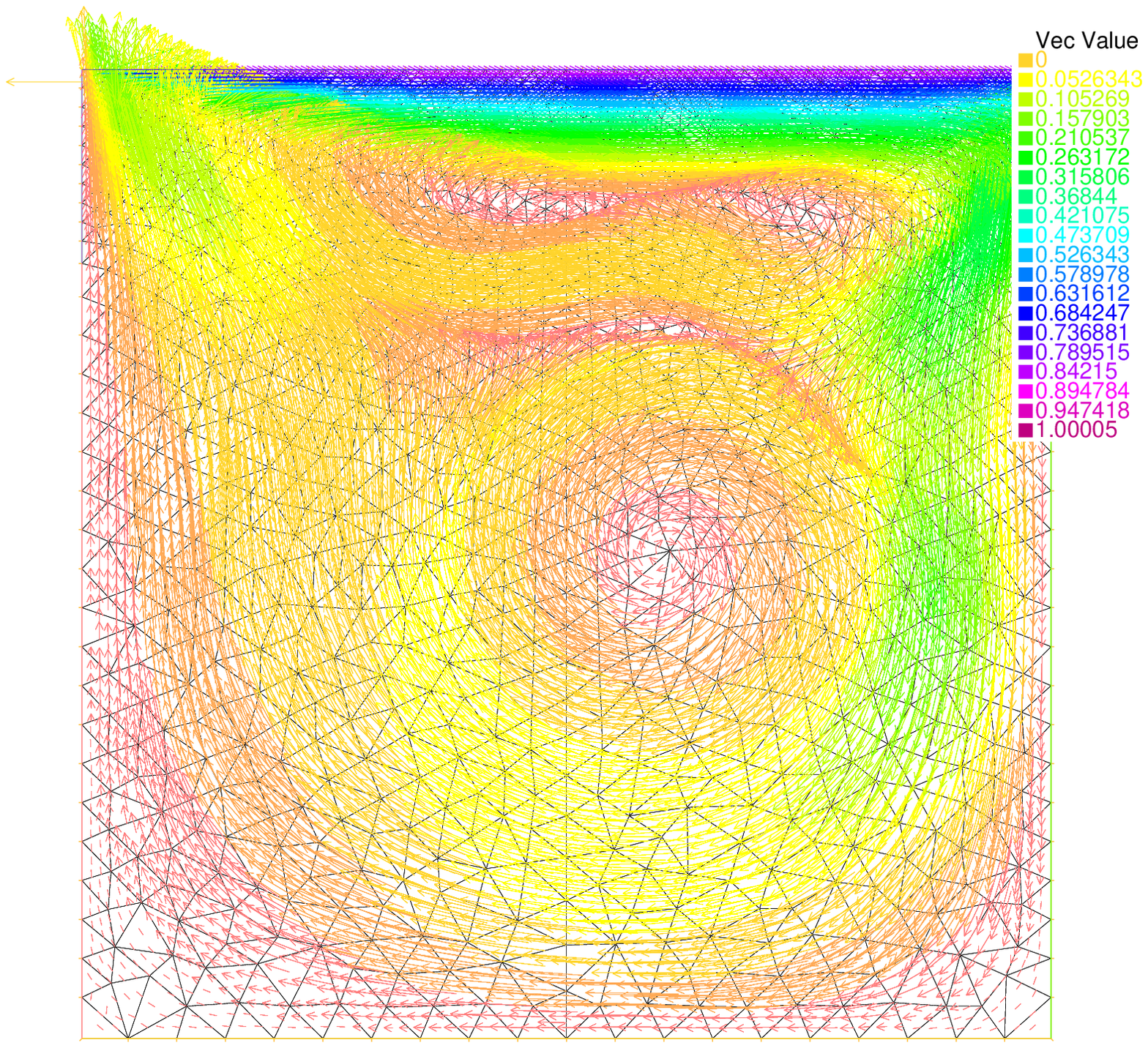}	
		\captionsetup{justification=centering}
		\caption* {\scriptsize (b) Veocity field shown by arrows.}
	\end{minipage}	
	\captionsetup{justification=centering}
	\caption {\scriptsize A control force holds and pushes the solid disc to rotate at its inital position; control at $t=5$ using a coarse mesh.} 
	\label{cavity_disc_roate}
\end{figure}

\begin{figure}[h!]
	\begin{minipage}[t]{0.5\linewidth}
		\centering  
		\includegraphics[width=2.8in,angle=0]{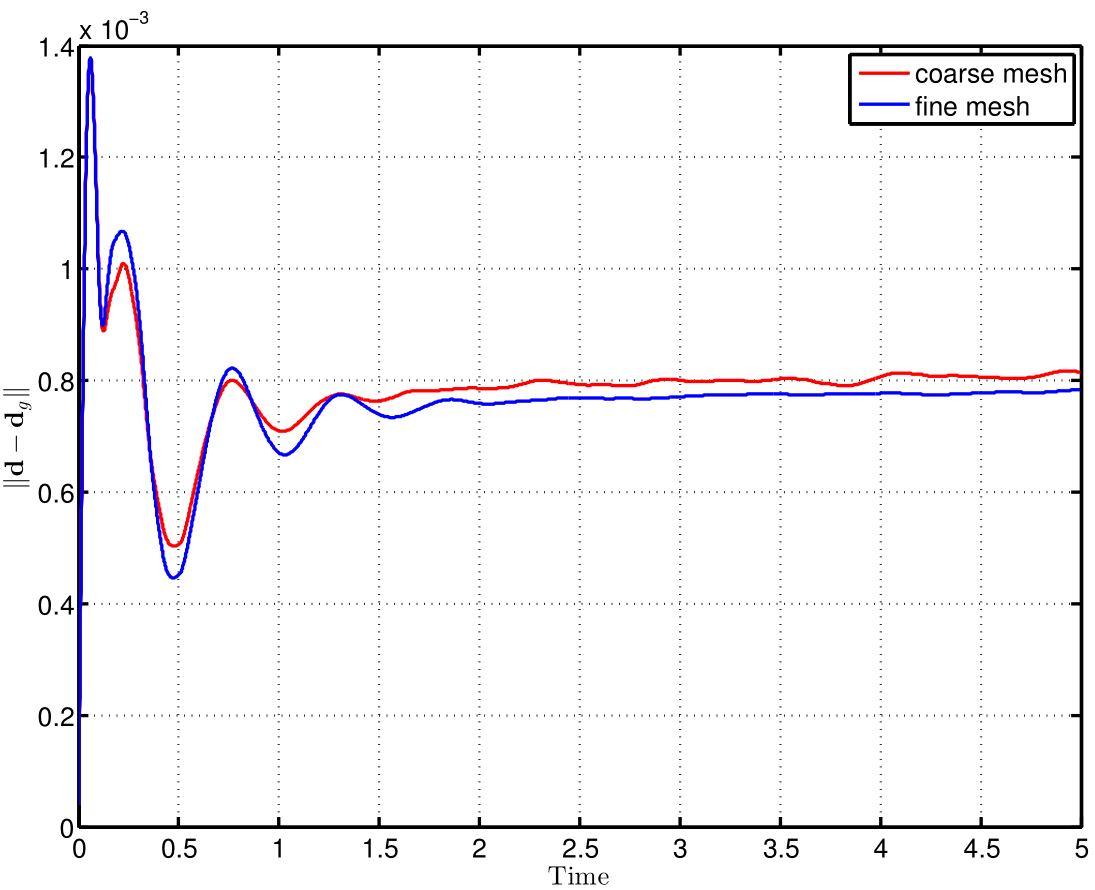}
		\captionsetup{justification=centering}
		\caption* {\scriptsize (a) Objective function as a function of time.}
	\end{minipage}     	
	\begin{minipage}[t]{0.5\linewidth}
		\centering  
		\includegraphics[width=2.8in,angle=0]{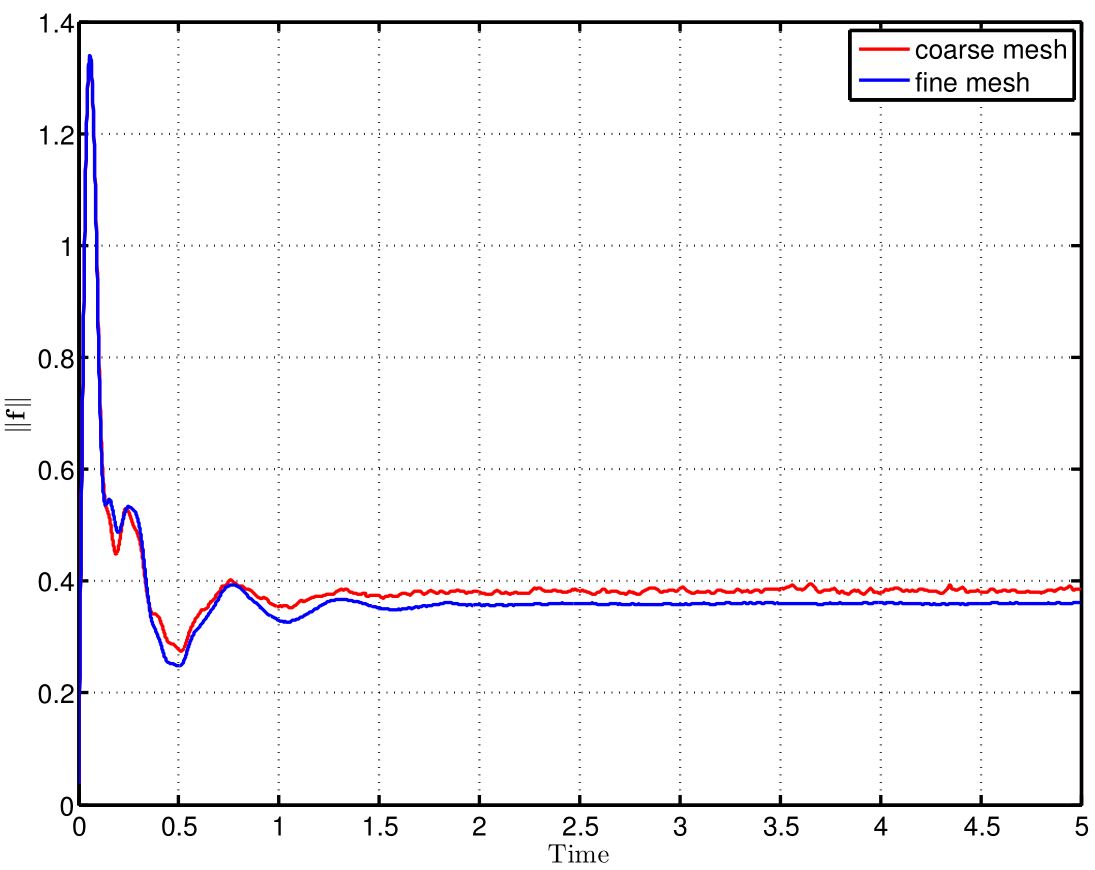}	
		\captionsetup{justification=centering}
		\caption* {\scriptsize (b) Control force as a function of time.}
	\end{minipage}	
	\captionsetup{justification=centering}
	\caption {\scriptsize A control force holds and pushes the solid disc to rotate at its inital position.} 
	\label{cavity_disc_err_f}
\end{figure}

Without any control, the movement of the solid is dominated by the surrounding fluid, which ends up rotating near the top of the cavity as shown in Figure \ref{cavity_disc_uncontrolled}. For this numerical test, we consider another challenging control of the solid: computing an appropriate force to hold the solid disc at its initial position $(x_0, y_0)$ and push it to rotate there without moving away. The objective displacement can be expressed as:
\[
\begin{pmatrix}
d_{x} \\
d_{y} 
\end{pmatrix}
= 
\left[
\begin{array}{ccc}
cos\left(\omega t\right) & -sin\left(\omega t\right)  \\
sin\left(\omega t\right) & cos\left(\omega t\right) \\
\end{array}
\right]
\begin{pmatrix}
x-x_0\\
y-y_0 \\
\end{pmatrix},
\]
with $\omega=-\pi/4$ being the angular velocity of the rotating disc we want to control. For this case, because the objective function is time dependent we find that a converged time step size is smaller: $\Delta t=0.001$. Using this time step, we presents the controlled velocity field at $t=2$ in Figure \ref{cavity_disc_roate}, from which it can be seen that the movement of the solid now dominates the cavity flow and a large vortex is created by the rotating disc. We also find that the solid disc gradually and slowly shifts away from its initial position using the previous coarse mesh as shown in Figure \ref{cavity_disc_err_f} (a), which presents the convergence of the objective function. However this shift becomes insignificant by using a finer mesh: 9618 triangles with 4950 vertices for the fluid, and 2570 triangles with 1346 vertices for the solid. We plot the $L^2$-norm of the control force in Figure \ref{cavity_disc_err_f} (b), from which it can be seen that the force dynamically responses to the error of the control and gradually approaches to a stable magnitude when the solid disc becomes tractable.

\section{Conclusion}
\label{sec_conclusion}
It is challenging to solve time-dependent FSI control problems with large solid deformations and very few examples have appeared in the literature. In this paper, we formulate a monolithic optimal control approach in the framework of piecewise-in-time control, which is stable for a range of regularisation parameters and efficient in reducing the displacement-tracking type of objective function; we consider an inequality constraint of the magnitude of the solid velocity, so that the solid speed can also be controlled when tracking its displacement; the proposed FSI control formulation, together with a reduced formulation for pure flow control problems, is first assessed by a typical flow control and a benchmark FSI problem, and then applied to a very challenging FSI control problem involving complicated movement and large deformation of the solid; all the numerical tests are implemented in open-source software package FreeFEM++ and shared via public Github site.

There are relevant topics which are interesting for future studies: it is traditionally proved that piecewise-in-time control is effective in dealing with velocity-tracking type of objectives \cite{abergel1990some,hou1997numerical} (now also displacement-tracking demonstrated in this paper), it is interesting to investigate other types of objective functions, such as reduction of drag force, based upon the proposed monolithic scheme; it is also worth investigating other types of control parameters, such as force distribution at the interface between the solid and surrounding fluids which would be very useful for accurate design and control of biologically inspired robots, such as swimming robots \cite{crespi2008controlling} or micro medical robots \cite{xiao2019classifications}.

\appendix
\section{Replication of results}
\label{sec_code}
The following FreeFEM code is an implementation of numerical test in Section \ref{subsec_flag}. A complete FreeFEM code for all the numerical tests can be found in the Github repository: \hyperlink{https://github.com/yongxingwang/}{https://github.com/yongxingwang/}.

~\\{\HighlightA }{\HighlightB //~FSI~control~using~two~meshes}{\HighlightA 
	\\}{\HighlightG real}{\HighlightA ~}{\HighlightF muf}{\HighlightA =1,}{\HighlightF c1}{\HighlightA =2.e6,}{\HighlightF g}{\HighlightA =-0,}{\HighlightF rhof}{\HighlightA =1.e3,}{\HighlightF rhos}{\HighlightA =1.e3,}{\HighlightF rhod}{\HighlightA =}{\HighlightF rhos}{\HighlightA -}{\HighlightF rhof}{\HighlightA ;
	\\}{\HighlightG real}{\HighlightA ~}{\HighlightF dt}{\HighlightA =0.001,~}{\HighlightF t}{\HighlightA ,~}{\HighlightF T0}{\HighlightA =3,~}{\HighlightF Tc}{\HighlightA =4,~}{\HighlightF mus}{\HighlightA =}{\HighlightF dt}{\HighlightA *}{\HighlightF c1}{\HighlightA -}{\HighlightF muf}{\HighlightA ;
	\\}{\HighlightG real}{\HighlightA ~}{\HighlightF x0}{\HighlightA =0.2,~}{\HighlightF y0}{\HighlightA =0.2,~}{\HighlightF r}{\HighlightA =0.05,~}{\HighlightF L}{\HighlightA =2.5,~}{\HighlightF H}{\HighlightA =0.41,~}{\HighlightF h}{\HighlightA =0.02,~}{\HighlightF l}{\HighlightA =0.35,~}{\HighlightF theta}{\HighlightA =0.2013579208;
	\\}{\HighlightG real}{\HighlightA ~}{\HighlightF x1}{\HighlightA =}{\HighlightF x0}{\HighlightA +}{\HighlightF r}{\HighlightA *}{\HighlightG cos}{\HighlightA (}{\HighlightF theta}{\HighlightA ),~}{\HighlightF y1}{\HighlightA =}{\HighlightF y0}{\HighlightA -}{\HighlightF h}{\HighlightA /2,~}{\HighlightF x2}{\HighlightA =}{\HighlightF x1}{\HighlightA ,~}{\HighlightF y2}{\HighlightA =}{\HighlightF y0}{\HighlightA +}{\HighlightF h}{\HighlightA /2;
	\\}{\HighlightG real}{\HighlightA ~}{\HighlightF alpha}{\HighlightA =1.e-17,~}{\HighlightF xtip}{\HighlightA =0.6,}{\HighlightF ytip}{\HighlightA =}{\HighlightF y0}{\HighlightA ,}{\HighlightF xotip}{\HighlightA =}{\HighlightF xtip}{\HighlightA ,}{\HighlightF yotip}{\HighlightA =}{\HighlightF ytip}{\HighlightA ;
	\\}{\HighlightG int}{\HighlightA ~}{\HighlightF mh}{\HighlightA =3,~}{\HighlightF m}{\HighlightA =20;
	\\}{\HighlightB //fluid~region}{\HighlightA 
	\\}{\HighlightG border}{\HighlightA ~}{\HighlightF a1}{\HighlightA (}{\HighlightF t}{\HighlightA =0,}{\HighlightF L}{\HighlightA )~\{}{\HighlightG x}{\HighlightA =}{\HighlightF t}{\HighlightA ;~}{\HighlightG y}{\HighlightA =0~;}{\HighlightG label}{\HighlightA =1;\};
	\\}{\HighlightG border}{\HighlightA ~}{\HighlightF a2}{\HighlightA (}{\HighlightF t}{\HighlightA =0,}{\HighlightF H}{\HighlightA )~\{}{\HighlightG x}{\HighlightA =}{\HighlightF L}{\HighlightA ;~}{\HighlightG y}{\HighlightA =}{\HighlightF t}{\HighlightA ~;}{\HighlightG label}{\HighlightA =2;\};
	\\}{\HighlightG border}{\HighlightA ~}{\HighlightF a3}{\HighlightA (}{\HighlightF t}{\HighlightA =}{\HighlightF L}{\HighlightA ,0)~\{}{\HighlightG x}{\HighlightA =}{\HighlightF t}{\HighlightA ;~}{\HighlightG y}{\HighlightA =}{\HighlightF H}{\HighlightA ~;}{\HighlightG label}{\HighlightA =1;\};
	\\}{\HighlightG border}{\HighlightA ~}{\HighlightF a4}{\HighlightA (}{\HighlightF t}{\HighlightA =}{\HighlightF H}{\HighlightA ,0)~\{}{\HighlightG x}{\HighlightA =0;~}{\HighlightG y}{\HighlightA =}{\HighlightF t}{\HighlightA ~;}{\HighlightG label}{\HighlightA =3;\};
	\\}{\HighlightB //hole}{\HighlightA 
	\\}{\HighlightG border}{\HighlightA ~}{\HighlightF disc}{\HighlightA (}{\HighlightF t}{\HighlightA =0,~2*}{\HighlightG pi}{\HighlightA )~~\{}{\HighlightG x}{\HighlightA =}{\HighlightF x0}{\HighlightA +}{\HighlightF r}{\HighlightA *}{\HighlightG cos}{\HighlightA (}{\HighlightF t}{\HighlightA );~}{\HighlightG y}{\HighlightA =}{\HighlightF y0}{\HighlightA +}{\HighlightF r}{\HighlightA *}{\HighlightG sin}{\HighlightA (}{\HighlightF t}{\HighlightA );~}{\HighlightG label}{\HighlightA =4;\};~~
	\\}{\HighlightB //lines~to~refine~mesh}{\HighlightA 
	\\}{\HighlightG border}{\HighlightA ~}{\HighlightF l0}{\HighlightA (}{\HighlightF t}{\HighlightA =0.25,0.6)~\{}{\HighlightG x}{\HighlightA =}{\HighlightF t}{\HighlightA ;~}{\HighlightG y}{\HighlightA =}{\HighlightF y0}{\HighlightA ;}{\HighlightG label}{\HighlightA =5;\};
	\\}{\HighlightG border}{\HighlightA ~}{\HighlightF l1}{\HighlightA (}{\HighlightF t}{\HighlightA =}{\HighlightF x2}{\HighlightA ,0.6)~\{}{\HighlightG x}{\HighlightA =}{\HighlightF t}{\HighlightA ;~}{\HighlightG y}{\HighlightA =0.04*(}{\HighlightG x}{\HighlightA -}{\HighlightF x2}{\HighlightA )/}{\HighlightF l}{\HighlightA +}{\HighlightF y2}{\HighlightA ;~~}{\HighlightG label}{\HighlightA =5;\};
	\\}{\HighlightG border}{\HighlightA ~}{\HighlightF l2}{\HighlightA (}{\HighlightF t}{\HighlightA =}{\HighlightF x1}{\HighlightA ,0.6)~\{}{\HighlightG x}{\HighlightA =}{\HighlightF t}{\HighlightA ;~}{\HighlightG y}{\HighlightA =-0.04*(}{\HighlightG x}{\HighlightA -}{\HighlightF x1}{\HighlightA )/}{\HighlightF l}{\HighlightA +}{\HighlightF y1}{\HighlightA ;~}{\HighlightG label}{\HighlightA =5;\};
	\\}{\HighlightG mesh}{\HighlightA ~}{\HighlightF Th}{\HighlightA ~=~}{\HighlightG buildmesh}{\HighlightA (}{\HighlightF a1}{\HighlightA (}{\HighlightF L}{\HighlightA *}{\HighlightF m}{\HighlightA /}{\HighlightF H}{\HighlightA )+}{\HighlightF a2}{\HighlightA (}{\HighlightF m}{\HighlightA )+}{\HighlightF a3}{\HighlightA (}{\HighlightF L}{\HighlightA *}{\HighlightF m}{\HighlightA /}{\HighlightF H}{\HighlightA )+}{\HighlightF a4}{\HighlightA (}{\HighlightF m}{\HighlightA )
	\\+}{\HighlightF disc}{\HighlightA (-}{\HighlightG pi}{\HighlightA *}{\HighlightF mh}{\HighlightA /}{\HighlightF theta}{\HighlightA )+}{\HighlightF l0}{\HighlightA (}{\HighlightF l}{\HighlightA *}{\HighlightF mh}{\HighlightA /}{\HighlightF h}{\HighlightA )+}{\HighlightF l1}{\HighlightA (}{\HighlightF l}{\HighlightA *}{\HighlightF mh}{\HighlightA /}{\HighlightF h}{\HighlightA )+}{\HighlightF l2}{\HighlightA (}{\HighlightF l}{\HighlightA *}{\HighlightF mh}{\HighlightA /}{\HighlightF h}{\HighlightA ));
	\\}{\HighlightB //solid~region}{\HighlightA 
	\\}{\HighlightG border}{\HighlightA ~}{\HighlightF b1}{\HighlightA (}{\HighlightF t}{\HighlightA =}{\HighlightF x1}{\HighlightA ,0.6)~\{}{\HighlightG x}{\HighlightA =}{\HighlightF t}{\HighlightA ;~}{\HighlightG y}{\HighlightA =}{\HighlightF y1}{\HighlightA ~;}{\HighlightG label}{\HighlightA =5;\};
	\\}{\HighlightG border}{\HighlightA ~}{\HighlightF b2}{\HighlightA (}{\HighlightF t}{\HighlightA =}{\HighlightF y1}{\HighlightA ,}{\HighlightF y2}{\HighlightA )~\{}{\HighlightG x}{\HighlightA =0.6;~}{\HighlightG y}{\HighlightA =}{\HighlightF t}{\HighlightA ~;}{\HighlightG label}{\HighlightA =5;\};
	\\}{\HighlightG border}{\HighlightA ~}{\HighlightF b3}{\HighlightA (}{\HighlightF t}{\HighlightA =0.6,}{\HighlightF x2}{\HighlightA )~\{}{\HighlightG x}{\HighlightA =}{\HighlightF t}{\HighlightA ;~}{\HighlightG y}{\HighlightA =}{\HighlightF y2}{\HighlightA ~;}{\HighlightG label}{\HighlightA =5;\};
	\\}{\HighlightG border}{\HighlightA ~}{\HighlightF b4}{\HighlightA (}{\HighlightF t}{\HighlightA =}{\HighlightF theta}{\HighlightA ,-}{\HighlightF theta}{\HighlightA )~\{}{\HighlightG x}{\HighlightA =}{\HighlightF x0}{\HighlightA +}{\HighlightF r}{\HighlightA *}{\HighlightG cos}{\HighlightA (}{\HighlightF t}{\HighlightA );~}{\HighlightG y}{\HighlightA =}{\HighlightF y0}{\HighlightA +}{\HighlightF r}{\HighlightA *}{\HighlightG sin}{\HighlightA (}{\HighlightF t}{\HighlightA );}{\HighlightG label}{\HighlightA =5;\};
	\\}{\HighlightG mesh}{\HighlightA ~}{\HighlightF Ths}{\HighlightA ~=~}{\HighlightG buildmesh}{\HighlightA (}{\HighlightF b1}{\HighlightA (}{\HighlightF l}{\HighlightA *}{\HighlightF mh}{\HighlightA /}{\HighlightF h}{\HighlightA )+}{\HighlightF b2}{\HighlightA (}{\HighlightF mh}{\HighlightA )+}{\HighlightF b3}{\HighlightA (}{\HighlightF l}{\HighlightA *}{\HighlightF mh}{\HighlightA /}{\HighlightF h}{\HighlightA )+}{\HighlightF b4}{\HighlightA (}{\HighlightF mh}{\HighlightA ));
	\\}{\HighlightG plot}{\HighlightA (}{\HighlightF Th}{\HighlightA ,~}{\HighlightF Ths}{\HighlightA ,~}{\HighlightG wait}{\HighlightA =1);
	\\
	\\}{\HighlightG mesh}{\HighlightA ~}{\HighlightF Thso}{\HighlightA =}{\HighlightF Ths}{\HighlightA ,~}{\HighlightF Ths0}{\HighlightA =}{\HighlightF Ths}{\HighlightA ;
	\\}{\HighlightG fespace}{\HighlightA ~}{\HighlightE Vh}{\HighlightA (}{\HighlightF Th}{\HighlightA ,}{\HighlightG P2}{\HighlightA );
	\\}{\HighlightG fespace}{\HighlightA ~}{\HighlightE Ph}{\HighlightA (}{\HighlightF Th}{\HighlightA ,}{\HighlightG P1}{\HighlightA );
	\\}{\HighlightG fespace}{\HighlightA ~}{\HighlightE Vhs}{\HighlightA (}{\HighlightF Ths}{\HighlightA ,}{\HighlightG P2}{\HighlightA );
	\\}{\HighlightG fespace}{\HighlightA ~}{\HighlightE Vhso}{\HighlightA (}{\HighlightF Thso}{\HighlightA ,}{\HighlightG P2}{\HighlightA );
	\\}{\HighlightG fespace}{\HighlightA ~}{\HighlightE Rh}{\HighlightA (}{\HighlightF Th}{\HighlightA ,$[$}{\HighlightG P2}{\HighlightA ,}{\HighlightG P2}{\HighlightA ,}{\HighlightG P1}{\HighlightA $]$);
	\\}{\HighlightG fespace}{\HighlightA ~}{\HighlightE RhAdj}{\HighlightA (}{\HighlightF Th}{\HighlightA ,$[$}{\HighlightG P2}{\HighlightA ,}{\HighlightG P2}{\HighlightA ,}{\HighlightG P2}{\HighlightA ,}{\HighlightG P2}{\HighlightA ,}{\HighlightG P1}{\HighlightA ,}{\HighlightG P1}{\HighlightA $]$);
	\\}{\HighlightG fespace}{\HighlightA ~}{\HighlightE Rhs}{\HighlightA (}{\HighlightF Ths}{\HighlightA ,$[$}{\HighlightG P2}{\HighlightA ,}{\HighlightG P2}{\HighlightA $]$);
	\\}{\HighlightG fespace}{\HighlightA ~}{\HighlightE RhsAdj}{\HighlightA (}{\HighlightF Ths}{\HighlightA ,$[$}{\HighlightG P2}{\HighlightA ,}{\HighlightG P2}{\HighlightA ,}{\HighlightG P2}{\HighlightA ,}{\HighlightG P2}{\HighlightA $]$);
	\\
	\\}{\HighlightE Ph}{\HighlightA ~}{\HighlightF p}{\HighlightA ,}{\HighlightF ph}{\HighlightA ,}{\HighlightF phat}{\HighlightA ,}{\HighlightF phath}{\HighlightA ;
	\\}{\HighlightE Vh}{\HighlightA ~}{\HighlightF u}{\HighlightA ,}{\HighlightF v}{\HighlightA ,}{\HighlightF uhat}{\HighlightA ,}{\HighlightF vhat}{\HighlightA ,}{\HighlightF uh}{\HighlightA ,}{\HighlightF vh}{\HighlightA ,}{\HighlightF uhath}{\HighlightA ,}{\HighlightF vhath}{\HighlightA ,}{\HighlightF uold}{\HighlightA =0,}{\HighlightF vold}{\HighlightA =0,}{\HighlightF uu}{\HighlightA ;
	\\}{\HighlightE Vhs}{\HighlightA ~}{\HighlightF us}{\HighlightA ,}{\HighlightF vs}{\HighlightA ,}{\HighlightF ushat}{\HighlightA =0,}{\HighlightF vshat}{\HighlightA =0,}{\HighlightF ush}{\HighlightA ,}{\HighlightF vsh}{\HighlightA ,}{\HighlightF ushath}{\HighlightA ,}{\HighlightF vshath}{\HighlightA ,}{\HighlightF usold}{\HighlightA =0,}{\HighlightF vsold}{\HighlightA =0,}{\HighlightF d1}{\HighlightA =0,}{\HighlightF d2}{\HighlightA =0,}{\HighlightF dg1}{\HighlightA =0,}{\HighlightF dg2}{\HighlightA =0;~
	\\}{\HighlightE Vhso}{\HighlightA ~}{\HighlightF uso}{\HighlightA ,}{\HighlightF vso}{\HighlightA ,}{\HighlightF usohat}{\HighlightA ,}{\HighlightF vsohat}{\HighlightA ,}{\HighlightF do1}{\HighlightA ,}{\HighlightF do2}{\HighlightA ;;
	\\
	\\}{\HighlightC macro~div(u,v)~(~dx(u)+dy(v)~)~//~EOM}{\HighlightA 
	\\}{\HighlightC macro~DD(u,v)~~$[$$[$2*dx(u),div(v,u)$]$,$[$div(v,u),2*dy(v)$]$$]$~//~EOM}{\HighlightA 
	\\}{\HighlightC macro~Grad(u,v)$[$$[$dx(u),dy(u)$]$,$[$dx(v),dy(v)$]$$]$~//~EOM}{\HighlightA 
	\\
	\\}{\HighlightG varf}{\HighlightA ~}{\HighlightF fluid}{\HighlightA ($[$}{\HighlightF u}{\HighlightA ,}{\HighlightF v}{\HighlightA ,}{\HighlightF p}{\HighlightA $]$,$[$}{\HighlightF uh}{\HighlightA ,}{\HighlightF vh}{\HighlightA ,}{\HighlightF ph}{\HighlightA $]$)~=
	\\	}{\HighlightG int2d}{\HighlightA (}{\HighlightF Th}{\HighlightA )(}{\HighlightF rhof}{\HighlightA *$[$}{\HighlightF u}{\HighlightA ,}{\HighlightF v}{\HighlightA $]$'*$[$}{\HighlightF uh}{\HighlightA ,}{\HighlightF vh}{\HighlightA $]$/}{\HighlightF dt}{\HighlightA -}{\HighlightC div(uh,vh)}{\HighlightA *}{\HighlightF p}{\HighlightA -}{\HighlightC div(u,v)}{\HighlightA *}{\HighlightF ph}{\HighlightA ~
	\\	+~}{\HighlightF muf}{\HighlightA /2*}{\HighlightG trace}{\HighlightA (}{\HighlightC DD(u,v)}{\HighlightA '*}{\HighlightC DD(uh,vh)}{\HighlightA ))
	\\	+~}{\HighlightG on}{\HighlightA (1,}{\HighlightF u}{\HighlightA =0,~}{\HighlightF v}{\HighlightA =0)~+~}{\HighlightG on}{\HighlightA (3,}{\HighlightF u}{\HighlightA =12*}{\HighlightG y}{\HighlightA *(}{\HighlightF H}{\HighlightA -}{\HighlightG y}{\HighlightA )/}{\HighlightF H}{\HighlightA /}{\HighlightF H}{\HighlightA ,}{\HighlightF v}{\HighlightA =0)~+~}{\HighlightG on}{\HighlightA (4,}{\HighlightF u}{\HighlightA =0,}{\HighlightF v}{\HighlightA =0);
	\\
	\\}{\HighlightG varf}{\HighlightA ~}{\HighlightF resf}{\HighlightA ($[$}{\HighlightF u}{\HighlightA ,}{\HighlightF v}{\HighlightA ,}{\HighlightF p}{\HighlightA $]$,$[$}{\HighlightF uh}{\HighlightA ,}{\HighlightF vh}{\HighlightA ,}{\HighlightF ph}{\HighlightA $]$)~=
	\\	}{\HighlightG int2d}{\HighlightA (}{\HighlightF Th}{\HighlightA )(}{\HighlightF g}{\HighlightA *}{\HighlightF rhof}{\HighlightA *}{\HighlightF vh}{\HighlightA +}{\HighlightF rhof}{\HighlightA *$[$}{\HighlightG convect}{\HighlightA ($[$}{\HighlightF uold}{\HighlightA ,}{\HighlightF vold}{\HighlightA $]$,-}{\HighlightF dt}{\HighlightA ,}{\HighlightF uold}{\HighlightA ),
	\\	}{\HighlightG convect}{\HighlightA ($[$}{\HighlightF uold}{\HighlightA ,}{\HighlightF vold}{\HighlightA $]$,-}{\HighlightF dt}{\HighlightA ,}{\HighlightF vold}{\HighlightA )$]$'*$[$}{\HighlightF uh}{\HighlightA ,}{\HighlightF vh}{\HighlightA $]$/}{\HighlightF dt}{\HighlightA ~)
	\\	+~}{\HighlightG on}{\HighlightA (1,}{\HighlightF u}{\HighlightA =0,~}{\HighlightF v}{\HighlightA =0)~+~}{\HighlightG on}{\HighlightA (3,}{\HighlightF u}{\HighlightA =12*}{\HighlightG y}{\HighlightA *(}{\HighlightF H}{\HighlightA -}{\HighlightG y}{\HighlightA )/}{\HighlightF H}{\HighlightA /}{\HighlightF H}{\HighlightA ,}{\HighlightF v}{\HighlightA =0)~+~}{\HighlightG on}{\HighlightA (4,}{\HighlightF u}{\HighlightA =0,}{\HighlightF v}{\HighlightA =0);
	\\
	\\}{\HighlightG varf}{\HighlightA ~}{\HighlightF solid}{\HighlightA ($[$}{\HighlightF us}{\HighlightA ,}{\HighlightF vs}{\HighlightA $]$,$[$}{\HighlightF ush}{\HighlightA ,}{\HighlightF vsh}{\HighlightA $]$)~=
	\\	}{\HighlightG int2d}{\HighlightA (}{\HighlightF Ths}{\HighlightA )(~}{\HighlightF rhod}{\HighlightA *$[$}{\HighlightF us}{\HighlightA ,}{\HighlightF vs}{\HighlightA $]$'*$[$}{\HighlightF ush}{\HighlightA ,}{\HighlightF vsh}{\HighlightA $]$/}{\HighlightF dt}{\HighlightA 
	\\	+}{\HighlightF mus}{\HighlightA /2*}{\HighlightG trace}{\HighlightA (}{\HighlightC DD(us,vs)}{\HighlightA '*}{\HighlightC DD(ush,vsh)}{\HighlightA )
	\\	-}{\HighlightF dt}{\HighlightA *}{\HighlightF c1}{\HighlightA *}{\HighlightG trace}{\HighlightA ((}{\HighlightC Grad(us,vs)}{\HighlightA '*}{\HighlightC Grad(d1,d2)}{\HighlightA +}{\HighlightC Grad(d1,d2)}{\HighlightA '*}{\HighlightC Grad(us,vs)}{\HighlightA )*}{\HighlightC Grad(ush,vsh)}{\HighlightA '))~;
	\\
	\\}{\HighlightG varf}{\HighlightA ~}{\HighlightF ress}{\HighlightA ($[$}{\HighlightF us}{\HighlightA ,}{\HighlightF vs}{\HighlightA $]$,$[$}{\HighlightF ush}{\HighlightA ,}{\HighlightF vsh}{\HighlightA $]$)~=	
	\\	}{\HighlightG int2d}{\HighlightA (}{\HighlightF Ths}{\HighlightA )(~}{\HighlightF g}{\HighlightA *}{\HighlightF rhod}{\HighlightA *}{\HighlightF vsh}{\HighlightA -}{\HighlightF c1}{\HighlightA *}{\HighlightG trace}{\HighlightA ((}{\HighlightC DD(d1,d2)}{\HighlightA -}{\HighlightC Grad(d1,d2)}{\HighlightA '*}{\HighlightC Grad(d1,d2)}{\HighlightA )*}{\HighlightC Grad(ush,vsh)}{\HighlightA ')~~
	\\	+~}{\HighlightF rhod}{\HighlightA *$[$}{\HighlightF usold}{\HighlightA ,}{\HighlightF vsold}{\HighlightA $]$'*$[$}{\HighlightF ush}{\HighlightA ,}{\HighlightF vsh}{\HighlightA $]$/}{\HighlightF dt}{\HighlightA ~);
	\\
	\\}{\HighlightG matrix}{\HighlightA ~}{\HighlightF A}{\HighlightA ~=~}{\HighlightF fluid}{\HighlightA (}{\HighlightE Rh}{\HighlightA ,}{\HighlightE Rh}{\HighlightA );
	\\
	\\}{\HighlightG ofstream}{\HighlightA ~}{\HighlightF file0}{\HighlightA (}{\HighlightD "tip\_disp.txt"}{\HighlightA );
	\\}{\HighlightF file0}{\HighlightA .}{\HighlightG precision}{\HighlightA (16);
	\\}{\HighlightG for}{\HighlightA (}{\HighlightF t}{\HighlightA =}{\HighlightF dt}{\HighlightA ;}{\HighlightF t}{\HighlightA $<$}{\HighlightF T0}{\HighlightA ;}{\HighlightF t}{\HighlightA +=}{\HighlightF dt}{\HighlightA )\{
	\\	}{\HighlightG real}{\HighlightA $[$}{\HighlightG int}{\HighlightA $]$~}{\HighlightF rhs1}{\HighlightA ~=~}{\HighlightF resf}{\HighlightA (0,}{\HighlightE Rh}{\HighlightA );
	\\	}{\HighlightG real}{\HighlightA $[$}{\HighlightG int}{\HighlightA $]$~}{\HighlightF rhs2}{\HighlightA ~=~}{\HighlightF ress}{\HighlightA (0,}{\HighlightE Rhs}{\HighlightA );
	\\	}{\HighlightG matrix}{\HighlightA ~}{\HighlightF A}{\HighlightA ~=~}{\HighlightF fluid}{\HighlightA (}{\HighlightE Rh}{\HighlightA ,}{\HighlightE Rh}{\HighlightA );
	\\	}{\HighlightG matrix}{\HighlightA ~}{\HighlightF B}{\HighlightA ~=~}{\HighlightF solid}{\HighlightA (}{\HighlightE Rhs}{\HighlightA ,}{\HighlightE Rhs}{\HighlightA );
	\\	}{\HighlightG matrix}{\HighlightA ~}{\HighlightF P}{\HighlightA ~=~}{\HighlightG interpolate}{\HighlightA (}{\HighlightE Rhs}{\HighlightA ,}{\HighlightE Rh}{\HighlightA );
	\\	}{\HighlightG real}{\HighlightA $[$}{\HighlightG int}{\HighlightA $]$~}{\HighlightF rhs}{\HighlightA ~=~}{\HighlightF P}{\HighlightA '*}{\HighlightF rhs2}{\HighlightA ;
	\\	}{\HighlightF rhs}{\HighlightA ~+=~}{\HighlightF rhs1}{\HighlightA ;
	\\	}{\HighlightG matrix}{\HighlightA ~}{\HighlightF T}{\HighlightA ~=~}{\HighlightF P}{\HighlightA '*}{\HighlightF B}{\HighlightA ;
	\\	}{\HighlightG matrix}{\HighlightA ~}{\HighlightF AB}{\HighlightA ~=~}{\HighlightF T}{\HighlightA *}{\HighlightF P}{\HighlightA ;
	\\	}{\HighlightF AB}{\HighlightA +=}{\HighlightF A}{\HighlightA ;
	\\	
	\\	}{\HighlightG set}{\HighlightA (}{\HighlightF AB}{\HighlightA ,}{\HighlightG solver}{\HighlightA =}{\HighlightG UMFPACK}{\HighlightA );
	\\	}{\HighlightE Rh}{\HighlightA ~$[$}{\HighlightF w1}{\HighlightA ,~}{\HighlightF w2}{\HighlightA ,~}{\HighlightF wp}{\HighlightA $]$;
	\\	}{\HighlightG real}{\HighlightA $[$}{\HighlightG int}{\HighlightA $]$~}{\HighlightF sol}{\HighlightA (}{\HighlightE Rh}{\HighlightA .}{\HighlightG ndof}{\HighlightA );
	\\	}{\HighlightF sol}{\HighlightA =~}{\HighlightF w1}{\HighlightA $[$$]$;~}{\HighlightF sol}{\HighlightA ~=~}{\HighlightF AB}{\HighlightA \textasciicircum{}-1~*~}{\HighlightF rhs}{\HighlightA ;
	\\	}{\HighlightF w1}{\HighlightA $[$$]$=}{\HighlightF sol}{\HighlightA ;~}{\HighlightF u}{\HighlightA =}{\HighlightF w1}{\HighlightA ;~}{\HighlightF v}{\HighlightA =~}{\HighlightF w2}{\HighlightA ;~}{\HighlightF p}{\HighlightA =}{\HighlightF wp}{\HighlightA ;
	\\
	\\	}{\HighlightF Thso}{\HighlightA =}{\HighlightF Ths}{\HighlightA ;
	\\	}{\HighlightF uso}{\HighlightA =}{\HighlightF u}{\HighlightA ;~}{\HighlightF vso}{\HighlightA =}{\HighlightF v}{\HighlightA ;~}{\HighlightF do1}{\HighlightA =}{\HighlightF d1}{\HighlightA ;~}{\HighlightF do2}{\HighlightA =}{\HighlightF d2}{\HighlightA ;
	\\	
	\\	}{\HighlightF xtip}{\HighlightA ~+=~}{\HighlightF uso}{\HighlightA (}{\HighlightF xotip}{\HighlightA ,}{\HighlightF yotip}{\HighlightA )*}{\HighlightF dt}{\HighlightA ;~}{\HighlightF ytip}{\HighlightA ~+=~}{\HighlightF vso}{\HighlightA (}{\HighlightF xotip}{\HighlightA ,}{\HighlightF yotip}{\HighlightA )*}{\HighlightF dt}{\HighlightA ;
	\\	}{\HighlightF xotip}{\HighlightA =}{\HighlightF xtip}{\HighlightA ;}{\HighlightF yotip}{\HighlightA =}{\HighlightF ytip}{\HighlightA ;		
	\\	
	\\	}{\HighlightF Ths}{\HighlightA ~=~}{\HighlightG movemesh}{\HighlightA (}{\HighlightF Ths}{\HighlightA ,~$[$}{\HighlightG x}{\HighlightA +}{\HighlightF us}{\HighlightA *}{\HighlightF dt}{\HighlightA ,~}{\HighlightG y}{\HighlightA +}{\HighlightF vs}{\HighlightA *}{\HighlightF dt}{\HighlightA $]$);
	\\
	\\	}{\HighlightF d1}{\HighlightA =0;~~}{\HighlightF d1}{\HighlightA $[$$]$=}{\HighlightF do1}{\HighlightA $[$$]$+}{\HighlightF uso}{\HighlightA $[$$]$*}{\HighlightF dt}{\HighlightA ;~
	\\	}{\HighlightF d2}{\HighlightA =0;~~}{\HighlightF d2}{\HighlightA $[$$]$=}{\HighlightF do2}{\HighlightA $[$$]$+}{\HighlightF vso}{\HighlightA $[$$]$*}{\HighlightF dt}{\HighlightA ;~
	\\	}{\HighlightF us}{\HighlightA =0;~~}{\HighlightF us}{\HighlightA $[$$]$=}{\HighlightF uso}{\HighlightA $[$$]$;
	\\	}{\HighlightF vs}{\HighlightA =0;~~}{\HighlightF vs}{\HighlightA $[$$]$=}{\HighlightF vso}{\HighlightA $[$$]$;
	\\	
	\\	}{\HighlightF uold}{\HighlightA =}{\HighlightF u}{\HighlightA ;}{\HighlightF vold}{\HighlightA =}{\HighlightF v}{\HighlightA ;~}{\HighlightF usold}{\HighlightA =}{\HighlightF us}{\HighlightA ;}{\HighlightF vsold}{\HighlightA =}{\HighlightF vs}{\HighlightA ;
	\\	
	\\	}{\HighlightF uu}{\HighlightA =}{\HighlightG sqrt}{\HighlightA (}{\HighlightF u}{\HighlightA \textasciicircum{}2+}{\HighlightF v}{\HighlightA \textasciicircum{}2);
	\\	}{\HighlightG plot}{\HighlightA (}{\HighlightF uu}{\HighlightA ,}{\HighlightF Ths}{\HighlightA ,}{\HighlightG coef}{\HighlightA =0.1,}{\HighlightG fill}{\HighlightA =1,}{\HighlightG value}{\HighlightA =1,}{\HighlightG wait}{\HighlightA =0);
	\\	
	\\	}{\HighlightF file0}{\HighlightA ~$<$$<$}{\HighlightF d2}{\HighlightA (}{\HighlightF xtip}{\HighlightA ,}{\HighlightF ytip}{\HighlightA )$<$$<$~}{\HighlightG endl}{\HighlightA ;	
	\\	}{\HighlightG cout}{\HighlightA $<$$<$}{\HighlightD "NS~Time:~"}{\HighlightA $<$$<$}{\HighlightF t}{\HighlightA $<$$<$}{\HighlightG endl}{\HighlightA ;
	\\\}
	\\
	\\}{\HighlightG varf}{\HighlightA ~}{\HighlightF NSAdj}{\HighlightA ($[$}{\HighlightF u}{\HighlightA ,}{\HighlightF v}{\HighlightA ,}{\HighlightF uhat}{\HighlightA ,}{\HighlightF vhat}{\HighlightA ,}{\HighlightF p}{\HighlightA ,}{\HighlightF phat}{\HighlightA $]$,$[$}{\HighlightF uh}{\HighlightA ,}{\HighlightF vh}{\HighlightA ,}{\HighlightF uhath}{\HighlightA ,}{\HighlightF vhath}{\HighlightA ,}{\HighlightF ph}{\HighlightA ,}{\HighlightF phath}{\HighlightA $]$)~=
	\\	}{\HighlightG int2d}{\HighlightA (}{\HighlightF Th}{\HighlightA )(}{\HighlightF rhof}{\HighlightA *$[$}{\HighlightF u}{\HighlightA ,}{\HighlightF v}{\HighlightA $]$'*$[$}{\HighlightF uh}{\HighlightA ,}{\HighlightF vh}{\HighlightA $]$/}{\HighlightF dt}{\HighlightA -}{\HighlightC div(uh,vh)}{\HighlightA *}{\HighlightF p}{\HighlightA -}{\HighlightC div(u,v)}{\HighlightA *}{\HighlightF ph}{\HighlightA 
	\\	+}{\HighlightF rhof}{\HighlightA *$[$}{\HighlightF uhat}{\HighlightA ,}{\HighlightF vhat}{\HighlightA $]$'*$[$}{\HighlightF uhath}{\HighlightA ,}{\HighlightF vhath}{\HighlightA $]$/}{\HighlightF dt}{\HighlightA -}{\HighlightC div(uhath,vhath)}{\HighlightA *}{\HighlightF phat}{\HighlightA -}{\HighlightC div(uhat,vhat)}{\HighlightA *}{\HighlightF phath}{\HighlightA 
	\\	+}{\HighlightF rhof}{\HighlightA *$[$}{\HighlightF uhath}{\HighlightA ,}{\HighlightF vhath}{\HighlightA $]$'*(}{\HighlightC Grad(uold,vold)}{\HighlightA '*$[$}{\HighlightF uhat}{\HighlightA ,}{\HighlightF vhat}{\HighlightA $]$)
	\\	-}{\HighlightF rhof}{\HighlightA *$[$}{\HighlightF uhath}{\HighlightA ,}{\HighlightF vhath}{\HighlightA $]$'*(}{\HighlightC Grad(uhat,vhat)}{\HighlightA ~*$[$}{\HighlightF uold}{\HighlightA ,}{\HighlightF vold}{\HighlightA $]$)	
	\\	+}{\HighlightF muf}{\HighlightA /2*}{\HighlightG trace}{\HighlightA (}{\HighlightC DD(u,v)}{\HighlightA '*}{\HighlightC DD(uh,vh)}{\HighlightA )
	\\	+}{\HighlightF muf}{\HighlightA /2*}{\HighlightG trace}{\HighlightA (}{\HighlightC DD(uhat,vhat)}{\HighlightA '*}{\HighlightC DD(uhath,vhath)}{\HighlightA ))
	\\	+}{\HighlightG on}{\HighlightA (1,}{\HighlightF u}{\HighlightA =0,}{\HighlightF v}{\HighlightA =0,}{\HighlightF uhat}{\HighlightA =0,}{\HighlightF vhat}{\HighlightA =0)~+~}{\HighlightG on}{\HighlightA (3,}{\HighlightF u}{\HighlightA =12*}{\HighlightG y}{\HighlightA *(}{\HighlightF H}{\HighlightA -}{\HighlightG y}{\HighlightA )/}{\HighlightF H}{\HighlightA /}{\HighlightF H}{\HighlightA ,}{\HighlightF v}{\HighlightA =0,}{\HighlightF uhat}{\HighlightA =0,}{\HighlightF vhat}{\HighlightA =0)~
	\\	+}{\HighlightG on}{\HighlightA (4,}{\HighlightF u}{\HighlightA =0,}{\HighlightF v}{\HighlightA =0,}{\HighlightF uhat}{\HighlightA =0,}{\HighlightF vhat}{\HighlightA =0);
	\\
	\\}{\HighlightG varf}{\HighlightA ~}{\HighlightF resNSAdj}{\HighlightA ($[$}{\HighlightF u}{\HighlightA ,}{\HighlightF v}{\HighlightA ,}{\HighlightF uhat}{\HighlightA ,}{\HighlightF vhat}{\HighlightA ,}{\HighlightF p}{\HighlightA ,}{\HighlightF phat}{\HighlightA $]$,$[$}{\HighlightF uh}{\HighlightA ,}{\HighlightF vh}{\HighlightA ,}{\HighlightF uhath}{\HighlightA ,}{\HighlightF vhath}{\HighlightA ,}{\HighlightF ph}{\HighlightA ,}{\HighlightF phath}{\HighlightA $]$)~=
	\\	}{\HighlightG int2d}{\HighlightA (}{\HighlightF Th}{\HighlightA )(}{\HighlightF g}{\HighlightA *}{\HighlightF rhof}{\HighlightA *}{\HighlightF vh}{\HighlightA +}{\HighlightF rhof}{\HighlightA *$[$}{\HighlightG convect}{\HighlightA ($[$}{\HighlightF uold}{\HighlightA ,}{\HighlightF vold}{\HighlightA $]$,-}{\HighlightF dt}{\HighlightA ,}{\HighlightF uold}{\HighlightA ),
	\\	}{\HighlightG convect}{\HighlightA ($[$}{\HighlightF uold}{\HighlightA ,}{\HighlightF vold}{\HighlightA $]$,-}{\HighlightF dt}{\HighlightA ,}{\HighlightF vold}{\HighlightA )$]$'*$[$}{\HighlightF uh}{\HighlightA ,}{\HighlightF vh}{\HighlightA $]$/}{\HighlightF dt}{\HighlightA )
	\\	+}{\HighlightG on}{\HighlightA (1,}{\HighlightF u}{\HighlightA =0,}{\HighlightF v}{\HighlightA =0,}{\HighlightF uhat}{\HighlightA =0,}{\HighlightF vhat}{\HighlightA =0)~+~}{\HighlightG on}{\HighlightA (3,}{\HighlightF u}{\HighlightA =12*}{\HighlightG y}{\HighlightA *(}{\HighlightF H}{\HighlightA -}{\HighlightG y}{\HighlightA )/}{\HighlightF H}{\HighlightA /}{\HighlightF H}{\HighlightA ,}{\HighlightF v}{\HighlightA =0,}{\HighlightF uhat}{\HighlightA =0,}{\HighlightF vhat}{\HighlightA =0)~
	\\	+}{\HighlightG on}{\HighlightA (4,}{\HighlightF u}{\HighlightA =0,}{\HighlightF v}{\HighlightA =0,}{\HighlightF uhat}{\HighlightA =0,}{\HighlightF vhat}{\HighlightA =0);
	\\
	\\}{\HighlightG varf}{\HighlightA ~}{\HighlightF solidAdj}{\HighlightA ($[$}{\HighlightF us}{\HighlightA ,}{\HighlightF vs}{\HighlightA ,}{\HighlightF ushat}{\HighlightA ,}{\HighlightF vshat}{\HighlightA $]$,$[$}{\HighlightF ush}{\HighlightA ,}{\HighlightF vsh}{\HighlightA ,}{\HighlightF ushath}{\HighlightA ,}{\HighlightF vshath}{\HighlightA $]$)~=
	\\	}{\HighlightG int2d}{\HighlightA (}{\HighlightF Ths}{\HighlightA )(}{\HighlightF rhod}{\HighlightA *$[$}{\HighlightF us}{\HighlightA ,}{\HighlightF vs}{\HighlightA $]$'*$[$}{\HighlightF ush}{\HighlightA ,}{\HighlightF vsh}{\HighlightA $]$/}{\HighlightF dt}{\HighlightA ~
	\\	+}{\HighlightF rhod}{\HighlightA *$[$}{\HighlightF ushath}{\HighlightA ,}{\HighlightF vshath}{\HighlightA $]$'*$[$}{\HighlightF ushat}{\HighlightA ,}{\HighlightF vshat}{\HighlightA $]$/}{\HighlightF dt}{\HighlightA 
	\\	+}{\HighlightF mus}{\HighlightA /2*}{\HighlightG trace}{\HighlightA (}{\HighlightC DD(us,vs)}{\HighlightA '*}{\HighlightC DD(ush,vsh)}{\HighlightA )
	\\	+}{\HighlightF mus}{\HighlightA /2*}{\HighlightG trace}{\HighlightA (}{\HighlightC DD(ushat,vshat)}{\HighlightA '*}{\HighlightC DD(ushath,vshath)}{\HighlightA )	
	\\	-}{\HighlightF dt}{\HighlightA *}{\HighlightF c1}{\HighlightA *}{\HighlightG trace}{\HighlightA ((}{\HighlightC Grad(us,vs)}{\HighlightA '*}{\HighlightC Grad(d1,d2)}{\HighlightA +}{\HighlightC Grad(d1,d2)}{\HighlightA '*}{\HighlightC Grad(us,vs)}{\HighlightA )*}{\HighlightC Grad(ush,vsh)}{\HighlightA ')
	\\	-}{\HighlightF dt}{\HighlightA *}{\HighlightF c1}{\HighlightA *}{\HighlightG trace}{\HighlightA ((}{\HighlightC Grad(ushath,vshath)}{\HighlightA '*}{\HighlightC Grad(d1,d2)}{\HighlightA +}{\HighlightC Grad(d1,d2)}{\HighlightA '
	\\	*}{\HighlightC Grad(ushath,vshath)}{\HighlightA )*}{\HighlightC Grad(ushat,vshat)}{\HighlightA ')
	\\	-$[$}{\HighlightF ushat}{\HighlightA ,}{\HighlightF vshat}{\HighlightA $]$'*$[$}{\HighlightF ush}{\HighlightA ,}{\HighlightF vsh}{\HighlightA $]$/}{\HighlightF alpha}{\HighlightA );
	\\
	\\}{\HighlightG varf}{\HighlightA ~}{\HighlightF ressAdj}{\HighlightA ($[$}{\HighlightF us}{\HighlightA ,}{\HighlightF vs}{\HighlightA ,}{\HighlightF ushat}{\HighlightA ,}{\HighlightF vshat}{\HighlightA $]$,$[$}{\HighlightF ush}{\HighlightA ,}{\HighlightF vsh}{\HighlightA ,}{\HighlightF ushath}{\HighlightA ,}{\HighlightF vshath}{\HighlightA $]$)~=
	\\	}{\HighlightG int2d}{\HighlightA (}{\HighlightF Ths}{\HighlightA )(}{\HighlightF g}{\HighlightA *}{\HighlightF rhod}{\HighlightA *}{\HighlightF vsh}{\HighlightA ~+~}{\HighlightF c1}{\HighlightA *}{\HighlightG trace}{\HighlightA ((}{\HighlightC Grad(d1,d2)}{\HighlightA '*}{\HighlightC Grad(d1,d2)}{\HighlightA )*}{\HighlightC Grad(ush,vsh)}{\HighlightA ')~~
	\\	-0.5*}{\HighlightF c1}{\HighlightA *}{\HighlightG trace}{\HighlightA (}{\HighlightC DD(d1,d2)}{\HighlightA *}{\HighlightC DD(ush,vsh)}{\HighlightA ')
	\\	+}{\HighlightF rhod}{\HighlightA *$[$}{\HighlightF usold}{\HighlightA ,}{\HighlightF vsold}{\HighlightA $]$'*$[$}{\HighlightF ush}{\HighlightA ,}{\HighlightF vsh}{\HighlightA $]$/}{\HighlightF dt}{\HighlightA ~
	\\	-}{\HighlightF dt}{\HighlightA *$[$}{\HighlightF d1}{\HighlightA -}{\HighlightF dg1}{\HighlightA ,}{\HighlightF d2}{\HighlightA -}{\HighlightF dg2}{\HighlightA $]$'*$[$}{\HighlightF ushath}{\HighlightA ,}{\HighlightF vshath}{\HighlightA $]$);
	\\
	\\}{\HighlightG matrix}{\HighlightA ~}{\HighlightF Aadj}{\HighlightA ~=~}{\HighlightF NSAdj}{\HighlightA (}{\HighlightE RhAdj}{\HighlightA ,}{\HighlightE RhAdj}{\HighlightA );
	\\
	\\}{\HighlightG ofstream}{\HighlightA ~}{\HighlightF file1}{\HighlightA (}{\HighlightD "objective\_force.txt"}{\HighlightA );
	\\}{\HighlightF file1}{\HighlightA .}{\HighlightG precision}{\HighlightA (16);
	\\}{\HighlightG for}{\HighlightA (}{\HighlightF t}{\HighlightA =}{\HighlightF T0}{\HighlightA ;}{\HighlightF t}{\HighlightA $<$}{\HighlightF Tc}{\HighlightA ;}{\HighlightF t}{\HighlightA +=}{\HighlightF dt}{\HighlightA )\{
	\\	}{\HighlightG real}{\HighlightA $[$}{\HighlightG int}{\HighlightA $]$~}{\HighlightF rhs1}{\HighlightA ~=~}{\HighlightF resNSAdj}{\HighlightA (0,}{\HighlightE RhAdj}{\HighlightA );
	\\	}{\HighlightG matrix}{\HighlightA ~}{\HighlightF Badj}{\HighlightA ~=~}{\HighlightF solidAdj}{\HighlightA (}{\HighlightE RhsAdj}{\HighlightA ,}{\HighlightE RhsAdj}{\HighlightA );
	\\	}{\HighlightG real}{\HighlightA $[$}{\HighlightG int}{\HighlightA $]$~}{\HighlightF rhs2}{\HighlightA ~=~}{\HighlightF ressAdj}{\HighlightA (0,}{\HighlightE RhsAdj}{\HighlightA );
	\\	}{\HighlightG matrix}{\HighlightA ~}{\HighlightF P}{\HighlightA ~=~}{\HighlightG interpolate}{\HighlightA (}{\HighlightE RhsAdj}{\HighlightA ,}{\HighlightE RhAdj}{\HighlightA );
	\\	}{\HighlightG real}{\HighlightA $[$}{\HighlightG int}{\HighlightA $]$~}{\HighlightF rhs}{\HighlightA ~=~}{\HighlightF P}{\HighlightA '*}{\HighlightF rhs2}{\HighlightA ;
	\\	}{\HighlightF rhs}{\HighlightA ~+=~}{\HighlightF rhs1}{\HighlightA ;
	\\	}{\HighlightG matrix}{\HighlightA ~}{\HighlightF T}{\HighlightA ~=~}{\HighlightF P}{\HighlightA '*}{\HighlightF Badj}{\HighlightA ;
	\\	}{\HighlightG matrix}{\HighlightA ~}{\HighlightF AB}{\HighlightA ~=~}{\HighlightF T}{\HighlightA *}{\HighlightF P}{\HighlightA ;
	\\	}{\HighlightF AB}{\HighlightA ~+=~}{\HighlightF Aadj}{\HighlightA ;	
	\\	
	\\	}{\HighlightG set}{\HighlightA (}{\HighlightF AB}{\HighlightA ,}{\HighlightG solver}{\HighlightA =}{\HighlightG UMFPACK}{\HighlightA );
	\\	}{\HighlightE RhAdj}{\HighlightA ~$[$}{\HighlightF w1}{\HighlightA ,~}{\HighlightF w2}{\HighlightA ,~}{\HighlightF s1}{\HighlightA ,~}{\HighlightF s2}{\HighlightA ,~}{\HighlightF wp}{\HighlightA ,~}{\HighlightF sp}{\HighlightA $]$;
	\\	}{\HighlightG real}{\HighlightA $[$}{\HighlightG int}{\HighlightA $]$~}{\HighlightF sol}{\HighlightA (}{\HighlightE RhAdj}{\HighlightA .}{\HighlightG ndof}{\HighlightA );
	\\	}{\HighlightF sol}{\HighlightA =~}{\HighlightF w1}{\HighlightA $[$$]$;~}{\HighlightF sol}{\HighlightA ~=~}{\HighlightF AB}{\HighlightA \textasciicircum{}-1~*~}{\HighlightF rhs}{\HighlightA ;~}{\HighlightF w1}{\HighlightA $[$$]$=}{\HighlightF sol}{\HighlightA ;~	
	\\	}{\HighlightF u}{\HighlightA =}{\HighlightF w1}{\HighlightA ;~}{\HighlightF v}{\HighlightA =}{\HighlightF w2}{\HighlightA ;~}{\HighlightF uhat}{\HighlightA =}{\HighlightF s1}{\HighlightA ;~}{\HighlightF vhat}{\HighlightA =~}{\HighlightF s2}{\HighlightA ;~}{\HighlightF p}{\HighlightA =}{\HighlightF wp}{\HighlightA ;~}{\HighlightF phat}{\HighlightA =}{\HighlightF sp}{\HighlightA ;
	\\
	\\	}{\HighlightF Thso}{\HighlightA =}{\HighlightF Ths}{\HighlightA ;
	\\	}{\HighlightF uso}{\HighlightA =}{\HighlightF u}{\HighlightA ;~}{\HighlightF vso}{\HighlightA =}{\HighlightF v}{\HighlightA ;~}{\HighlightF usohat}{\HighlightA =}{\HighlightF uhat}{\HighlightA ;~}{\HighlightF vsohat}{\HighlightA =}{\HighlightF vhat}{\HighlightA ;
	\\	}{\HighlightF do1}{\HighlightA =}{\HighlightF d1}{\HighlightA ;~}{\HighlightF do2}{\HighlightA =}{\HighlightF d2}{\HighlightA ;
	\\	
	\\	}{\HighlightF xtip}{\HighlightA ~+=~}{\HighlightF uso}{\HighlightA (}{\HighlightF xotip}{\HighlightA ,}{\HighlightF yotip}{\HighlightA )*}{\HighlightF dt}{\HighlightA ;~}{\HighlightF ytip}{\HighlightA ~+=~}{\HighlightF vso}{\HighlightA (}{\HighlightF xotip}{\HighlightA ,}{\HighlightF yotip}{\HighlightA )*}{\HighlightF dt}{\HighlightA ;
	\\	}{\HighlightF xotip}{\HighlightA =}{\HighlightF xtip}{\HighlightA ;}{\HighlightF yotip}{\HighlightA =}{\HighlightF ytip}{\HighlightA ;			
	\\	
	\\	}{\HighlightF Ths}{\HighlightA ~=~}{\HighlightG movemesh}{\HighlightA (}{\HighlightF Ths}{\HighlightA ,~$[$}{\HighlightG x}{\HighlightA +}{\HighlightF us}{\HighlightA *}{\HighlightF dt}{\HighlightA ,~}{\HighlightG y}{\HighlightA +}{\HighlightF vs}{\HighlightA *}{\HighlightF dt}{\HighlightA $]$);
	\\	
	\\	}{\HighlightF d1}{\HighlightA =0;~~}{\HighlightF d1}{\HighlightA $[$$]$=}{\HighlightF do1}{\HighlightA $[$$]$+}{\HighlightF uso}{\HighlightA $[$$]$*}{\HighlightF dt}{\HighlightA ;~
	\\	}{\HighlightF d2}{\HighlightA =0;~~}{\HighlightF d2}{\HighlightA $[$$]$=}{\HighlightF do2}{\HighlightA $[$$]$+}{\HighlightF vso}{\HighlightA $[$$]$*}{\HighlightF dt}{\HighlightA ;~
	\\	}{\HighlightF us}{\HighlightA =0;~~}{\HighlightF us}{\HighlightA $[$$]$=}{\HighlightF uso}{\HighlightA $[$$]$;
	\\	}{\HighlightF vs}{\HighlightA =0;~~}{\HighlightF vs}{\HighlightA $[$$]$=}{\HighlightF vso}{\HighlightA $[$$]$;		
	\\	}{\HighlightF ushat}{\HighlightA =0;~}{\HighlightF ushat}{\HighlightA $[$$]$=}{\HighlightF usohat}{\HighlightA $[$$]$;~
	\\	}{\HighlightF vshat}{\HighlightA =0;~}{\HighlightF vshat}{\HighlightA $[$$]$=}{\HighlightF vsohat}{\HighlightA $[$$]$;
	\\	
	\\	}{\HighlightF uold}{\HighlightA =}{\HighlightF u}{\HighlightA ;}{\HighlightF vold}{\HighlightA =}{\HighlightF v}{\HighlightA ;}{\HighlightF usold}{\HighlightA =}{\HighlightF us}{\HighlightA ;}{\HighlightF vsold}{\HighlightA =}{\HighlightF vs}{\HighlightA ;
	\\
	\\	}{\HighlightF uu}{\HighlightA =}{\HighlightG sqrt}{\HighlightA (}{\HighlightF u}{\HighlightA \textasciicircum{}2+}{\HighlightF v}{\HighlightA \textasciicircum{}2);
	\\	}{\HighlightG plot}{\HighlightA (}{\HighlightF uu}{\HighlightA ,}{\HighlightF Ths}{\HighlightA ,~}{\HighlightG coef}{\HighlightA =10,}{\HighlightG fill}{\HighlightA =1,}{\HighlightG value}{\HighlightA =1,}{\HighlightG wait}{\HighlightA =0);
	\\	
	\\	}{\HighlightG real}{\HighlightA ~}{\HighlightF error}{\HighlightA =}{\HighlightG sqrt}{\HighlightA (}{\HighlightG int2d}{\HighlightA (}{\HighlightF Ths}{\HighlightA )((}{\HighlightF d1}{\HighlightA -}{\HighlightF dg1}{\HighlightA )\textasciicircum{}2+(}{\HighlightF d2}{\HighlightA -}{\HighlightF dg2}{\HighlightA )\textasciicircum{}2));
	\\	}{\HighlightG real}{\HighlightA ~}{\HighlightF force}{\HighlightA =}{\HighlightG sqrt}{\HighlightA (}{\HighlightG int2d}{\HighlightA (}{\HighlightF Ths}{\HighlightA )(}{\HighlightF ushat}{\HighlightA \textasciicircum{}2+}{\HighlightF vshat}{\HighlightA \textasciicircum{}2))/}{\HighlightF alpha}{\HighlightA ;
	\\	}{\HighlightF file1}{\HighlightA ~$<$$<$~}{\HighlightF error}{\HighlightA ~$<$$<$}{\HighlightD "~"}{\HighlightA $<$$<$~}{\HighlightF force}{\HighlightA ~$<$$<$~}{\HighlightG endl}{\HighlightA ;
	\\	}{\HighlightF file0}{\HighlightA ~$<$$<$}{\HighlightF d2}{\HighlightA (}{\HighlightF xtip}{\HighlightA ,}{\HighlightF ytip}{\HighlightA )$<$$<$~}{\HighlightG endl}{\HighlightA ;		
	\\	}{\HighlightG cout}{\HighlightA $<$$<$}{\HighlightD "Control~Time:~"}{\HighlightA $<$$<$}{\HighlightF t}{\HighlightA $<$$<$}{\HighlightD "~~"}{\HighlightA $<$$<$}{\HighlightF error}{\HighlightA $<$$<$}{\HighlightG endl}{\HighlightA ;
	\\\}
	\\}{\HighlightF file0}{\HighlightA .}{\HighlightG flush}{\HighlightA ;
	\\}{\HighlightF file1}{\HighlightA .}{\HighlightG flush}{\HighlightA ;}

\bibliography{mybibfile}
\end{document}